\documentclass{jpsj3}

\usepackage{endnotes}
\usepackage{color}
\renewcommand{\Re}{\mathop{\textrm{Re}}}

\title{Resonant Spectrum Analysis of the Conductance\\of an Open Quantum System
and Three Types of  Fano Parameter}

\author{Keita SASADA$^1$,
Naomichi HATANO$^2$\thanks{E-mail address: hatano@iis.u-tokyo.ac.jp} and
Gonzalo ORDONEZ$^3$\thanks{E-mail address: gordonez@butler.edu}}

\inst{$^1$Department of Physics and $^2$Institute of Industrial Science, University of Tokyo,\\
Komaba, Meguro, Tokyo 153-8505, Japan\\
and\\
$^3$Department of Physics and Astronomy, Butler University,\\
4600 Sunset Ave., Indianapolis, IN 46208, USA.}

\abst{We explain the Fano peak (an asymmetric resonance peak) as an interference effect involving resonant states.
We reveal that there are three types of Fano asymmetry according to their origins:
the interference between a resonant state and an anti-resonant state, that between a resonant state and a bound state, and that between two resonant states.
We show that the last two show the asymmetric energy dependence given by Fano, but the first one shows a slightly different form.
In order to show the above, we analytically and microscopically derive a formula where we express the conductance purely in terms of the summation over all discrete eigenstates including resonant states and anti-resonant states, without any background integrals.
We thereby obtain microscopic expressions of the Fano parameters that describe the three types of the Fano asymmetry.
One of the expressions indicates that the corresponding Fano parameter becomes complex under an external magnetic field.
}

\kword{Resonant state, Fano parameter, Landauer formula, Aharonov-Bohm effect}
\begin{document}
\maketitle

\section{Introduction}\label{sec:intro}
The electronic conduction in nano-scale systems has been studied extensively in recent years~\cite{TB1993,Datta95,RW2001,CWB2001,LLZ2005,Chakrabarti2006,JHS2007,FN2008,KAKI2002,KAKI2003,SAKKI2004,SAKKI2005,GGHKSMM2000,ZGGKK2001,KKLPSKKYK2003,BS2004,Brisker08,SH2005,HSNP2007}.
The resonant transport is one of its interesting phenomena, where resonant states affect the conductance in its ballistic transport regime.
Resonance is an intrinsic feature of open systems~\cite{Gamow28,Siegert1939,Peierls59,leCouteur60,Zeldovich60,Hokkyo65,Romo68,Berggren70,Gyarmati71,Romo80,Berggren82,Berggren96,Madrid05,Eckart30,Bethe36,Hulthen42a,Hulthen42b,Jost51,Vogt54,Wigner55,Corinaldesi56,Nussenzveig59,Fivel60,Humblet61,Rosenfeld61,Humblet62,Humblet64-1,Jeukenne64,Humblet64-2,Mahaux65,Rosenfeld65,Bhattacharjie62,Wojtczak63,Spector64,Bose64,Aly65,McVoy67,Bahethi71,Fuda71,Bawin74,Doolen78,Narnhofer81,Rittby82,Alhassid85,Massmann85,Colbert86,Benjamin86,Amrein87,Bohm89,Ginocchio94,Rakityansky96,Homma97,Masui99,Barkay01,Carvalho02,Razavy03,Ahmed04,Kelkar04,Jain05,Amrein06,SH2005,HSNP2007,Moiseyev08,Rotter09}.
When we use nano-devices, we inevitably attach leads to them.
Hence the devices are always open systems and have resonant states; an electron comes into the device through a lead, is trapped in the
confining potential of the device for a while with a finite lifetime, and may come out through another lead.

More specifically, resonance scattering has an effect on the electronic conduction through
the celebrated Landauer formula for a microscopic system.
The formula tells us that the conductance $\mathcal{G}$ of the system is proportional to the transmission probability $\mathcal{T}$ of the quantum scatterer:
\begin{align}
\mathcal{G}=\frac{2e^2}{h}\mathcal{T},
\end{align}
where $e$ is the charge unit and $h$ is the Planck constant.
This reduces the problem of the electronic conduction to the fundamental problem of quantum scattering.
Resonant states in the scattering problem thereby come into play in the electronic conduction.

Many textbook examples of the resonance peak of transmission probability $\mathcal{T}$ (or the conductance $\mathcal{G}$) are of the Lorentzian form.
However, Fano in his celebrated paper~\cite{Fano61} showed that the resonance peak can generally be of an asymmetric form
\begin{align}\label{eq-Fano}
\mathcal{G}(E)=\frac{2e^2}{h}\mathcal{T}(E)\sim\frac{(q+\tilde{E})^2}{1+\tilde{E}^2},
\end{align}
where $\tilde{E}$ is a dimensionless energy variable whose origin is set at the resonance point.
The newly introduced parameter $q$, which is now called the Fano parameter, specifies how asymmetric the peak is;
the peak reduces to the Lorentzian for $q=0$.
The asymmetric Fano resonance has been observed in various fields of physics.
For mesoscopic systems, K.~Kobayashi {\it et al.}\ observed Fano resonance peaks in the conductance through an Aharonov-Bohm system with a quantum dot~\cite{KAKI2002,KAKI2003} as well as through a T-shaped (or side-coupled) quantum dot~\cite{SAKKI2004,SAKKI2005}.

One of the main messages of the present paper is as follows:
we can explain the Fano asymmetry as an interference effect involving resonant states.
We will reveal in \S~\ref{subsec:T-shaped} and \S~\ref{subsec:Three} that there are in fact three types of Fano asymmetry according to their origins:
(i) the interference between a resonant state and an anti-resonant state;
(ii) the interference between a resonant-state pair (a resonant state and the corresponding anti-resonant state) and a bound state;
(iii) the interference between a resonant-state pair and another resonant-state pair.
We will claim in \S~\ref{subsec:T-shaped} that, though the second and the third types take the form of eq.~(\ref{eq-Fano}), the first type contains a term of a slightly different form,
\begin{align}\label{eq-Fano1}
\left(\frac{q+\tilde{E}}{1+\tilde{E}^2}\right)^2.
\end{align}

Fano's argument~\cite{Fano61} for the asymmetric resonance peaks was partially phenomenological in the following sense:
he considered a very general situation where one impurity level is coupled to a continuum in an arbitrary form and assumed that the system is diagonalized to produce one resonant state.
In strong contrast, we will microscopically analyze open quantum systems with a specific (but appropriately general) Hamiltonian in the present paper.
We are optimistic that the present argument may be generalized even to systems with interactions~\cite{Presilla97};
the Landauer formula has been partially extended to interacting cases with the use of many-body scattering states~\cite{Nishino09,Imamura09,Nishino11}.

There have been other approaches to the Fano asymmetry, for example, considering a matrix element between resonant states due to many-body interactions~\cite{Magunov03}.
In the present paper, however, we leave out many-body interactions and focus on the one-body problem.
In our work, the Fano asymmetry arises as a crossing term of the form $\mathop{\mathrm{Re}}A^\ast B$ in the square modulus $|A+B|^2$ of the sum of the wave function amplitudes $A$ and $B$.

In order to show the above point, in \S~\ref{sec:conductance} and \S~\ref{sec:spectrum}, we will analytically develop a resonant-state expansion of the conductance for open quantum-dot systems.
In other words, we will express the conductance purely in terms of the summation over all discrete eigenstates (eigenstates with point spectra), namely the resonant states, the anti-resonant states, the bound states and the anti-bound states.
(We will review these terminologies in \S~\ref{sec:resonant}.
Note that in some papers, the discrete states refer only to the bound states, but we do not use this terminology here.)
The summation over the discrete eigenstates comes into the conductance formula as squared and hence contains various crossing terms, or interference terms.
We will classify all the interference terms into the above three types.

We then realize that even a resonant state with a broad resonance width and equivalently with a short lifetime can manifest itself by causing the Fano asymmetry in neighboring resonant peaks;
an explicit example will be given in \S~\ref{subsec:Three}.
Such a very unstable resonant state is often ignored as unmeasurable.
The present result, however, suggests that we may be able to detect a broad, short-lived resonant state by analyzing the Fano asymmetry of nearby resonant peaks.

Another important message of the present paper is the following;
the resonant-state expansion that we will derive in \S~\ref{sec:spectrum} and use in the conductance formula in \S~\ref{sec:Fano}, does not contain any background integrals.
The expansion takes the following form (see \S~\ref{sec:spectrum} for details):
\begin{align}\label{eq03}
G^\textrm{R}(E)+G^\textrm{A}(E)=\sum_{n}\frac{|\psi_n\rangle\langle\tilde{\psi}_n|}{E-E_n},
\end{align}
where $G^\textrm{R}$ and $G^\textrm{A}$ on the left-hand side denote the retarded and advanced Green's functions, respectively, whereas the summation on the right-hand side is taken over all discrete eigenstates with $\psi_n$ and $\tilde{\psi}_n$ being the corresponding right- and left eigenvectors, respectively.

This is a remarkable fact from the viewpoint of common difficulties that the resonant expansion initiated by Berggren~\cite{Berggren68,Berggren70,Berggren96} usually faces.
The standard resonant-state expansion of a Green's function starts from the resolution of unity that R.G.~Newton proved~\cite{Newton1961}
\begin{align}\label{eq50}
1=\sum_p|\psi_p^\textrm{b}\rangle\langle\tilde{\psi}_p^\textrm{b}|+\int\frac{dk}{2\pi}|\psi_k\rangle\langle\tilde{\psi}_k|,
\end{align}
where, on the right-hand side, the summation is taken over the bound states and the integral is taken over an appropriate range.
When we apply the resolution of unity, eq.~\eqref{eq50}, to the Green's functions, we have
\begin{align}
G^\textrm{R/A}(E)=\sum_p\frac{|\psi_p^\textrm{b}\rangle\langle\tilde{\psi}_p^\textrm{b}|}{E-E_p\pm i\delta}
+\int\frac{dk}{2\pi}\frac{|\psi_k\rangle\langle\tilde{\psi}_k|}{E-E_k\pm i\delta}.
\end{align}
We can take into account some of the resonant states or the anti-resonant states by modifying the integration contour on the right-hand side in the complex $k$ plane~\cite{Berggren68,Berggren70,Berggren96}.
The integration, however, persists as a background integral no matter how we modify its contour.
Hence, most studies that consider the Fano asymmetry in eq.~\eqref{eq-Fano} had to use approximations by omitting the background integral at least.
In contrast, we will eliminate the background integral by summing up the retarded and advanced Green's functions.
Therefore, our treatment of the Fano asymmetry in \S~\ref{sec:Fano} is free from approximations, keeping all terms.

It would be useless, of course, if we were not able to express the conductance in terms of the {\it sum} of the retarded and advanced Green's functions.
In fact, we will derive from the Landauer formula, an expression of the conductance in terms of the two matrices given by
\begin{align}\label{eq06}
\Lambda&\equiv G^\textrm{R}+G^\textrm{A},
\\\label{eq07}
i\Gamma&\equiv \left(G^\textrm{R}\right)^{-1}-\left(G^\textrm{A}\right)^{-1},
\end{align}
not using each of $G^\textrm{R}$ and $G^\textrm{A}$ alone; see \S~\ref{sec:conductance} for details.
We will thereby be able to express the conductance in terms of the resonant-state expansion in eq.~\eqref{eq03}.
We again emphasize that there will be no background integrals in the expression that we will derive.
To our knowledge, this is the first example of such case (see ref.~\citen{endnote}).

The open quantum system that we will analyze hereafter is specific but general enough to account for various physically interesting systems.
For example, we can consider a system often called a ``T-shaped quantum dot" or a ``side-coupled quantum dot" as well as a side-coupled quantum-dot array~\cite{Kim01,Kikoin01,Kang01,Affleck01,Simon01,Affleck07,SAKKI2004,SAKKI2005,Torio02,Orellana03a,Orellana03b,Rodriguez03,MSU2004,Aligia04,Tanaka05,Lara05,Franco06,Wang06,Chakrabarti06,Zitko06,Li08}.
The T-shaped quantum dot has been experimentally realized and studied extensively.
A similar situation was also studied in the context of quantum wave guides~\cite{Porod92,Porod93,Shao94}.
One of the many interesting results is observation of Fano asymmetric peaks~\cite{SAKKI2004,SAKKI2005}.
This is one of the motivations of the present study.

The present paper is organized as follows.
In \S~\ref{sec:resonant}, we will review the theory of resonant states in open quantum systems.
We will introduce the terminologies such as the resonant states, the anti-resonant states, the bound states and the anti-bound states.
In \S~\ref{sec:conductance}, we will express the transmission probability (and hence the conductance) in terms of the two matrices $\Lambda$ and $\Gamma$ defined in eqs.~\eqref{eq06} and~\eqref{eq07}.
In order to do so, we will regard eqs.~\eqref{eq06} and~\eqref{eq07} as a set of simultaneous matrix equations and solve it with respect to $G^\textrm{R}$ and $G^\textrm{A}$.
In \S~\ref{sec:spectrum}, we will express, for an $N$-site open quantum-dot model, the retarded and advanced Green's functions in terms of the summation over all the discrete eigenstates.
Combining the results in \S~\ref{sec:conductance} and \S~\ref{sec:spectrum}, we will derive a conductance formula in terms of the summation over all discrete eigenstates without any background integrals.
In \S~\ref{sec:Fano}, we will show that the asymmetry of the Fano conductance peak arises from the interference between discrete states and classify them into the above-mentioned three types.
We will derive microscopic expressions of the Fano parameters that control the asymmetry of the Green's functions.
We will also argue that the thus-defined Fano parameter becomes complex in the presence of an external magnetic field that induces the Aharonov-Bohm effect.

\section{Resonant states}\label{sec:resonant}
As a preparation for the main part of the present paper, we review in this section mathematics of the resonant state as an eigenfunction of the time-independent Schr\"{o}dinger equation~\cite{HSNP2007}.
There are a dynamical view of resonance and a static one.
In the dynamical view, the resonance is described as follows;
a particle comes into a scattering potential, is captured for a while and escapes after a lifetime.
This time evolution is governed by the time-dependent Schr\"{o}dinger equation.
In the present paper, we focus on the static view, where resonance is described as an eigenstate of the time-independent Schr\"{o}dinger equation~\cite{HSNP2007}.

It is rather common in the static view to define a resonant state as a pole of the $S$ matrix.
In fact, there are mainly two ways of defining the resonant state in the static view.
(We have been notified that there is yet another way~\cite{Presilla96}.)
The definition based on the $S$ matrix may be called the indirect method~\cite{RBBM1997}.
We here use the direct method of its definition; that is, we describe it as an explicit eigenfunction of the Schr\"{o}dinger equation~\cite{Gamow28,Siegert1939,Peierls59,leCouteur60,Zeldovich60,Hokkyo65,Romo68,Berggren70,Gyarmati71,Landau77,Romo80,Berggren82,Berggren96,Madrid05,HSNP2007}.

Suppose that we have a scatterer with several semi-infinite leads attached to it.
For simplicity and concreteness, we hereafter restrict ourselves to the tight-binding model for the lead Hamiltonians.
The total Hamiltonian is of the form 
\begin{align}\label{eq90}
H=H_\textrm{d}+\sum_{\alpha}\left(H_\alpha +H_{\textrm{d}, \alpha}\right),
\end{align}
where $H_\textrm{d}$ is the one-body Hamiltonian of the scatterer (namely, the dot Hamiltonian), $H_{\alpha}$ is the Hamiltonian of a lead $\alpha$, and $H_{\textrm{d}, \alpha}$ is the coupling between the dot and the lead $\alpha$.
We assume that the leads are given by tight-binding models as
\begin{align}
H_{\alpha}=-t\sum_{x_{\alpha}=0}^{\infty}\left(|x_{\alpha}+1\rangle \langle x_{\alpha}|+|x_{\alpha}\rangle \langle x_{\alpha}+1|\right),
\end{align}
where $t>0$.
Therefore, the energy $E_k$ and the wave number $k$ of incoming and outgoing electrons are related through the dispersion relation
\begin{align}\label{eq110}
E_k\equiv-2t\cos k.
\end{align}
In other words, the band width is $4t$.
(The formulation throughout the present paper is basically unchanged in the wide-band limit, where we first shift the energy band to $[0,4t]$ and let $t$ to $\infty$.)
Specific examples of the system are given in \S~\ref{sec:Fano}.

\begin{figure}
\includegraphics[width=0.35\textwidth]{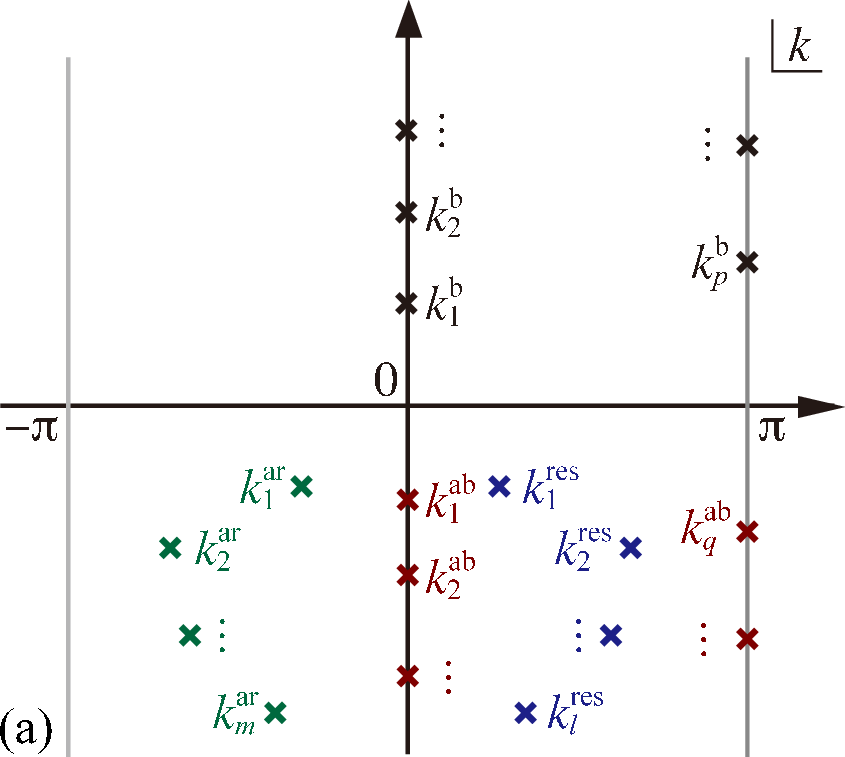}

\vspace*{\baselineskip}

\includegraphics[width=0.45\textwidth]{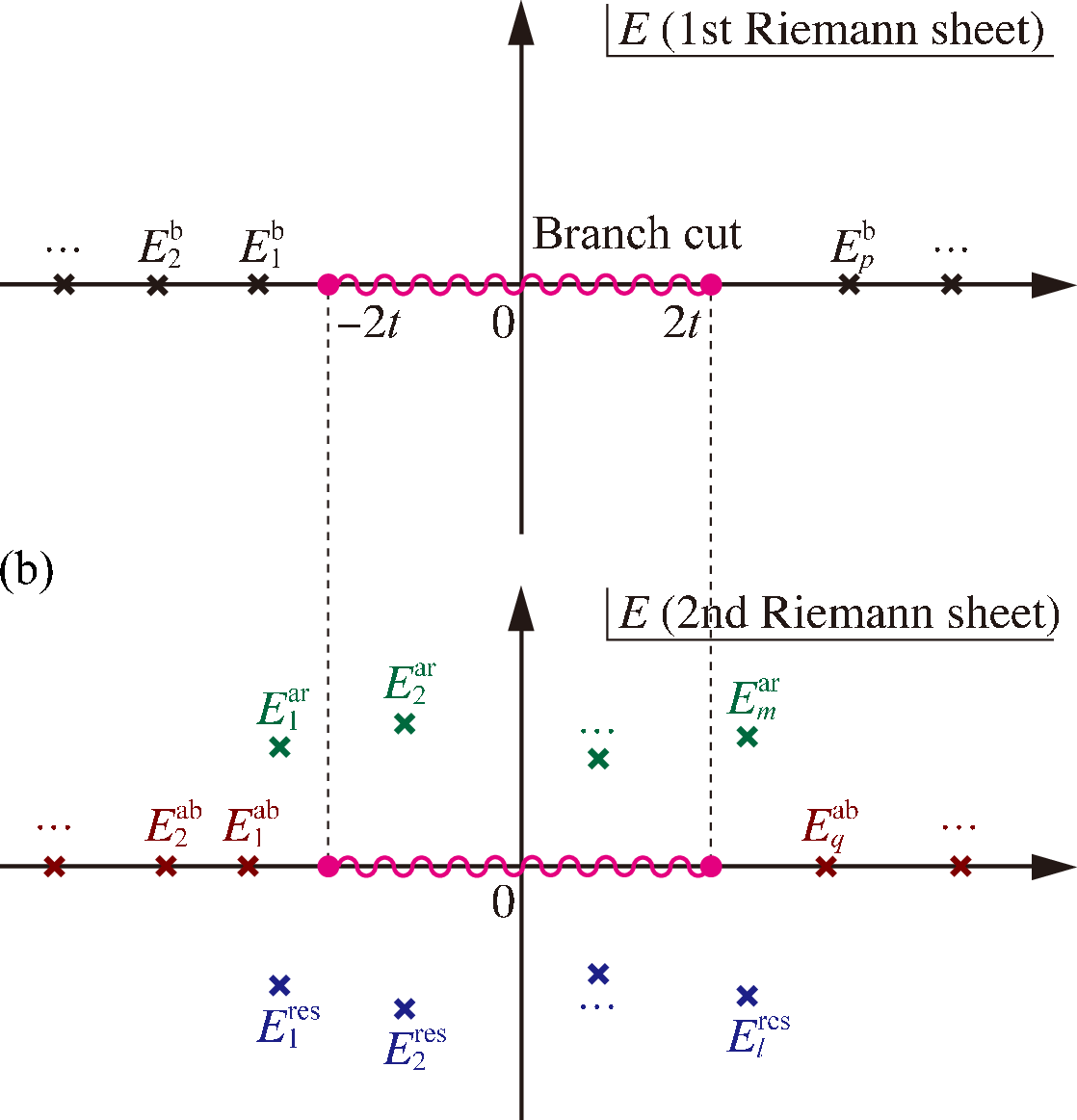}
\caption{(Color online) (a) Distribution of the eigen-wave-numbers $k^\textrm{b}_p$ of the bound states (black crosses), $k^\textrm{res}_l$ of the resonant states (blue crosses), $k^\textrm{ar}_m$ of the anti-resonant states (green crosses),
and $k_q^\textrm{ab}$ of the anti-bound states (red crosses) on the complex wave-number plane.
(b) Distribution of the eigenvalues $E^\textrm{b}_p$ of the bound states (black crosses), $E^\textrm{res}_l$ of the resonant states (blue crosses), $E^\textrm{ar}_m$ of the anti-resonant states (green crosses), and $E^\textrm{ab}_q$ of the anti-bound states (red crosses) on the complex energy plane.
The upper and lower halves of the $k$ plane respectively correspond to the first and second Riemann sheets of the $E$ plane.
A branch cut $-2t<E<2t$ accompanied by two branch points $E=\pm 2t$ connect the two Riemann sheets.}
\label{fig:distribution_eigen} 
\end{figure}

We can define the resonant state as a solution of the time-independent Schr\"{o}dinger equation for the whole Hamiltonian $H$ under the boundary conditions that the wave function has only out-going waves away from the scatterer~\cite{Gamow28,Siegert1939,Peierls59,HSNP2007}.
The condition is often called the Siegert condition~\cite{Siegert1939}.
More specifically, we seek discrete and generally complex eigenvalues $E_n$ of the whole system $H$,
\begin{align}
H|\psi_n\rangle&=E_n|\psi_n\rangle, \label{eq:shrodinger_right}\\
\langle \tilde{\psi}_n|H&=E_n\langle \tilde{\psi}_n|, \label{eq:shrodinger_left}
\end{align}
in the first Brillouin zone $-\pi<\Re k\leq\pi$
under the Siegert boundary condition as~\cite{HSNP2007,Shapiro06,Shapiro08} 
\begin{align}\label{eq6}
\langle x_{\alpha}|\psi_n\rangle=\langle\tilde{\psi}_n| x_\alpha\rangle&\propto \mathrm{e}^{ik_n |x_\alpha|}
\end{align}
for $x_{\alpha}$ on any lead $\alpha $,
where $|\psi_n\rangle$ is the right-eigenfunction and $\langle \tilde{\psi}_n|$ is the left-eigenfunction~\cite{Zeldovich60,Hokkyo65,Romo68,Berggren70,Gyarmati71,Romo80,Berggren82,Berggren96,Madrid05,Nakanishi58,PPT1991}.
(Note that $\langle \tilde{\psi}_n|^\dagger\neq|\psi_n\rangle$ in general.)
The thus-obtained eigen-wave-number
\begin{align}\label{eq15}
k_n\equiv k_{\textrm{r}n}+i\kappa_n
\end{align}
as well as the corresponding eigenenergy
\begin{align}
E_n&\equiv E_{\textrm{r}n}+i E_{\textrm{i}n} =-2t\cos k_n \label{eq:dispersion_resonant}
\end{align}
are generally complex numbers.
The complex eigenenergy is possible because the corresponding eigenfunction~\eqref{eq6} is outside the Hilbert space for the complex wave number~\eqref{eq15}.
Note here that we have two Riemann sheets of $E$ for the entire complex plane of $k$ (Fig.~\ref{fig:distribution_eigen}).
A branch cut $-2t<E<2t$ with two branch points $E=\pm 2t$ connects the two Riemann sheets.

\begin{table*}
\caption{Classification of the discrete eigenstates (eigenstates with point spectra).}
\label{tab1}
\begin{tabular}{lclclcrcl}
\hline
Bound states & \quad\quad
& ${k_\textrm{r}}^\textrm{b}_p=0$ &\quad\quad
& $\kappa^\textrm{b}_p>0$ & \quad\quad
& first Riemann sheet &\quad\quad
& $E^\textrm{b}_p<-2t$
\\ \cline{3-9}
 & \quad\quad
& ${k_\textrm{r}}^\textrm{b}_p=\pi$ &\quad\quad
& $\kappa^\textrm{b}_p>0$ & \quad\quad
& first Riemann sheet &\quad\quad
& $E^\textrm{b}_p>2t$
\\ \hline
Anti-bound states & \quad\quad
& ${k_\textrm{r}}^\textrm{ab}_q=0$ &\quad\quad
& $\kappa^\textrm{ab}_q<0$ &\quad\quad
& second Riemann sheet & \quad\quad
& $E^\textrm{ab}_q<-2t$
\\ \cline{3-9}
 & \quad\quad
& ${k_\textrm{r}}^\textrm{ab}_q=\pi$ &\quad\quad
& $\kappa^\textrm{ab}_q<0$ &\quad\quad
& second Riemann sheet & \quad\quad
& $E^\textrm{ab}_q>2t$
\\ \hline
Resonant states &\quad\quad
& ${k_\textrm{r}}^\textrm{res}_l>0$ &\quad\quad
& $\kappa^\textrm{res}_l<0$ &\quad\quad
& second Riemann sheet &\quad\quad
& ${E_\textrm{i}}^\textrm{res}_l<0$
\\ \hline
Anti-resonant states &\quad\quad
& ${k_\textrm{r}}^\textrm{ar}_m<0$ &\quad\quad
& $\kappa^\textrm{ar}_m<0$ &\quad\quad
& second Riemann sheet &\quad\quad
& ${E_\textrm{i}}^\textrm{ar}_l>0$
\\ \hline
\end{tabular}
\end{table*}
The discrete eigenstates thus obtained are classified as follows (Table~\ref{tab1} and Fig.~\ref{fig:distribution_eigen}).
First, the eigenstates with $\kappa_n>0$ are necessarily on the imaginary axis $\Re k=0$ or on the edge of the Brillouin zone $\Re k=\pi$.
(In systems with continuous space, the bound states exist only on the imaginary $k$ axis;
the bound states on the line $\Re k =\pi$ appear because the leads of the present system are lattice systems and hence the energy band~\eqref{eq110} has an upper bound.)
By putting $\kappa_n>0$ in eq.~(\ref{eq6}), we see that the eigenstates are in fact bound states.
Hereafter, we use the subscript $p$ and the superscript `b' for the bound states as in $k^\textrm{b}_p$ and $E^\textrm{b}_p$.
The bound states with ${k_\textrm{r}}^\textrm{b}_p=0$ have real negative eigenenergies $E_p^\textrm{b}<-2t$ while the bound states with ${k_\textrm{r}}^\textrm{b}_p=\pi$ have real positive ones $E_p^\textrm{b}>2t$.

Next, the eigenstates in the fourth quadrant of the $k$ plane are referred to as the resonant states.
Hereafter, we use the subscript $l$ and the superscript `res' for the resonant states as in $k^\textrm{res}_l$ and $E^\textrm{res}_l$.
The corresponding eigenenergies are in the lower half of the second Riemann sheet of the $E$ plane: ${E_\textrm{i}}^\textrm{res}_l<0$.

Third, the eigenstates in the third quadrant of the $k$ plane are referred to as the anti-resonant states.
(In the context of the condensed-matter physics, some refer to a resonance in the form of a dip of the conductance as an anti-resonance.
In the present terminology, this is
 still associated with a resonant state, which is different from the anti-resonant state here.)
Hereafter, we use the subscript $m$ and the superscript `ar' for the resonant states as in $k^\textrm{ar}_m$ and $E^\textrm{ar}_m$.
The corresponding eigenenergies are in the upper half of the second Riemann sheet of the $E$ plane: ${E_\textrm{i}}^\textrm{ar}_m>0$.
A resonant state and an anti-resonant state always appear in pair.
The states of a pair are related to each other as
\begin{align}\label{eq:eigenfunction_anti-resonant} 
|\psi^\textrm{ar}_m\rangle&= \langle\tilde{\psi}^\textrm{res}_l|^\dagger, \quad\mbox{and}\quad
\langle\tilde{\psi}^\textrm{ar}_m|= |\psi^\textrm{res}_l\rangle^\dagger, 
\\ \label{eq:eigenwavenumbers_anti-resonant}
k^\textrm{ar}_m &=-\left(k^\textrm{res}_l\right)^\ast,
\quad\mbox{or}\quad
\nonumber\\
{k_\textrm{r}}^\textrm{ar}_m&=-{k_\textrm{r}}^\textrm{res}_l
\quad\mbox{and}\quad
\kappa^\textrm{ar}_m=\kappa^\textrm{res}_l,
\\ \label{eq:eigenvalues_anti-resonant}
E^\textrm{ar}_m &= \left(E^\textrm{res}_l\right)^\ast,
\quad\mbox{or}\quad
\nonumber\\
{E_\textrm{r}}^\textrm{ar}_m&={E_\textrm{r}}^\textrm{res}_l
\quad\mbox{and}\quad
{E_\textrm{i}}^\textrm{ar}_m=-{E_\textrm{i}}^\textrm{res}_l;
\end{align}
see Appendix\ref{app:states}.
We refer to a pair of the resonant state and the corresponding anti-resonant state as a resonant-state pair.

Some systems have additional states on the negative part of the imaginary $k$ axis or on the negative part of the edge of the Brillouin zone $\Re k=\pi$.
Such states often appear when resonant and anti-resonant states of a pair collide on the $k=0$ or $k=\pi$ axis or when a bound state moves down into the lower $k$ plane on the $k=0$ or $k=\pi$ axis.
We refer to them as anti-bound states~\cite{Ohanian74} and use the subscript $q$ and the superscript `ab' as in $k^\textrm{ab}_q$ and $E^\textrm{ab}_q$.
Anti-bound states possess real eigenenergies 
(see Appendix\ref{app:states}) 
but on the second Riemann sheet and still have properties of the resonant states such as diverging wave functions.

For a practical method of finding all discrete states, see Appendices\ref{app:selfenergy} and\ref{app:calceig}, where we employ the method of an effective Hamiltonian with self-energies of the leads.
We can show that the model that we will introduce in \S~\ref{sec:conductance} has $2N$ discrete eigenstates in total, where $N$ is the number of sites of the tight-binding model for the quantum scatterer $H_\textrm{d}$;
see Appendix\ref{app:calceig}.

\section{Conductance formula for an open quantum $N$-level dot}\label{sec:conductance}
In the present section, we will consider an $N$-level extension of the Friedrichs-Fano (Newns-Anderson) model~\cite{Fano61,Friedrichs1948,Anderson1961,Sudershan1962,PPT1991,OPP2001,Miyamoto2004,Miyamoto2005}.
We will derive a simple conductance formula for the model.
Then in \S~\ref{sec:spectrum}, we will show that the formula is given by pure summation over all the discrete eigenstates without any background integrals.
We do not claim that this formula is advantageous in actual computation.
We rather emphasize the fact that the formula explicitly shows that the conductance contains interference terms between various discrete states listed in \S~\ref{sec:resonant}.

\begin{figure}
\includegraphics[width=0.38\textwidth]{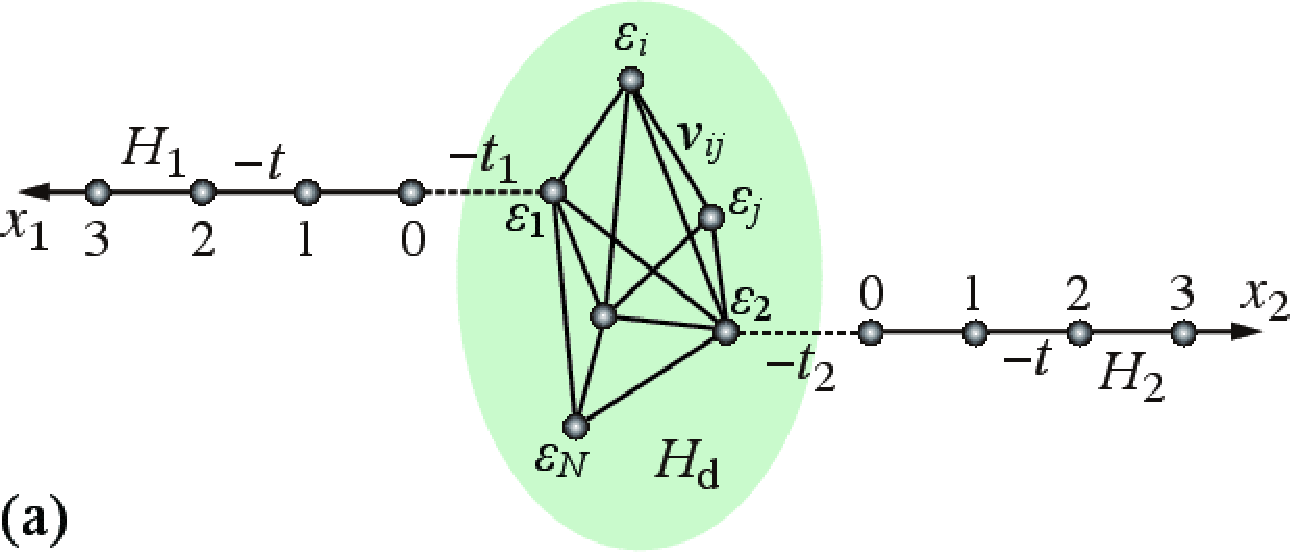}

\vspace*{\baselineskip}

\includegraphics[width=0.38\textwidth]{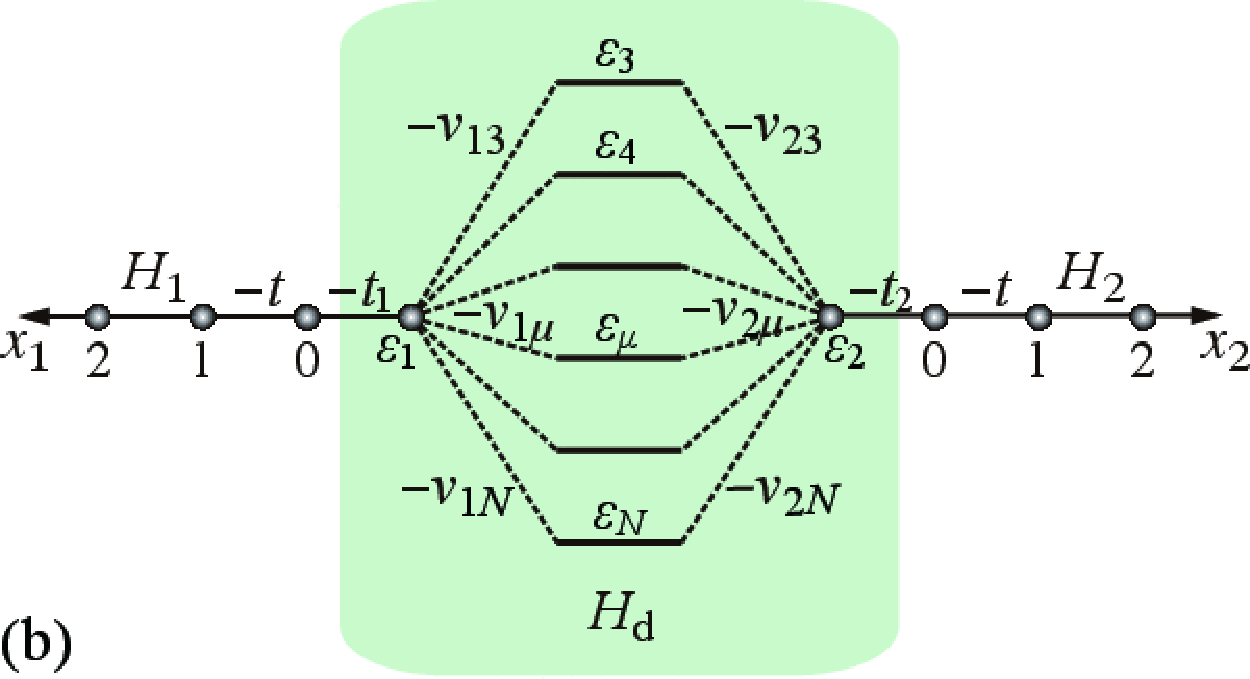}

\vspace*{\baselineskip}

\includegraphics[width=0.38\textwidth]{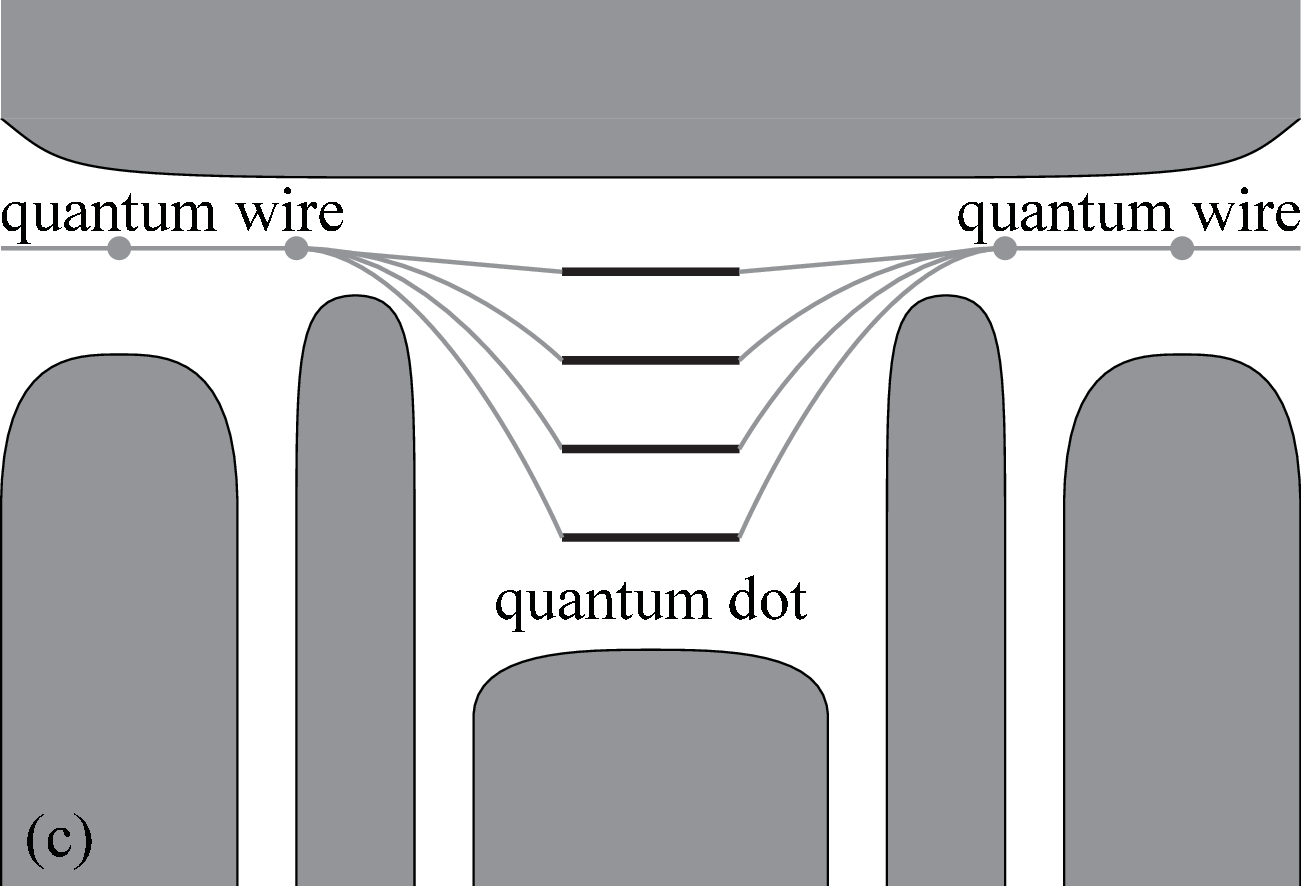}

\vspace*{\baselineskip}

\includegraphics[width=0.38\textwidth]{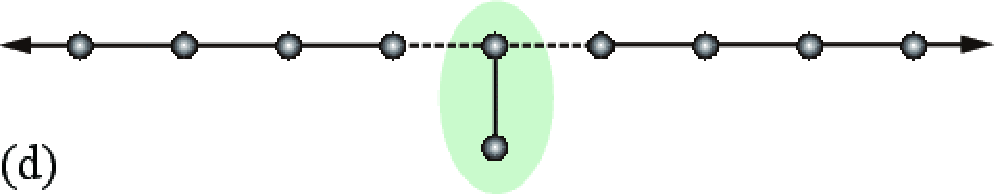}
\vspace*{\baselineskip}

\includegraphics[width=0.38\textwidth]{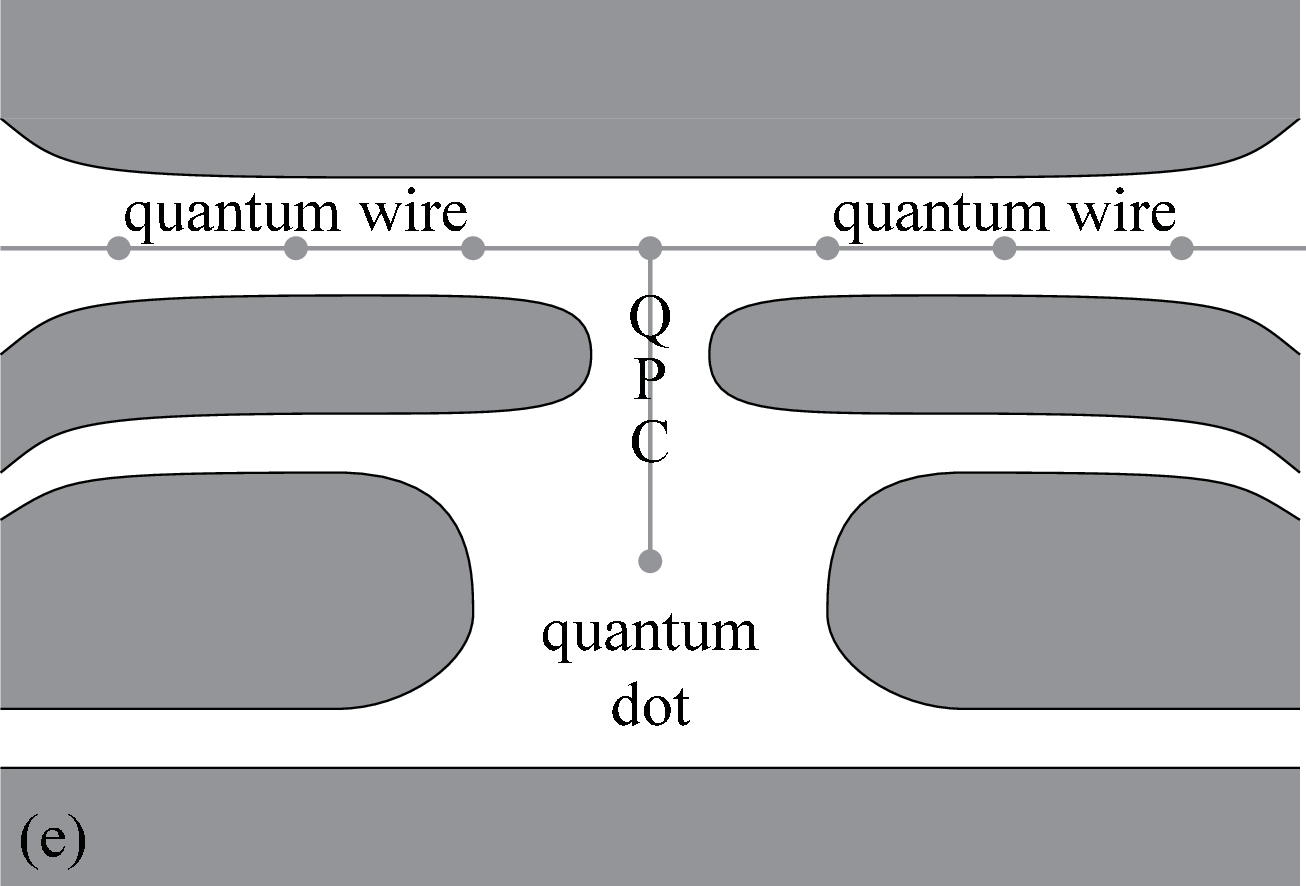}
\caption{(Color online) (a) The model that we consider in present paper, an open quantum $N$-``site" dot with two semi-infinite leads.
The couplings among the ``sites" as well as those between each lead and the contacting site of the dot can be arbitrary.
Therefore, the model includes the cases exemplified in (b) and (d).
The model may be achieved experimentally in a structure (c), where two quantum wires are coupled to multiple levels of a quantum dot, as well as in a structure (e), where a quantum point contact (QPC) couples an infinite quantum wire to a level of a quantum dot, a situation often called a side-coupled dot.}
\label{fig:NQD}
\end{figure}

The model that we discuss hereafter is a general one-body problem of an $N$-``site" (or $N$-level) dot with two semi-infinite leads attached to it, as illustrated in Fig.~\ref{fig:NQD}(a).
(We will show below that the ``site" is not necessarily an actual spatial position but can represent an energy level.)
The Hamiltonian is given by
\begin{align}\label{eq:Hamiltonian_original}
H= H_\textrm{d}+\sum_{\alpha=1,2}\left(H_{\alpha}+H_{\textrm{d},\alpha}\right), 
\end{align}
where the quantum dot is given by a tight-binding Hamiltonian of the form
\begin{align}\label{eqHd}
H_\textrm{d}\equiv&\sum_{i=1}^{N}\varepsilon_{i}|d_{i}\rangle\langle d_{i}|
\nonumber\\
&-\sum_{1\leqslant i<j\leqslant N}v_{ij}\left( |d_{i}\rangle\langle d_{j}|+|d_{j}\rangle\langle d_{i}|\right),
\end{align}
with $\{d_i\}$ denoting ``sites" (or levels) in the dot, $\varepsilon_i$ the potential at the site $i$, and $v_{ij}$ the hopping element between the sites $i$ and $j$ with 
$v_{ij}\in\mathbb{R}$ and 
$v_{ij}\equiv v_{ji}$, while each lead is given by the standard tight-binding Hamiltonian
\begin{align}\label{eqHalpha}
H_{\alpha}&\equiv-t\sum_{x_{\alpha}=0}^{\infty}\left(|x_{\alpha}+1\rangle\langle x_{\alpha}|+|x_{\alpha}\rangle\langle x_{\alpha}+1|\right)
\end{align}
with $t>0$.
The coupling between the dot and each lead is given by
\begin{align}\label{eqHdalpha}
H_{\textrm{d},\alpha}\equiv&-t_{\alpha}\left(|x_{\alpha}=0\rangle\langle d_\alpha |+| d_\alpha\rangle\langle x_{\alpha}=0| \right)
\end{align}
with $t_{\alpha}$ denoting the hopping element between the site $d_\alpha$, to which the lead $\alpha$ is attached, and the end point $x_\alpha=0$ of the lead $\alpha$.
As stated above and in eq.~\eqref{eq:Hamiltonian_original}, we consider the case where there are two leads, $\alpha=1,2$, attached to two contact sites $d_1$ and $d_2$.

The above system is an $N$-site (or $N$-level) extension of the Friedrichs-Fano model~\cite{Fano61,Friedrichs1948,Anderson1961,Sudershan1962,PPT1991,OPP2001,Miyamoto2004,Miyamoto2005}.
The model~(\ref{eq:Hamiltonian_original})--(\ref{eqHdalpha}) is so general that it can account for the system shown in Fig.~\ref{fig:NQD}(b) with the dot Hamiltonian of a partially diagonalized form
\begin{align}\label{eqHd-diag}
H_\mathrm{d}&\equiv \sum_{\mu=3}^{N} \varepsilon_\mu |d_\mu\rangle\langle d_\mu|
\nonumber\\
&-\sum_{\alpha=1,2}\sum_{\mu=3}^{N}v_{\alpha\mu}\left(|d_\alpha\rangle\langle d_\mu |+| d_\mu\rangle\langle d_\alpha| \right),
\end{align}
where $\varepsilon_\mu$ now denotes the energy of a one-particle level $\mu$, not necessarily a spatial site.
The model may be experimentally realized in the structure schematically shown in Fig.~\ref{fig:NQD}(c), where a quantum dot is sandwiched by two quantum wires.

As a simpler case, our model also includes the side-coupled (or T-shaped) quantum-dot system shown in Fig.~\ref{fig:NQD}(d).
The side-coupled quantum-dot systems have been intensively studied~\cite{Kim01,Kikoin01,Kang01,Affleck01,Simon01,Affleck07,SAKKI2004,SAKKI2005,Torio02,Orellana03a,Orellana03b,Rodriguez03,MSU2004,Aligia04,Tanaka05,Lara05,Franco06,Wang06,Chakrabarti06,Zitko06,Li08} in various contexts including the Kondo problem.
(Note, however, that we here do not take account of the electron-electron interactions.) 
The model may be experimentally realized in the structure schematically shown in Fig.~\ref{fig:NQD}(e), where a quantum dot is connected to a quantum wire through a quantum point contact.
A `quantum point contact' may be considered in fact as a narrow valley that couples both sides of the contact.
When the separation of the energy levels in the quantum dot is wide enough, the quantum point contact may couple the wire with only several levels in the quantum dot.
Such situations may be summarized in the form~(\ref{eq:Hamiltonian_original}).

We will obtain the conductance $\mathcal{G}(E)$ from the lead 1 to the lead 2 in the form
\begin{align} \label{eq:conductance_G}
\mathcal{G}_{12}(E)&=\frac{2e^2}{h}
\Gamma_{11}\Lambda_{12}\Gamma_{22}\Lambda_{21}
\nonumber\\
&\times
\frac{-(D-4)\pm\sqrt{(D+4)^2-4T^2}}{2(T^2-4D)},
\end{align}
where
$\Lambda$ and $\Gamma$ are $N$-by-$N$ matrices given by
\begin{align}\label{eq:Lambda}
\Lambda&\equiv G^\textrm{R}+G^\textrm{A}
\\ \label{eq:Gamma}
i\Gamma&\equiv \left(G^\textrm{R}\right)^{-1}-\left(G^\textrm{A}\right)^{-1}
\end{align}
with $G^\textrm{R}$ and $G^\textrm{A}$ being the retarded and advanced Green's functions of the total Hamiltonian $H$, respectively, and
\begin{align}\label{eq:T}
T&=\mathop{\textrm{Tr}}\check{\Gamma}\check{\Lambda},
\\\label{eq:D}
D&=\mathop{\textrm{det}}\check{\Gamma}\check{\Lambda}.
\end{align}
The matrices $\check{\Gamma}$ and $\check{\Lambda}$ are two-by-two matrices constructed from the $(1,1)$, $(1,2)$, $(2,1)$, and $(2,2)$ elements of the $N$-by-$N$ matrices $\Gamma$ and $\Lambda$, respectively, where $1$ and $2$ denote the contact sites; see Appendix\ref{app:selfenergy} and particularly eq.~\eqref{eqA310} for details.
Because the matrix $\Gamma$ has only diagonal elements at the contact sites $d_1$ and $d_2$ as shown below, we can reduce the Hilbert space to the two-dimensional space spanned by $|d_1\rangle$ and $|d_2\rangle$.
Then all calculations in eqs.~\eqref{eq:conductance_G}--\eqref{eq:D} can be carried out in terms of two-by-two matrices instead of original $N$-by-$N$ matrices.
The point to note here is that the conductance $\mathcal{G}_{12}$ (or the transmission probability $\mathcal{T}_{12}$) is given by the matrices $\Lambda$ and $\Gamma$ only, not in terms of each of $G^\textrm{R}$ or $G^\textrm{A}$.

The reason why we use the matrix $\Lambda$ is as follows.
In \S~\ref{sec:spectrum}, we will show that the matrix $\Lambda$ can be expanded purely in terms of all the discrete eigenstates as
\begin{align}\label{eq:expansionGRGA}
\Lambda=\sum_{n\in p,q,l,m}\frac{|\psi_n\rangle\langle\tilde{\psi}_n|}{E-E_n}
\end{align}
for the states of the central dot, $\{|d_i\rangle\}$, (and more specifically for the contact sites $|d_1\rangle$ and $|d_2\rangle$), where the subscripts $p$, $q$, $l$ and $m$ respectively denote sets of the bound states, the anti-bound states, the resonant states and the anti-resonant states, whereas $|\psi_n\rangle$ and $\langle\tilde{\psi}_n|$ are the right- and left-eigenvectors of each state, respectively; 
see Appendix\ref{app:states}.
The important point here is that the expansion does not contain any background integrals.
This shows that the  `background' of the conductance profile~\eqref{eq:conductance_G} is not a background, but in fact, just a sum of the tails of various resonance peaks.
(One could of course refer to the sum of the tails as a `background,' but we do not use this terminology in order to emphasize that we are free of background integrals.)
Note at this point that the expression~(\ref{eq:conductance_G}) has the factors of the form
$\Lambda_{\alpha\beta}\Lambda_{\beta\alpha} = \left|\Lambda_{\alpha\beta}\right|^2$, which contains crossing terms, or interferences, between various discrete states. 
(In the absence of a magnetic field $\Lambda_{\alpha\beta}$ is real.)

On the other hand, the matrix $\Gamma$ defined by eq.~\eqref{eq:Gamma}, or by
\begin{align}
G^\textrm{A}-G^\textrm{R}&=iG^\textrm{R}\Gamma G^\textrm{A}, \label{eq:GA-GR_gamma}
\end{align}
is given by
\begin{align}\label{eq:GA-GR_gamma2}
\Gamma&\equiv\sum_{\alpha=1,2} \Gamma^{(\alpha)}
\end{align}
with
\begin{align}
\label{eq28}
\Gamma^{(\alpha)}&\equiv
\frac{{t_\alpha}^2}{t^2}\sqrt{4t^2-E^2}|d_\alpha\rangle\langle d_\alpha|;
\end{align}
see Appendix\ref{app:selfenergy} for details.
The $N$-by-$N$ matrix $\Gamma$ is, in the two-dimensional Hilbert subspace of the contact sites $|d_1\rangle$ and $|d_2\rangle$, given in the form
\begin{align}\label{eq:Gamma0}
\check{\Gamma}=\frac{\sqrt{4t^2-E^2}}{t^2}\left(\begin{array}{cc}
{t_1}^2 & 0 \\
0 & {t_2}^2 
\end{array}
\right);
\end{align}
all other elements are zero.
Equation~(\ref{eq:expansionGRGA}) shows that the real part of the Green's function is given by the discrete eigenstates, while eq.~(\ref{eq:Gamma0}) shows that the imaginary part of the Green's function is given by the inverse of the van Hove singularities at the branch points $E=\pm 2t$.

The simultaneous matrix equations~(\ref{eq:Lambda}) and~(\ref{eq:GA-GR_gamma}) result in the matrix Riccati equations
\begin{align}\label{eq:RiccatiGR}
\langle d_i|&\left\{G^\textrm{R}\left(i\Gamma\right)G^\textrm{R}-G^\textrm{R}\left[2+\left( i\Gamma\right)\Lambda\right]+\Lambda\right\}|d_j\rangle=0,
\\
\label{eq:RiccatiGA}
\langle d_i|&\left\{G^\textrm{A}\left(- i\Gamma\right)G^\textrm{A}-\left[2+\Lambda\left(- i\Gamma\right)\right]G^\textrm{A}+\Lambda\right\}|d_j\rangle=0.
\end{align}
The solution gives each Green's function in terms of the contribution of the discrete eigenstates, $\Lambda$, and the contribution of the branch-point singularities, $\Gamma$.
We first solve eqs.~(\ref{eq:RiccatiGR}) and~(\ref{eq:RiccatiGA}) for the sites $d_1$ and $d_2$ and then use the solution in the Fisher-Lee relation~\cite{FL1981,Datta95}
\begin{align}\label{eq:conductance}
&\mathcal{G}_{12}(E)\equiv\frac{2e^2}{h}
\textrm{Tr}\left(\Gamma^{(1)}G^\textrm{R}\Gamma^{(2)}G^\textrm{A}\right) \nonumber\\
&
=\frac{2e^2}{h}\Gamma_{11}G^\textrm{R}_{12}\Gamma_{22}G^\textrm{A}_{21}.
\end{align}
For details of the solution, see Appendix\ref{app:sign}.
We thus arrive at the conductance $\mathcal{G}(E)$ between the lead 1 and the lead 2 in the form~(\ref{eq:conductance_G}).
The sign in front of the square root of eq.~(\ref{eq:conductance_G}) is chosen according to the rule that is also given in Appendix\ref{app:sign}.

The formula~\eqref{eq:conductance_G} may be less advantageous than the formula~\eqref{eq:conductance} in actual computation of the conductance.
It is not our aim here to find an easy-to-calculate formula.
Our aim is to show explicitly that the conductance contains the pure summation over all discrete eigenvalues, eq.~\eqref{eq:expansionGRGA}, not any background integrals.

Incidentally, eq.~(\ref{eq:conductance_G}) reduces to a much simpler formula particularly when the contact sites $d_1$ and $d_2$ are identical, in other words, when the two leads are attached to one site as in the T-shaped quantum dot of Fig.~\ref{fig:NQD}(d).
Hereafter, whenever the two leads are attached to one site, we will denote the one contact site as $d_0$.
In this case, the $\Gamma$ matrix has only the $(0,0)$ element
\begin{align}
\check{\Gamma}=\Gamma_{00}=\frac{\sqrt{4t^2-E^2}}{t^2}\left({t_1}^2+{t_2}^2\right).
\end{align}
Then, we modify the quantities in eq.~\eqref{eq:conductance_G} as
\begin{align}
\Gamma_{11}&\longrightarrow \frac{{t_1}^2}{{t_1}^2+{t_2}^2}\Gamma_{00},
\\
\Gamma_{22}&\longrightarrow \frac{{t_2}^2}{{t_1}^2+{t_2}^2}\Gamma_{00},
\\
\Lambda_{11},\Lambda_{22},\Lambda_{12},\Lambda_{21}&\longrightarrow \Lambda_{00},
\\
T=\Gamma_{11}\Lambda_{11}+\Gamma_{22}\Lambda_{22}&\longrightarrow \Gamma_{00}\Lambda_{00},
\\
D=\Gamma_{11}\Gamma_{22}\left(\Lambda_{11}\Lambda_{22}-\Lambda_{12}\Lambda_{21}\right)&\longrightarrow 0,
\end{align}
where $T$ and $D$ were defined in eqs.~\eqref{eq:T} and~\eqref{eq:D}.
The conductance in this case is thereby given by
\begin{align}
\mathcal{G}_{00}(E)
&=\displaystyle{\frac{e^2}{h}\left(\frac{2t_1t_2}{{t_1}^2+{t_2}^2}\right)^2\displaystyle{\left[1\pm \sqrt{1-\left(\frac{\Lambda_{00}\Gamma_{00}}{2}\right)^2}\right]}}. \label{eq:conductance_J}
\end{align}
Indeed, we can confirm this expression, following the derivation in Appendix\ref{app:sign} for the two-contact case with its simplification to the one-contact case.

\section{Resonant-state expansion of the Green's functions}\label{sec:spectrum}

As we mentioned in \S~\ref{sec:conductance}, the conductance formula~(\ref{eq:conductance_G}) contains the matrix elements of $\Lambda$ squared, whereas the matrix $\Lambda$ is given by the summation over all discrete eigenstates as eq.~(\ref{eq:expansionGRGA}).
We thereby have interferences between the discrete eigenstates.
This is the main point of the present paper.

\begin{figure}
\includegraphics[width=0.35\textwidth]{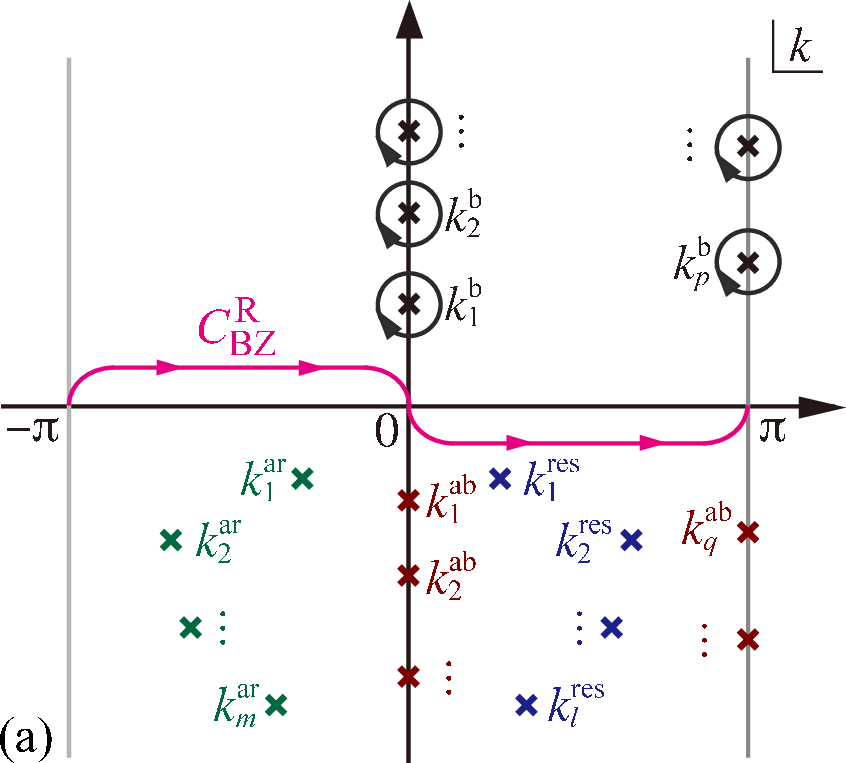}

\vspace*{\baselineskip}

\includegraphics[width=0.35\textwidth]{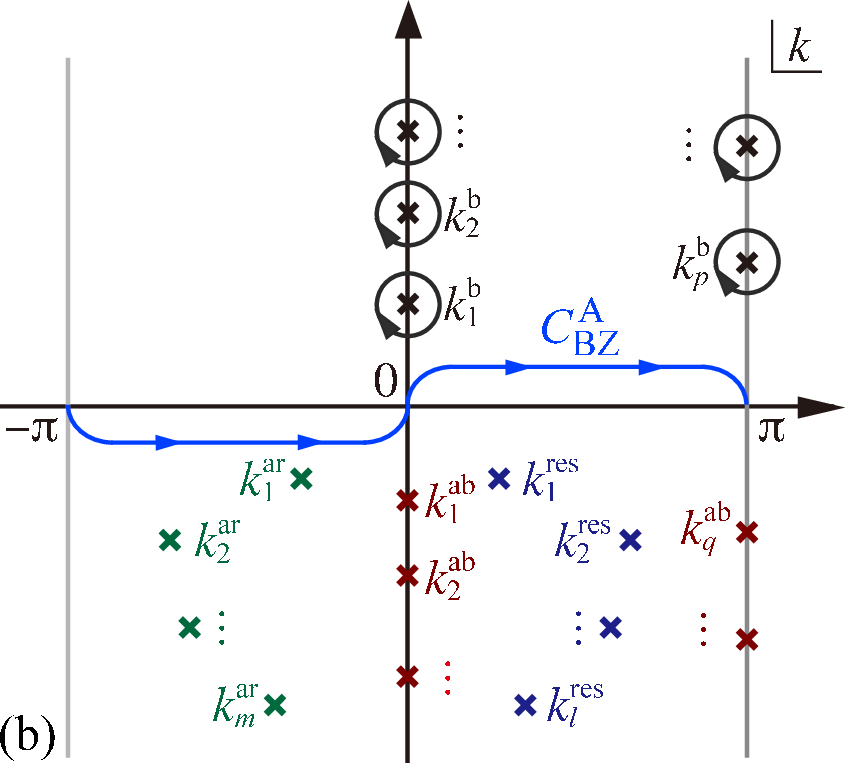}
\caption{(Color online) (a) The integration contour $C^\textrm{R}_\textrm{BZ}$ (purple curve) for the retarded Green's function $G^\textrm{R}(E)$ and (b) the integration contour $C^\textrm{A}_\textrm{BZ}$ (cyan curve) for the advanced Green's function $G^\textrm{A}(E)$,
with the circular (gray) contours extracting bound states in the complex wave-number plane.}
\label{fig:G@kspace_N_second} 
\end{figure}

Let us describe the derivation of the resonant-state expansion~(\ref{eq:expansionGRGA}) hereafter.
We can obtain the exact expression of the scattering states of the system~(\ref{eq:Hamiltonian_original}), namely the Friedrichs solution~\cite{Friedrichs1948} $|\psi_{k}\rangle$, with the eigenvalue $E_k=-2t\cos k$; see Appendix\ref{app:Friedrichs}.
The completeness with respect to the scattering states is given by~\cite{Newton1961}
\begin{align}\label{eq450}
1=\sum_p|\psi^\textrm{b}_p\rangle\langle\tilde{\psi}^\textrm{b}_p| +\int_\textrm{BZ}\frac{dk}{2\pi}|\psi_{k}\rangle\langle\tilde{\psi}_{k}|,
\end{align}
where $|\psi^\textrm{b}_p\rangle$ is a bound state and $|\psi_{k}\rangle$ is a scattering state given in Appendix\ref{app:Friedrichs}, and we used the notation
\begin{equation}
|\psi_k\rangle \langle\tilde{\psi}_k|\equiv \sum_\alpha|\psi_{k,\alpha}\rangle \langle\tilde{\psi}_{k,\alpha}|.
\end{equation}
We first express the retarded and advanced Green's functions in the spectral representation;
\begin{align}
G^\textrm{R}(E)&=\displaystyle{\sum_p\frac{|\psi^\textrm{b}_p\rangle\langle\tilde{\psi}^\textrm{b}_p|}{E-E^\textrm{b}_p}
+\int_{C^\textrm{R}_\textrm{BZ}}\frac{dk}{2 \pi}\frac{|\psi_{k}\rangle\langle\tilde{\psi}_{k}|}{E-E_k}} , \\
G^\textrm{A}(E)&=\displaystyle{\sum_p\frac{|\psi^\textrm{b}_p\rangle\langle\tilde{\psi}^\textrm{b}_p|}{E-E^\textrm{b}_p}
+\int_{C^\textrm{A}_\textrm{BZ}}\frac{dk}{2 \pi}\frac{|\psi_{k}\rangle\langle\tilde{\psi}_{k}|}{E-E_k}},
\end{align}
where the integration contours $C^\textrm{R}_\textrm{BZ}$ and $C^\textrm{A}_\textrm{BZ}$ cover the Brillouin zone as indicated in Fig.~\ref{fig:G@kspace_N_second}.

\begin{figure}
\includegraphics[width=0.35\textwidth]{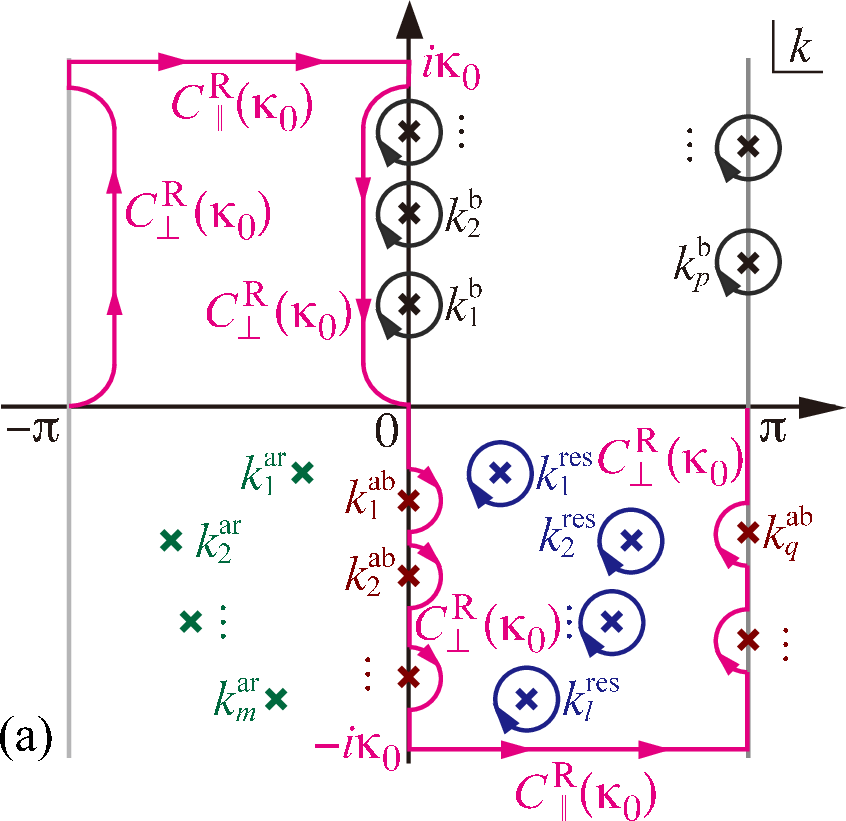}

\vspace*{\baselineskip}

\includegraphics[width=0.35\textwidth]{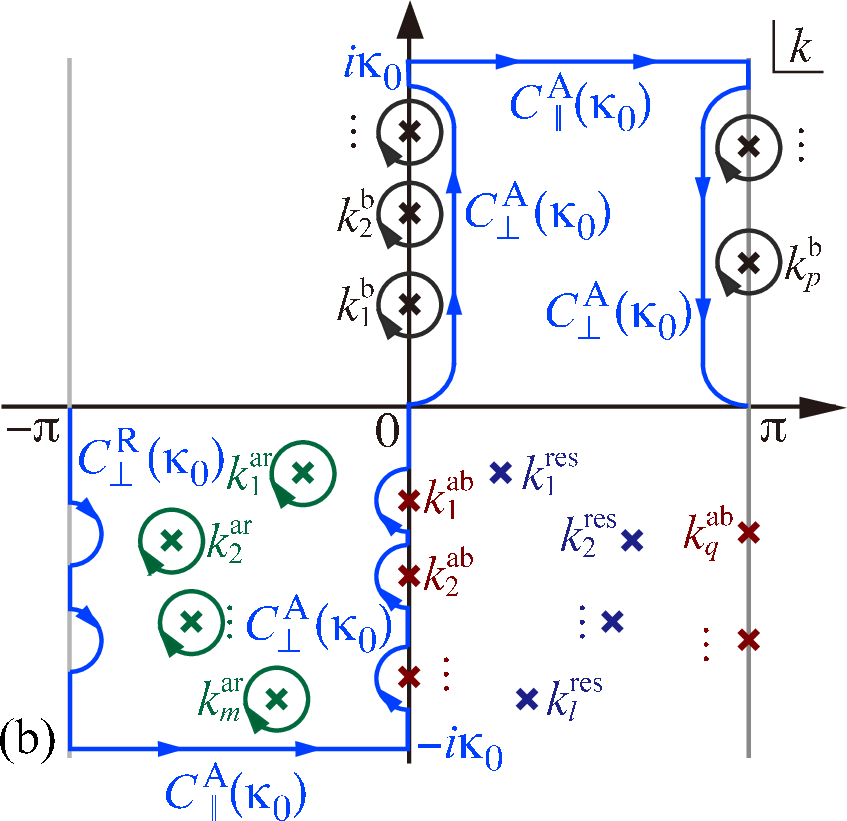}
\caption{(Color online) (a) The integration contours $C^\textrm{R}_\perp(\kappa_0)$ and $C^\textrm{R}_\parallel(\kappa_0)$ (purple curves) for the retarded Green's function $G^\textrm{R}(E)$, modified for extracting the resonant states (blue crosses) in the complex wave-number plane.
(b) The integration contours $C^\textrm{A}_\perp(\kappa_0)$ and $C^\textrm{A}_\parallel(\kappa_0)$ (cyan curves) for the advanced Green's function $G^\textrm{A}(E)$, modified for extracting the anti-resonant states (green crosses).}
\label{fig:G@kspace_N_third} 
\end{figure}

Next, we replace the integration contours $C^\textrm{R}_\textrm{BZ}$ and $C^\textrm{A}_\textrm{BZ}$ with the ones shown in Fig.~\ref{fig:G@kspace_N_third}.
Then the retarded Green's function $G^\textrm{R}(E)$ acquires the residual integrals of the resonant states $k^\textrm{res}_l$, which lie in the fourth quadrant, while the advanced Green's function $G^\textrm{A}(E)$ acquires the residual integrals of  the anti-resonant states $k^\textrm{ar}_m$, which lie in the third quadrant:
\begin{align}
\oint_{C(k=k^\textrm{res}_l )}\frac{dk}{2 \pi}\frac{|\psi_{k}\rangle\langle\tilde{\psi}_{k}|}{E-E_k}
&=\frac{|\psi^\textrm{res}_l\rangle\langle\tilde{\psi}^\textrm{res}_l|}{E-E^\textrm{res}_l}, \\
\oint_{ C(k=k^\textrm{ar}_m)}\frac{dk}{2 \pi}\frac{|\psi_{k}\rangle\langle\tilde{\psi}_{k}|}{E-E_k}
&=\frac{|\psi^\textrm{ar}_m\rangle\langle\tilde{\psi}^\textrm{ar}_m|}{E-E^\textrm{ar}_m}.
\end{align}
We then have
\begin{align}
G^\textrm{R}(E)=&\displaystyle{\sum_p\frac{|\psi^\textrm{b}_p\rangle\langle\tilde{\psi}^\textrm{b}_p|}{E-E^\textrm{b}_p}
+\sum_l\frac{|\psi^\textrm{res}_l\rangle\langle\tilde{\psi}^\textrm{res}_l|}{E-E^\textrm{res}_l} }\nonumber \\
&\displaystyle{+\lim_{\kappa_0\rightarrow +\infty}\int_{C^\textrm{R}_{\perp}(\kappa_0)+C^\textrm{R}_{\parallel}(\kappa_0)}\frac{dk}{2 \pi}\frac{|\psi_{k}\rangle\langle\tilde{\psi}_{k}|}{E-E_k}}, \label{eq:GR_diagonalized} \\
G^\textrm{A}(E)=&\displaystyle{\sum_p\frac{|\psi^\textrm{b}_p\rangle\langle\tilde{\psi}^\textrm{b}_p|}{E-E^\textrm{b}_p}
+\sum_m\frac{|\psi^\textrm{ar}_m\rangle\langle\tilde{\psi}^\textrm{ar}_m|}{E-E^\textrm{ar}_m}}\nonumber \\
&\displaystyle{+\lim_{\kappa_0 \rightarrow +\infty}\int_{C^\textrm{A}_{\perp}(\kappa_0)+C^\textrm{A}_{\parallel}(\kappa_0)}\frac{dk}{2 \pi}\frac{|\psi_{k}\rangle\langle\tilde{\psi}_{k}|}{E-E_k}}. \label{eq:GA_diagonalized}
\end{align}
Here $C^\textrm{R}_{\parallel}(\kappa_0)$ indicates the sum of the paths parallel to the real axis and 
$C^\textrm{R}_{\perp}(\kappa_0)$ the sum of the paths perpendicular to the real axis including the contributions from the anti-bound states; 
see Fig.~\ref{fig:G@kspace_N_third} for the definitions of these contours.
Note that $\kappa_0$ of the modified integration contour must be positive and greater than the imaginary parts of all the resonant eigen-wave-numbers.

A comment is in order here.
Many studies on resonant-state expansions stop here and hence have background integrals, that is, the third terms in eqs.~(\ref{eq:GR_diagonalized}) and~(\ref{eq:GA_diagonalized}), respectively.
The key point of our algebra is to cancel these background integrals by summing up the two.

Let us sum up the retarded and advanced Green's functions;
\begin{align}
&G^\textrm{R}(E)+G^\textrm{A}(E) \nonumber \\
=&\displaystyle{2 \sum_p\frac{|\psi^\textrm{b}_p\rangle\langle\tilde{\psi}^\textrm{b}_p|}{E-E^\textrm{b}_p}+\sum_l\frac{|\psi^\textrm{res}_l\rangle\langle\tilde{\psi}^\textrm{res}_l|}{E-E^\textrm{res}_l}} +\displaystyle{\sum_m\frac{|\psi^\textrm{ar}_m\rangle\langle\tilde{\psi}^\textrm{ar}_m|}{E-E^\textrm{ar}_m}}\nonumber \\
&+\lim_{\kappa_{0}\rightarrow \infty}\displaystyle{\int_{C^\textrm{R}_{\perp}(\kappa_0)+C^\textrm{A}_{\perp}(\kappa_0)}\frac{dk}{2 \pi}\frac{|\psi_{k}\rangle\langle\tilde{\psi}_{k}|}{E-E_k}}\nonumber \\
&+\lim_{\kappa_{0}\rightarrow \infty}\displaystyle{\int_{C^\textrm{R}_{\parallel}(\kappa_0)+C^\textrm{A}_{\parallel}(\kappa_0)}\frac{dk}{2 \pi}\frac{|\psi_{k}\rangle\langle\tilde{\psi}_{k}|}{E-E_k}}
\label{eq:GA+GR}.
\end{align}
The sum of the contributions of the integration contour $C^\textrm{R}_{\perp}(\kappa_0)$ and $C^\textrm{A}_{\perp}(\kappa_0)$ is equal to the contribution of the bound states and anti-bound states
except for the sign;
\begin{align}
\lim_{\kappa_{0}\rightarrow \infty}&\displaystyle{\int_{C^\textrm{R}_{\perp}(\kappa_0)+C^\textrm{A}_{\perp}(\kappa_0)}\frac{dk}{2 \pi}\frac{|\psi_{k}\rangle\langle\tilde{\psi}_{k}|}{E-E_k}}
\nonumber \\
=&\displaystyle{-\sum_p\frac{|\psi^\textrm{b}_p\rangle\langle\tilde{\psi}^\textrm{b}_p|}{E-E^\textrm{b}_p}+\sum_q\frac{|\psi_q^\textrm{ab}\rangle\langle\tilde{\psi}_q^\textrm{ab}|}{E-E_q^\textrm{ab}}}\label{eq:GR+GA_perp}.
\end{align}
On the other hand, we proved that the contributions of the parallel integration contours $C^\textrm{R}_{\parallel}(\kappa_0)$ and $C^\textrm{A}_{\parallel}(\kappa_0)$ vanish for the states on the central dot; i.e. for $|d_i\rangle$ and $|d_j\rangle$ with any $i$ and $j$, we have
\begin{align}
\lim_{\kappa_0\rightarrow +\infty}&\int_{C^\textrm{R}_{\parallel}(\kappa_0)+C^\textrm{A}_{\parallel}(\kappa_0)}\frac{dk}{2 \pi}\frac{\langle d_i|\psi_{k}\rangle\langle\tilde{\psi}_{k}|d_j\rangle}{E-E_k}=0.
\label{eq:GA+GR_parallel}
\end{align}
See Appendix~\ref{appendix:A} for the proof.

\begin{figure}
\includegraphics[width=0.35\textwidth]{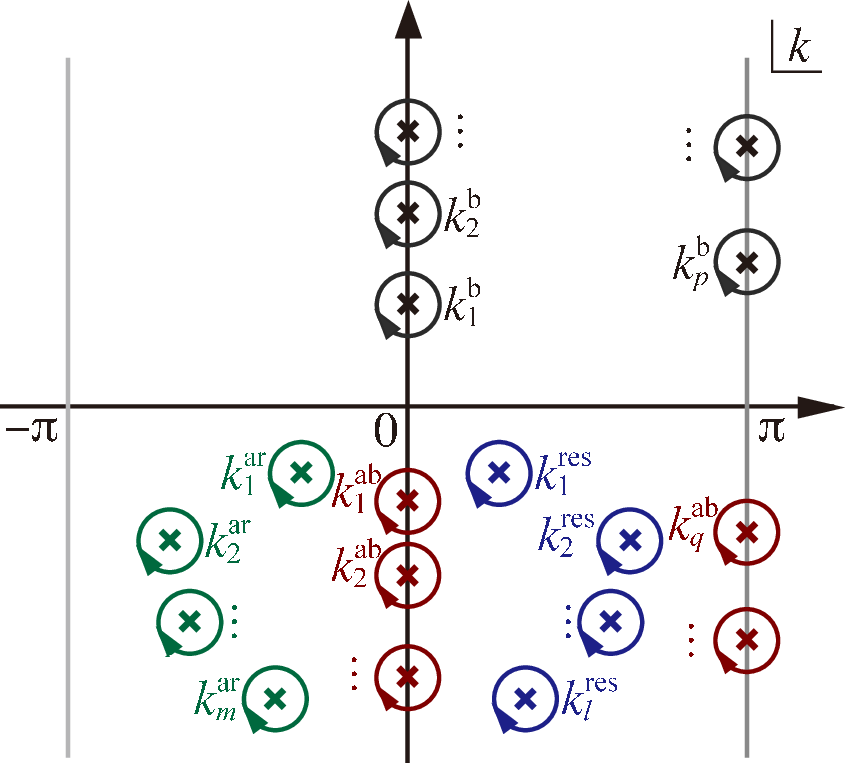}
\caption{(Color online) Summation of the modified integration contours for the retarded and advanced Green's functions in the complex wave-number plane results in the sum of the contributions of all discrete eigenstates containing the bound states (black crosses), the resonant states (blue crosses), the anti-resonant states (green crosses) and the anti-bound states (red crosses).}
\label{fig:GR+GA@kspace_N}
\end{figure}

Thus we find that the sum of the retarded and advanced Green's functions is equal to the contributions of only the discrete eigenstates
for the states on the central dot, $\{|d_i\rangle\}$,  (Fig.~\ref{fig:GR+GA@kspace_N});
\begin{align}
\langle d_i|\left(G^\textrm{A}(E)+G^\textrm{R}(E)\right)|d_j\rangle&=\langle d_i|\Lambda(E)|d_j\rangle, \label{eq:GA+GR_spectrum}
\end{align}
where
\begin{align}\label{eq5100}
\Lambda(E)&\equiv\displaystyle{\sum_{n \in p,q,l,m}\frac{|\psi_n\rangle\langle\tilde{\psi}_n|}{E-E_n}}.
\end{align}
We will show in Appendix\ref{app:calceig} that there are $2N$ pieces of discrete eigenvalues in total.

The factor~\eqref{eq:GA+GR_spectrum} comes into the conductance formula~(\ref{eq:conductance_G}) and causes resonance peaks as well as interferences among them in the conductance profile.
In other words, we revealed the effect of resonant states and bound states on the conductance explicitly and rigorously.
To our knowledge, this is for the first time the conductance is exactly given in terms of the summation over simple poles of the discrete eigenstates (see ref.~\citen{endnote}).
This became possible because we succeeded in canceling out the background integrals by summing up the retarded and advanced Green's functions.

Incidentally, we found that, when the couplings between the dot and each lead, $t_1$ and $t_2$, are equal to the hopping amplitude $t$, two of the discrete eigenvalues become infinite; see Appendix\ref{app:inf}.
This contribution may be regarded as a background integral, although it is purely a constant independent of the energy $E$.

Finally, we make a comment from another point of view.
The background integrals shown in Fig.~\ref{fig:G@kspace_N_second}, when mapped to the complex energy plane, would have given rise to branch-point singularities (or the van Hove singularities) at the band edges $E=\pm 2t$.
While the resonance poles yield the Markovian dynamics (exponential decay), the branch-point singularities are known to yield non-Markovian dynamics (\textit{i.e.}, power-law decay) with no characteristic time or length scales, which cause deviations from exponential decay for both long time 
scales~\cite{Khalfin} and short time scales~\cite{Misra,Petrosky}.
The reason why we have been able to exclude the background integrals from the conductance formula is that the Landauer conductance does not have the branch-point singularities at the band edges.
The branch-point singularities are eliminated by the factor $\sqrt{4t^2-E^2}$ in the numerator of the matrix $\Gamma$ given in eq.~\eqref{eq28}.
Therefore, the Landauer conductance continually converges to zero at the band edges, without having any van Hove singularities.
This lack of the branch-point singularities is the reason why we can express the Landauer formula purely in terms of the resonance poles.

\section{Quantum interference effect of discrete eigenstates}\label{sec:Fano}
In the present section, we argue that the Fano conductance arises as a result of interference between discrete eigenstates.
The conductance formulae~(\ref{eq:conductance_G}) and~(\ref{eq:conductance_J}) have a square of the summation over the discrete eigenstates.
Therefore, we have crossing terms within a resonant-state pair (between a resonant state and an anti-resonant state), between two resonant-state pairs (two sets of a resonant state and an anti-resonant state), and between a resonant-state pair and a bound state.
We show in the present section that discrete eigenvalues decide the symmetry or the asymmetry of the conductance peaks in addition to the location of the conductance peaks, using several examples.
We thereby microscopically derive the Fano parameters that control the asymmetry of $\Lambda^2$.
We then relate the parameters to the Fano parameter that controls the asymmetry of the conductance as well as to Fano parameter that appears in Fano's original argument.\cite{Fano61}

We will stress two more points in the present section.
We will show that a sharp resonance peak due to a resonant state with a small imaginary part is strongly asymmetrized by a broad resonance peak nearby due to a resonant state with a large imaginary part.
A broad resonance peak is often left out from consideration as a background.
The present result shows that broad resonances can manifest themselves as the asymmetry of nearby sharp resonances and suggests that we may be able to detect a broad one from the Fano parameter of a sharp one.

The third point of the present section is the effect of an applied magnetic field.
We will show that an external magnetic field that causes an Aharonov-Bohm phase in the dot makes the Fano asymmetric parameter complex.
This result is indeed consistent with recent experimental observation.\cite{KAKI2002,KAKI2003,SAKKI2004}

In \S~\ref{subsec:point}, \S~\ref{subsec:T-shaped} and \S~\ref{subsec:Three}, we consider the system~(\ref{eq:Hamiltonian_original}) with the following restrictions: the two semi-infinite leads are attached to one site, which is denoted by $d_0$; the coupling $t_1=t_2=t$; the number of sites in the dot $N=1,2,3$ 
(and therefore, according to the argument developed in Appendix\ref{app:calceig}, the number of discrete eigenvalues $2N=2,4,6$). 
Because the two semi-infinite leads are attached to one site ``$0$" in these cases, the conductance is given by the simpler formula~\eqref{eq:conductance_J} and the system is free from the problem described in Appendix\ref{app:inf}.
We will microscopically derive the three types of the Fano parameter that controls the asymmetry of $\Lambda^2$.
In \S~\ref{subsec:FanoG} we then reveal the relation between the Fano parameter for $\Lambda^2$ and the Fano parameter for the conductance.

We consider in \S~\ref{subsec:two-contacts} a case where the two semi-infinite leads are attached to different sites $d_1$ and $d_2$ of a triangle (\textit{i.e.}\ $N=3$).
In this particular case, we set $t_1=t_2=t/2$ and thereby have six discrete eigenvalues; if we set $t_1=t_2=t$, two eigenvalues would tend to infinity as we describe in Appendix\ref{app:inf}.
We then apply an external magnetic field in \S~\ref{sec:mag} to the system in \S~\ref{subsec:two-contacts}.
This causes an Aharonov-Bohm phase in the triangle of the dot.
We will then find that the Fano parameter becomes complex under a magnetic field.

We finally consider in \S~\ref{subsec:effect of leads} the effect of changing $t_\alpha$ in \S~\ref{subsec:effect of leads}.
Throughout the present section, we computed the conductance using the Fisher-Lee relation~(\ref{eq:conductance}) and obtained all discrete eigenvalues solving eq.~(\ref{eqB171}).

\subsection{Point contact system: $N=1$}\label{subsec:point}
\begin{figure}[h]
\includegraphics[width=0.4\textwidth]{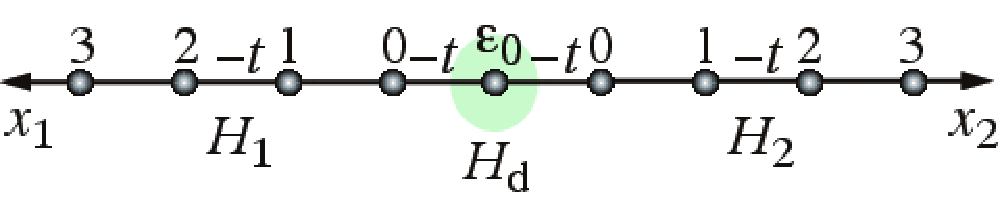}
\caption{(Color online) A point contact $d_0$.}
\label{fig:1QD}
\end{figure}

\begin{figure}[h]
\includegraphics[width=0.45\textwidth]{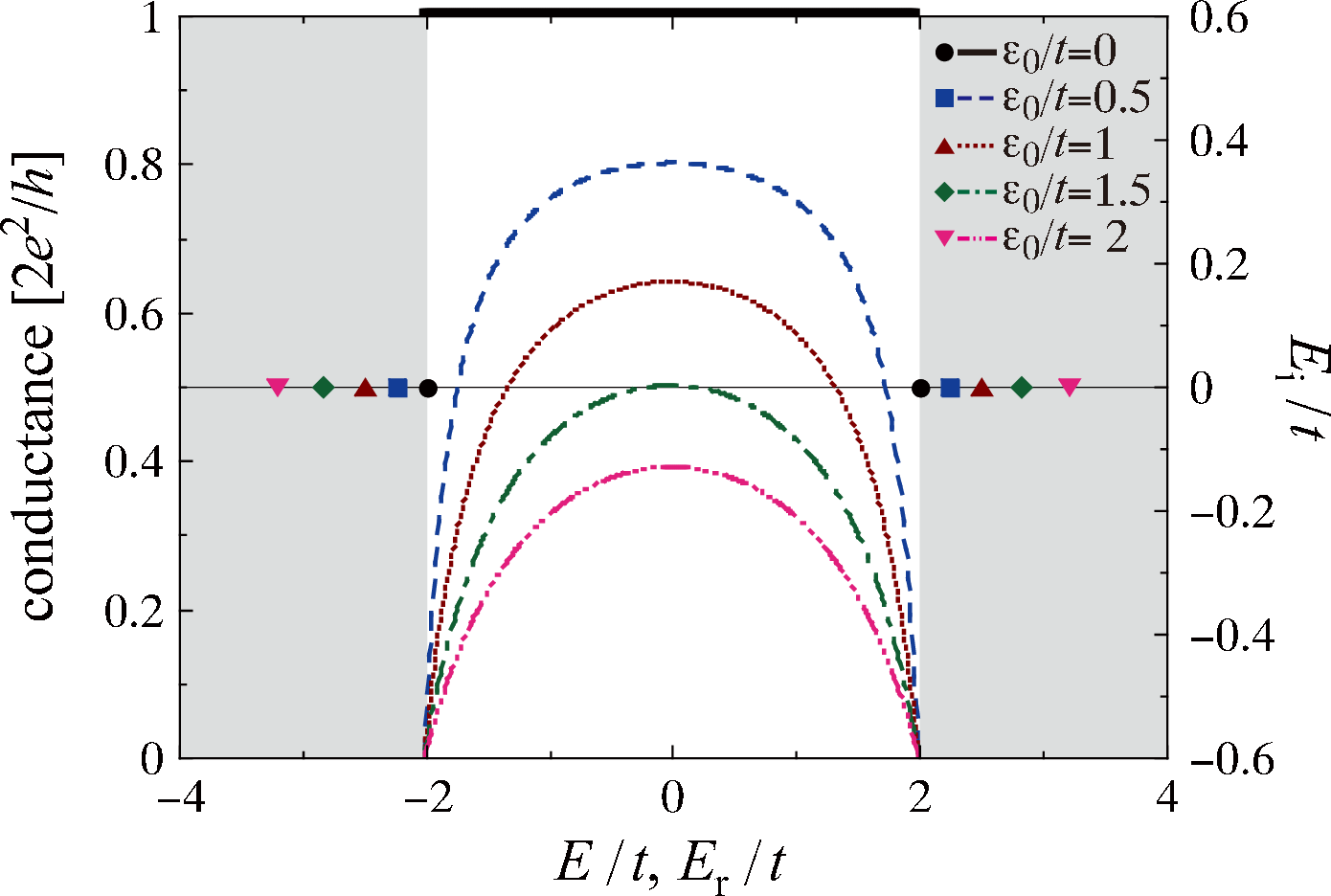}
\caption{(Color online) The energy dependence of the conductance (the left axis) and the discrete eigenvalues of the bound states (the right axis) for the one-site dot with $\varepsilon_0/t=0$, 1, 1.5, 2, 2.5.}
\label{fig:1QD_conductance} 
\end{figure}

First we show the conductance as well as the discrete eigenvalues of the one-site dot, namely the point contact shown in Fig.~\ref{fig:1QD}.
There are only two bound states and no resonant state.
We plot in Fig.~\ref{fig:1QD_conductance} the conductance with the eigenvalues of the two bound states for $\varepsilon_0/t=0$, 1, 1.5, 2, 2.5.
The conductance of the point contact has no peculiar behavior such as the Breit-Wigner peak or the Fano peak.
Upon increasing the potential $\varepsilon_0$, the eigenvalues of the two bound states move away from the branch points $E=\pm 2t$.
This decreases the contribution of the quantity
\begin{align}
\Lambda_{00}(E)=\sum_{p=1,2}\frac{\langle d_0|\psi_p^\textrm{b}\rangle\langle\tilde{\psi}_p^\textrm{b}|d_0\rangle}%
{E-E_p^\textrm{p}}
\end{align}
and hence deflates the conductance gradually.

\subsection{T-shaped quantum-dot system: $N=2$}\label{subsec:T-shaped}
\begin{figure}[h]
\includegraphics[width=0.4\textwidth]{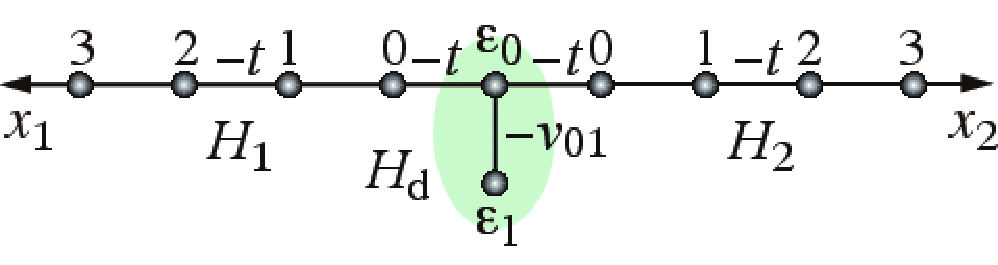}
\caption{(Color online) A two-site quantum dot.}
\label{fig:2QD}
\end{figure}

We next show the conductance and the discrete eigenvalues of the two-site quantum dot, namely a T-shaped (or side-coupled) quantum dot shown in Fig.~\ref{fig:2QD}.
This may be realized when the quantum point contact depicted in Fig.~\ref{fig:NQD} couples the quantum wire with an energy level in the quantum dot.
This system is a minimal model that possesses a resonant-state pair (a resonant state and the corresponding anti-resonant state) and is directly related to Fano's original argument~\cite{Fano61}.

\begin{figure}[!h]
\includegraphics[width=0.4\textwidth]{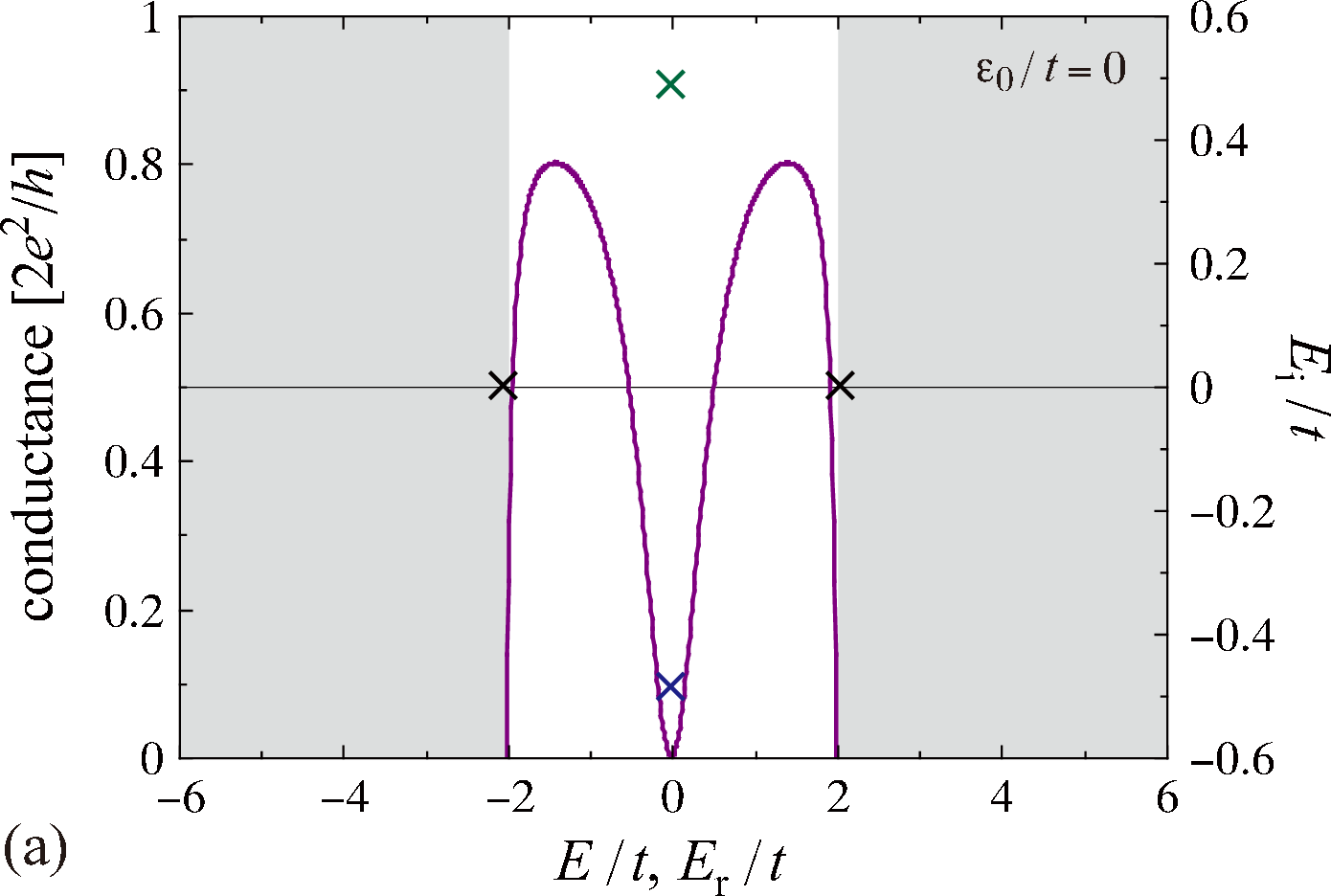}

\vspace*{\baselineskip}

\includegraphics[width=0.4\textwidth]{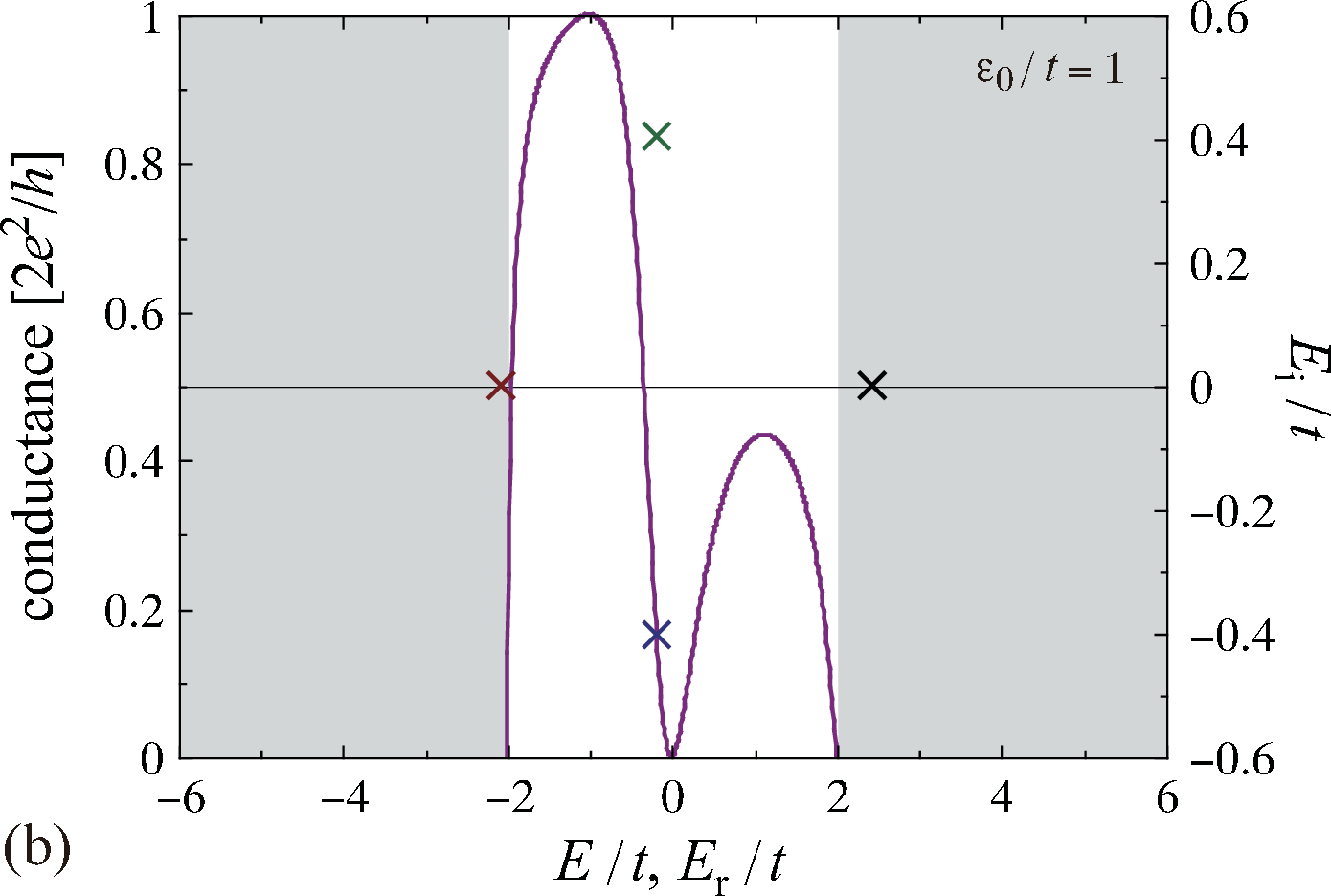}

\vspace*{\baselineskip}

\includegraphics[width=0.4\textwidth]{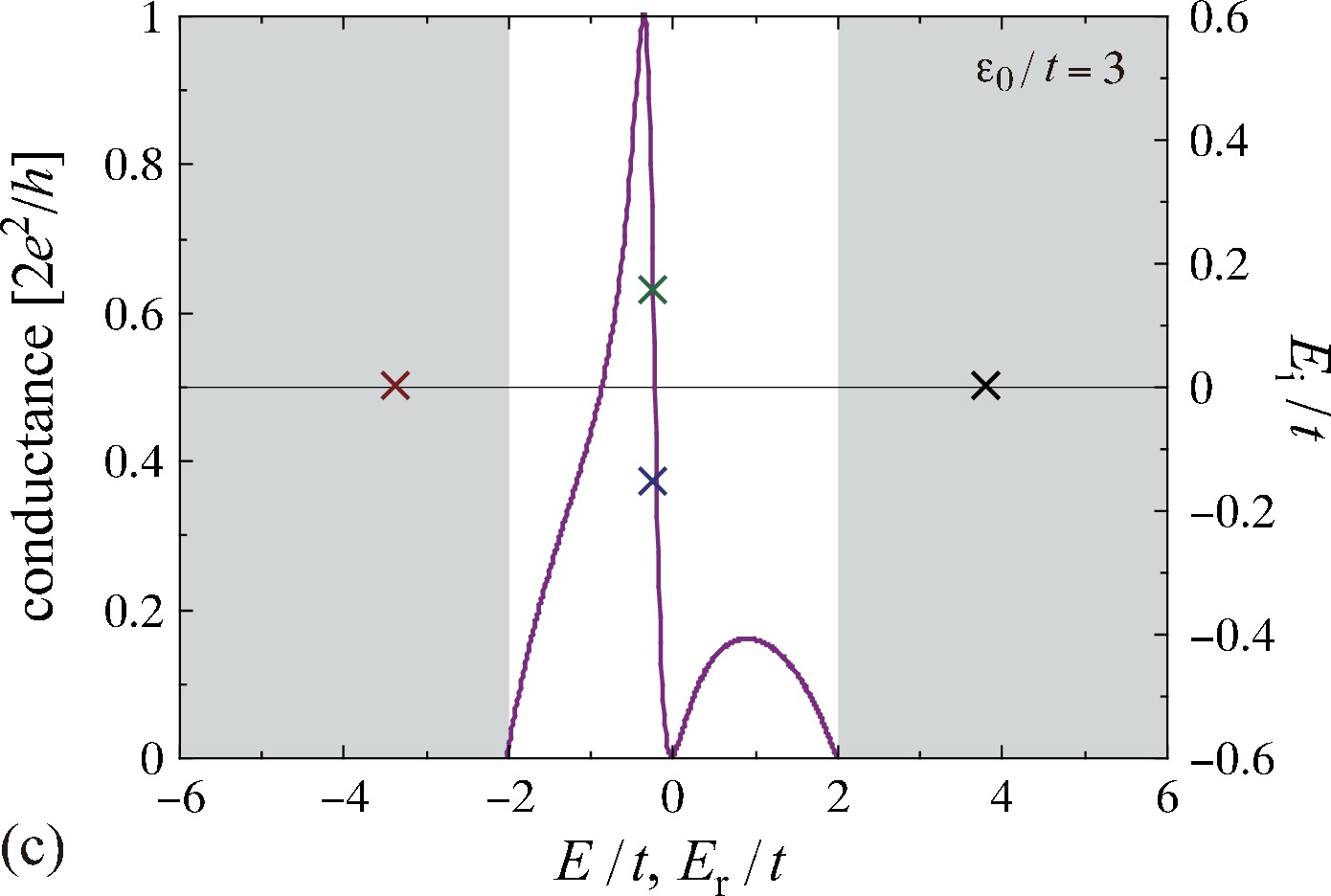}

\vspace*{\baselineskip}

\includegraphics[width=0.4\textwidth]{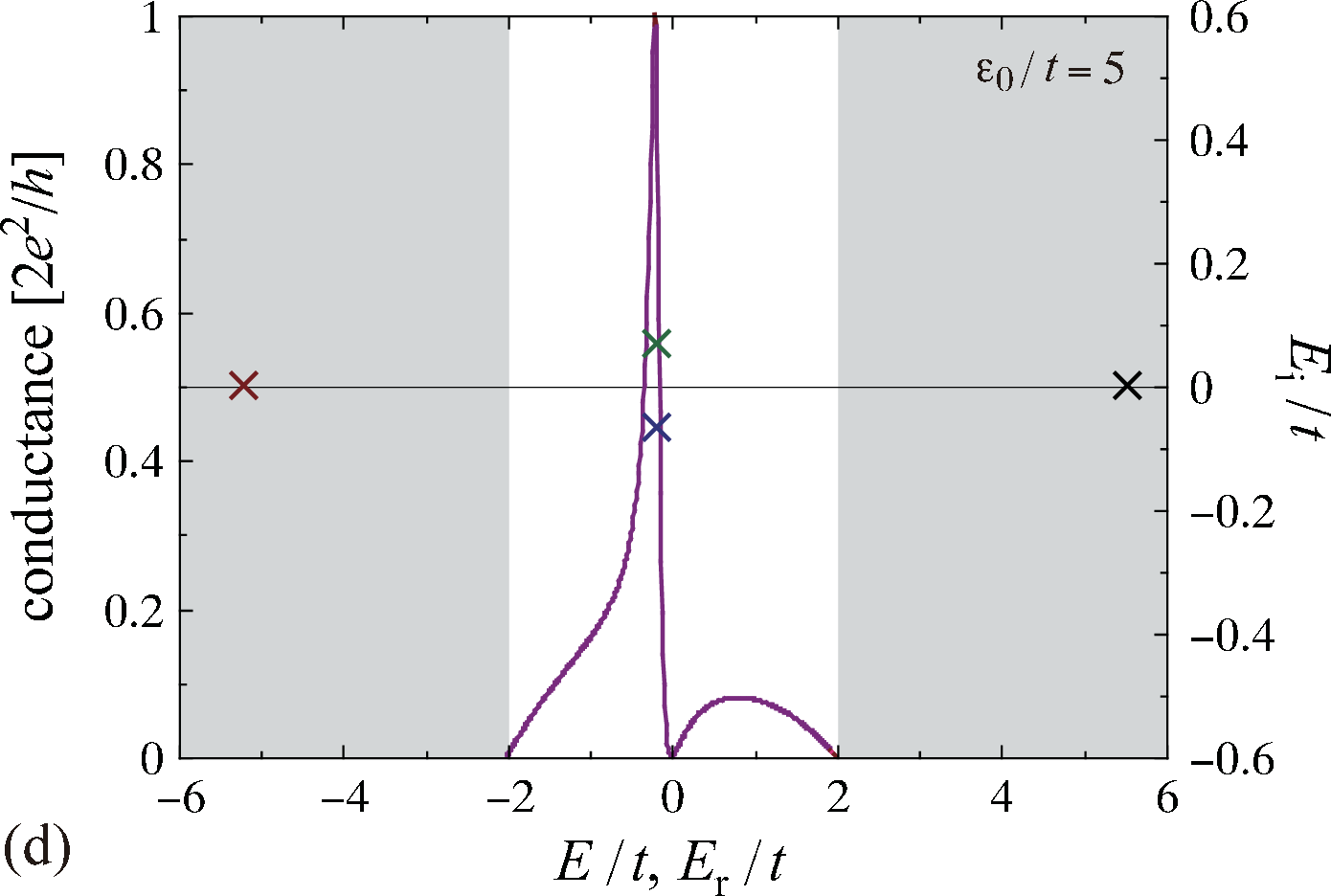}
\caption{(Color online) (a) The $\varepsilon_0$ dependence of the conductance (the left axis) and the discrete eigenvalues (the right axis)
for the two-site dot with (a) $\varepsilon_0/t=0$, (b) $\varepsilon_0/t=1$, (c) $\varepsilon_0/t=3$ and (d) $\varepsilon_0/t=5$.
Here we fixed $\varepsilon_1/t=0$ and $v_{01}/t=v_{10}/t=1$.}
\label{fig:2QD_eigen} 
\end{figure}

We plot in Fig.~\ref{fig:2QD_eigen} the conductance and all the discrete eigenvalues.
Since the dot Hamiltonian contains $N=2$ sites, the system has $2N=4$ pieces of discrete eigenstates, two of which are the resonant-state pair, $E^\textrm{res}$ and $E^\textrm{ar}$.
The other two eigenstates are both bound states, $E^\textrm{b}_1$ and $E^\textrm{b}_2$, for the parameter set in Fig.~\ref{fig:2QD_eigen}(a), that is, for $\varepsilon_0/t=0$, $\varepsilon_1=0$ and $v_{01}/t=v_{10}/t=1$.
As we increase $\varepsilon_0$, however, one of the bound states moves down on the $k=0$ axis onto the lower $k$ plane and becomes an anti-bound state, $E^\textrm{ab}$, whereas the other remains a bound state, $E^\textrm{b}$, in Fig.~\ref{fig:2QD_eigen}(b--d) for $\varepsilon/t=1,3,5$.

We have a Breit-Wigner dip for $\varepsilon_0=0$, but for $\varepsilon_0\neq 0$, we have an asymmetric peak, namely a Fano conductance peak.
Maruyama {\it et al.}~\cite{MSU2004} claimed that the asymmetry of the conductance peak of the T-shaped quantum dot is proportional to $\varepsilon_0$.
We here discuss the asymmetry from the viewpoint of interference among the discrete eigenstates.

\begin{figure}[t]
\includegraphics[width=0.45\textwidth]{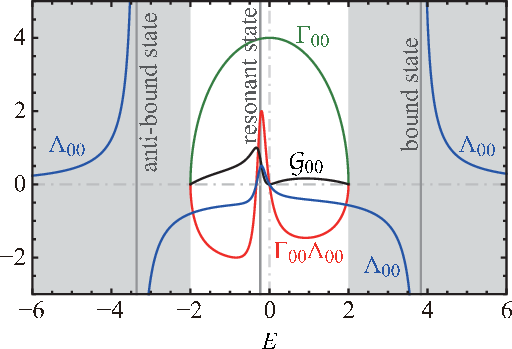}
\caption{(Color online) $\Lambda_{00}(E)$ (blue curve in the whole range with two singularities at the bound and anti-bound states), $\Gamma_{00}(E)$ (green oval curve), $\Gamma_{00}(E)\Lambda_{00}(E)$ (red curve) and the conductance $\mathcal{G}_{00}(E)$ (black curve in the range $|E|<2t$) for the two-site dot.
The parameter values are the same as in Fig.~\ref{fig:2QD_eigen}(c).
The gray vertical lines indicate the bound and anti-bound states ($E^\textrm{ab}=-3.36212\ldots$, $E^\textrm{b}=3.82578\ldots$) as well as the real part of the resonant-state pair ($E^\textrm{res/ar}=-0.23183\ldots\mp i0.154915\ldots$). 
}
\label{fig:DOS1}
\end{figure}

The conductance formula~(\ref{eq:conductance_J}) contains the square of the sum over the discrete eigenvalues of the form
\begin{align}\label{eq45}
\Lambda_{00}(E)^2=&\left(\Lambda^{\textrm{b}+\textrm{ab}}(E)+\Lambda^\textrm{pair}(E)\right)^2,
\end{align}
where
\begin{align}
\Lambda^{\textrm{b}+\textrm{ab}}(E)\equiv&\frac{\langle d_0|\psi^\textrm{b}\rangle\langle\tilde{\psi}^\textrm{b} |d_0\rangle}{E-E^\textrm{b}}+\frac{\langle d_0|\psi^\textrm{ab}\rangle\langle\tilde{\psi}^\textrm{ab} |d_0\rangle}{E-E^\textrm{ab}},
\\ \label{eq:rho_cross_N=2}
\Lambda^\textrm{pair}(E)\equiv&\frac{\langle d_0|\psi^\textrm{res}\rangle\langle\tilde{\psi}^\textrm{res} |d_0\rangle}{E-E^\textrm{res}}
+\frac{\langle d_0|\psi^\textrm{ar}\rangle\langle\tilde{\psi}^\textrm{ar} |d_0\rangle}{E-E^\textrm{ar}}.
\end{align}
Since the conductance formula~(\ref{eq:conductance_J}) is given in the form
\begin{align}
\mathcal{G}_{00}\propto\frac{{\Lambda_{00}}^2}{1\mp\sqrt{1-\left(\frac{\Gamma_{00}\Lambda_{00}}{2}\right)^2}},
\end{align}
the symmetry or the asymmetry of the quantity $\Lambda_{00}(E)^2$ is directly reflected on the symmetry or the asymmetry of the conductance peak; 
see Fig.~\ref{fig:DOS1}.
Equation~(\ref{eq45}) therefore implies that the symmetry or the asymmetry of the conductance peak is strongly affected by crossing terms, or the interference between states with discrete eigenvalues.
We hereafter show that the Fano conductance peak arises from two types of interference, or two types of crossing terms.
First, we have a crossing term within the resonant-state pair, or the interference between the resonant state and the anti-resonant state. 
Second, we have a crossing term between the bound and anti-bound states and the resonant-state pair.

We compare in Fig.~\ref{fig:N=2_ta=1_lead=2_e0=5_cross} the following quantities:
\begin{figure}[h]
\includegraphics[width=0.45\textwidth]{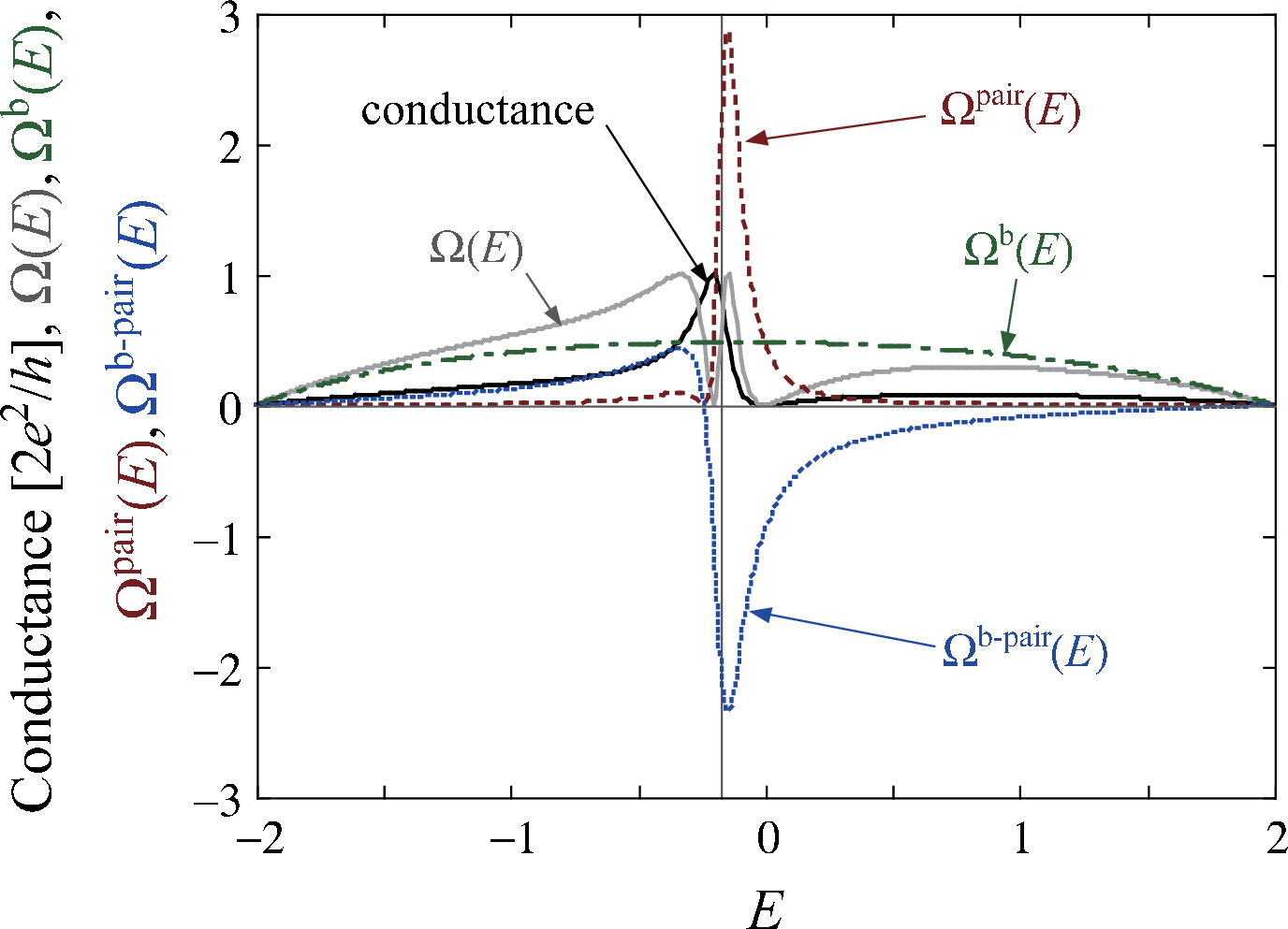}
\caption{(Color online) The quantities $\Omega(E)$ (gray curve), $\Omega^\textrm{b}(E)$ (chained green curve), $\Omega^\textrm{pair}(E)$ (broken red curve) and $\Omega^\textrm{b-pair}(E)$ (dotted blue curve), defined in eqs.~(\ref{eq45})--(\ref{eq48}), plotted with the conductance (solid black curve), eq.~(\ref{eq:conductance}), or eq.~(\ref{eq:conductance_J}).
The system is the two-site quantum dot.
We fixed $\varepsilon_0/t=5$, $\varepsilon_1/t=0$ and $v_{01}/t=v_{10}/t=1$.
The gray vertical line indicates the real part of the resonant eigenvalue $E^\textrm{res}_\textrm{r}=-0.172712\ldots$.}
\label{fig:N=2_ta=1_lead=2_e0=5_cross} 
\end{figure}
\begin{align} 
\Omega(E)\equiv&\left(\frac{{\Gamma_{00}}{\Lambda_{00}}}{2}\right)^2
\nonumber\\
=&\frac{{\Gamma_{00}}^2}{4}\left(\Lambda^{\textrm{b}+\textrm{ab}}(E)+\Lambda^\textrm{pair}(E)\right)^2,
\\\label{eq46}
\Omega^\textrm{b}(E)\equiv&\frac{{\Gamma_{00}}^2}{4}\Lambda^{\textrm{b}+\textrm{ab}}(E)^2,
\\ \label{eq47}
\Omega^\textrm{pair}(E)\equiv&\frac{{\Gamma_{00}}^2}{4}\Lambda^\textrm{pair}(E)^2,
\\ \label{eq48}
\Omega^\textrm{b-pair}(E)\equiv&\frac{{\Gamma_{00}}^2}{2}\Lambda^{\textrm{b}+\textrm{ab}}(E)\Lambda^\textrm{pair}(E).
\end{align}
Note that the third quantity~(\ref{eq47}) contains a crossing term between the resonant state and the anti-resonant state.
The fourth quantity~(\ref{eq48}) contains crossing terms between the resonant state and a bound state as well as crossing terms between the anti-resonant state and a bound state.
We can see in Fig.~\ref{fig:N=2_ta=1_lead=2_e0=5_cross} that the asymmetry of the conductance peak comes partly from the asymmetry of the term $\Omega^\textrm{pair}(E)$ and partly from the crossing term $\Omega^\textrm{b-pair}(E)$.
The quantity $\Omega^\textrm{b}(E)$ is almost symmetric.

In order to derive the Fano parameters for the asymmetry of the two terms  $\Omega^\textrm{pair}(E)$ and $\Omega^\textrm{b-pair}(E)$ microscopically, we expand the terms~(\ref{eq47}) and~(\ref{eq48}) in the neighborhood of $E=E_\textrm{r}^\textrm{res}=E_\textrm{r}^\textrm{ar}$ by using the normalized energy
\begin{align}\label{eq57-1}
\tilde{E}\equiv \frac{E-E_\textrm{r}^\textrm{res}}{\left|E_\textrm{i}^\textrm{res}\right|}.
\end{align} 
We first rewrite $\Lambda^\textrm{pair}(E)$ in the forms
\begin{align}
\Lambda^\textrm{pair}(E)=\frac{\tilde{N}\mathrm{e}^{i\theta}}{E-\left(E_\textrm{r}^\textrm{res}+iE_\textrm{i}^\textrm{res}\right)}+\textrm{c.c.},
\end{align}
where we express the coefficient of the local density of the resonant state with the amplitude $\tilde{N}$ and the phase $\theta$:
\begin{align}\label{eqNtheta}
\tilde{N}\mathrm{e}^{i\theta}\equiv\langle d_0|\psi^\textrm{res}\rangle\langle\tilde{\psi}^\textrm{res}|d_0\rangle.
\end{align}
Note that this is generally a complex number because the left-eigenvector $\langle\tilde{\psi}^\textrm{res}|$ is not generally Hermitian conjugate to the right-eigenvector $|\psi^\textrm{res}\rangle$ for a resonant state (see eq.~(\ref{eq:eigenfunction_anti-resonant})).
We then rewrite the local density of the resonant-state pair in the form
\begin{align}
\Lambda^\textrm{pair}(E)&=2\tilde{N}\frac{(E-E_\textrm{r}^\textrm{res})\cos \theta +\left|E_\textrm{i}^\textrm{res}\right|\sin\theta}{\left(E-E_\textrm{r}^\textrm{res}\right)^2+\left|E_\textrm{i}^\textrm{res}\right|^2}  \nonumber \\
&=2\frac{\tilde{N}}{\left|E_\textrm{i}^\textrm{res}\right|}\frac{\sin\theta+\tilde{E}\cos \theta  }{1+\tilde{E}^2},
\label{LamPair}
\end{align}
or
\begin{align}\label{eq550}
\Omega^\textrm{pair}(E)\equiv&\frac{{\Gamma_{00}}^2}{4}\Lambda^\textrm{pair}(E)^2
\propto \left(\frac{q^\textrm{pair}+\tilde{E}}{1+\tilde{E}^2}\right)^2,
\end{align}
where
\begin{equation}\label{eq560}
q^\textrm{pair}\equiv\tan\theta.
\end{equation}
The parameter~(\ref{eq560}) controls the asymmetry of the term~(\ref{eq47}) and hence may be called the Fano parameter, although eq.~(\ref{eq550}) is different from the form originally derived by Fano~\cite{Fano61}:
\begin{equation}\label{eq:Fano}
\mathcal{G}(E) \sim \frac{\left(q+\tilde{E}\right)^2}{1+\tilde{E}^2}.
\end{equation}
Indeed, the asymmetry caused by the above interference between a resonant state and the corresponding anti-resonant state is not seen in Fano's argument. (We will come back to this point in \S~\ref{subsec:FanoG}.)

On the other hand, the crossing term~(\ref{eq48}) produces asymmetry of Fano's original form~(\ref{eq:Fano}).
In order to see this, we approximate the local density of the bound and anti-bound states as
\begin{align}
 \Lambda^{\textrm{b}+\textrm{ab}}(E)&\simeq\Lambda^{\textrm{b}+\textrm{ab}}(E_\textrm{r}^\textrm{res})+{\Lambda^{\textrm{b}+\textrm{ab}}}'(E_\textrm{r}^\textrm{res}) 
\left|E_\textrm{i}^\textrm{res}\right| \tilde{E} \nonumber\\
&+ \frac{1}{2}{\Lambda^{\textrm{b}+\textrm{ab}}}''(E_\textrm{r}^\textrm{res}) 
\left|E_\textrm{i}^\textrm{res}\right|^2\tilde{E}^2
\end{align}
in the neighborhood of $E=E_\textrm{r}^\textrm{res}$.
We therefore have the crossing term between the resonant-state pair and the the bound and anti-bound states as
\begin{align}\label{eq56}
\Omega^\textrm{b-pair}(E)\equiv&\frac{{\Gamma_{00}}^2}{2}\Lambda^{\textrm{b}+\textrm{ab}}(E)\Lambda^\textrm{pair}(E)\sim\frac{r +s\tilde{E} +t\tilde{E}^2}{1+\tilde{E}^2},
\end{align}
where
\begin{align}
r&\equiv
\frac{\Lambda^{\textrm{b}+\textrm{ab}}(E_\textrm{r}^\textrm{res})}{\left|E_\textrm{i}^\textrm{res}\right|}\sin\theta, \\
s&\equiv
\frac{\Lambda^{\textrm{b}+\textrm{ab}}(E_\textrm{r}^\textrm{res})}{\left|E_\textrm{i}^\textrm{res}\right|}\cos\theta
+{\Lambda^{\textrm{b}+\textrm{ab}}}'(E_\textrm{r}^\textrm{res})\sin\theta, \label{eq:s}\\
t&\equiv
{\Lambda^{\textrm{b}+\textrm{ab}}}'(E_\textrm{r}^\textrm{res})\cos \theta  + \frac{1}{2}{\Lambda^{\textrm{b}+\textrm{ab}}}''(E_\textrm{r}^\textrm{res}) 
\left|E_\textrm{i}^\textrm{res}\right|\sin\theta
\end{align}
In order to derive a Fano parameter $q^\textrm{b-pair}$ that controls the asymmetry of the term $\Omega^\textrm{b-pair}(E)$, we extract the form on the right-hand side of eq.~(\ref{eq:Fano}) by putting
\begin{equation}\label{eq7100}
\frac{r +s\tilde{E} +t\tilde{E}^2}{1+\tilde{E}^2}=a+b\frac{\left(q^\textrm{b-pair}+\tilde{E}\right)^2}{1+\tilde{E}^2},
\end{equation}
We obtain the Fano parameter $q^\textrm{b-pair}$ by solving the equation
\begin{equation}\label{eq63-1}
s\left(q^\textrm{b-pair}\right)^2-2(r-t)q^\textrm{b-pair}-s=0
\end{equation}
and choose the solution in the range $-1<q^\textrm{b-pair}<1$.
This controls the asymmetry of the term~(\ref{eq48}), a Fano parameter that is different from the one given by eq.~(\ref{eq560}), but that conforms to Fano's original form~(\ref{eq:Fano}).

\begin{figure}[h]
\includegraphics[width=0.45\textwidth]{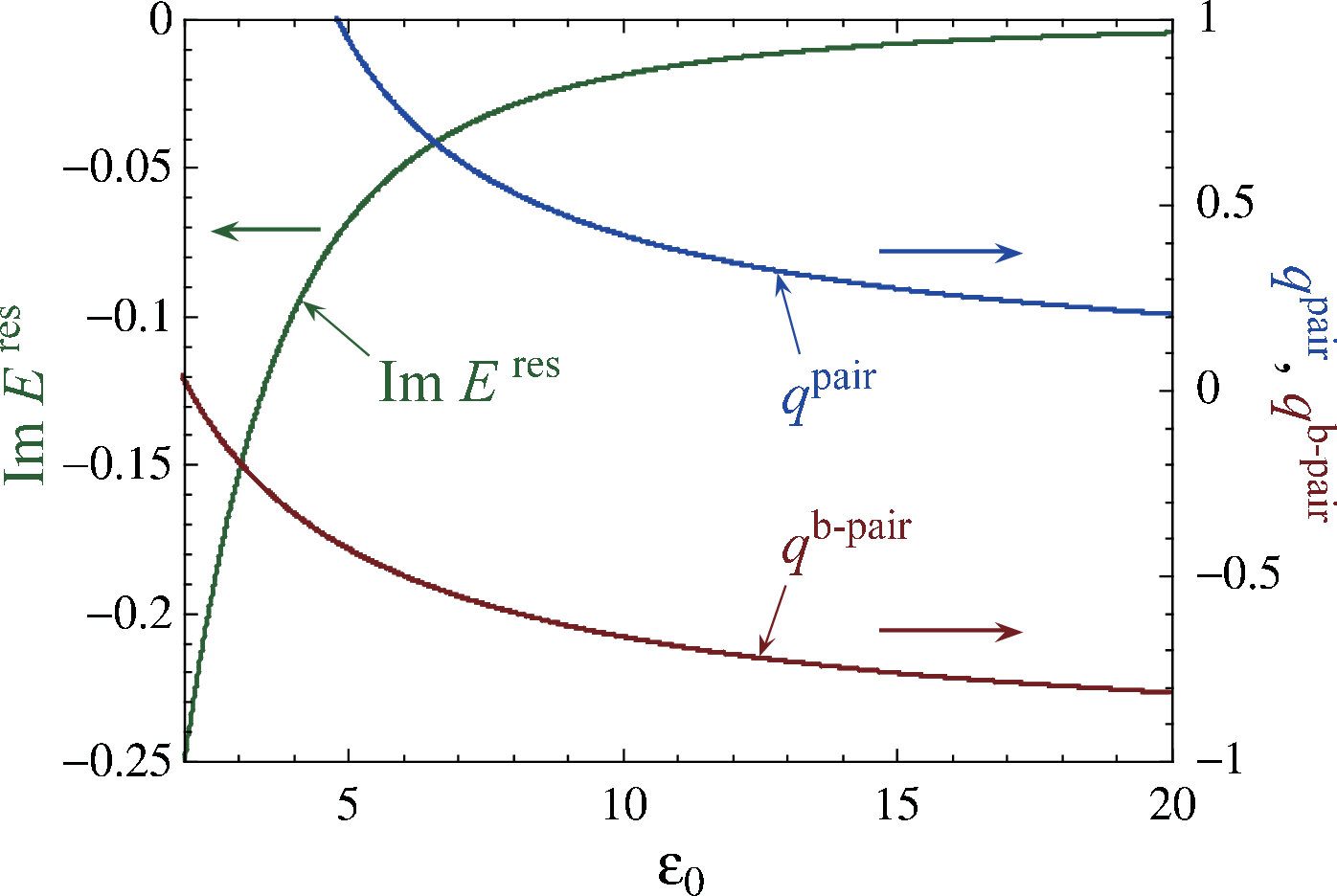}
\caption{(Color online) The Fano parameters $q^\textrm{pair}$ (blue curve) and $q^\textrm{b-pair}$ (red curve) for the two-site dot, plotted with the imaginary part of the resonant eigenvalue $E_\textrm{i}^\textrm{res}$.
Use the right axis for the Fano parameters and the left axis for the resonant eigenvalue.
(Note that the horizontal axis does not start from $\varepsilon_0=0$.
We omitted the part $0\leq\varepsilon_0<2$ to avoid confusion because the structure of the spectrum is drastically different in that region.)
We fixed $\varepsilon_1/t=0$ and $v_{01}/t=v_{10}/t=1$.}
\label{fig12}
\end{figure}

We show in Fig.~\ref{fig12} how the two Fano parameters $q^\textrm{pair}$ and $q^\textrm{b-pair}$ depend on the system parameter $\varepsilon_0$.
In the particular case of Fig.~\ref{fig12}, $q^\textrm{b-pair}$ tends to dominate over $q^\textrm{pair}$ as we increase the system parameter $\varepsilon_0$.
In fact, comparing the two Fano profiles
\begin{align}\label{eq:fanoprofiles}
\frac{\left(q+\tilde{E}\right)^2}{1+\tilde{E}^2}
\quad\mbox{and}\quad
\frac{\left(q+\tilde{E}\right)^2}{\left(1+\tilde{E}^2\right)^2},
\end{align}
we see that the latter for $\Omega^\textrm{pair}$ in eq.~\eqref{eq550} is more localized than the former for $\Omega^\textrm{b-pair}$ in eq.~\eqref{eq7100}, because in the limit $|\tilde{E}|\to\infty$, the former goes to unity but the latter goes to zero.
As $\varepsilon_0$ increases, therefore, the former Fano profile with a negative $q^\textrm{b-pair}$ determines the resulting conductance profile in Fig.~\ref{fig:2QD_eigen}(d).

The development of the Fano profile is in coordination with the decrease of $\left|E_\textrm{i}^\textrm{res}\right|$.
We can see in eq.~(\ref{eq:s}) that a small imaginary part $\left|E_\textrm{i}^\textrm{res}\right|$ causes a particularly strong asymmetry of the term $\Omega^\textrm{b-pair}(E)$.
This is indeed demonstrated in Fig.~\ref{fig:2QD_eigen}, where, as we increase $\varepsilon_0$, the asymmetry rapidly develops while the the resonant eigenvalue approaches the real axis.

Incidentally, the present system has the particle-hole symmetry $E\leftrightarrow -E$ for $\varepsilon_0=\varepsilon_1=0$, and hence $q^\textrm{pair}=q^\textrm{b-pair}= 0$,
for which the resonance peak takes the form of a symmetric Lorentzian as shown in Fig.~\ref{fig:2QD_eigen}(a).

\subsection{Relating the Fano parameter of the conductance to the microscopic parameter $q^{\rm pair}$.}
\label{subsec:FanoG}

\begin{figure}
\includegraphics[width=0.45\textwidth]{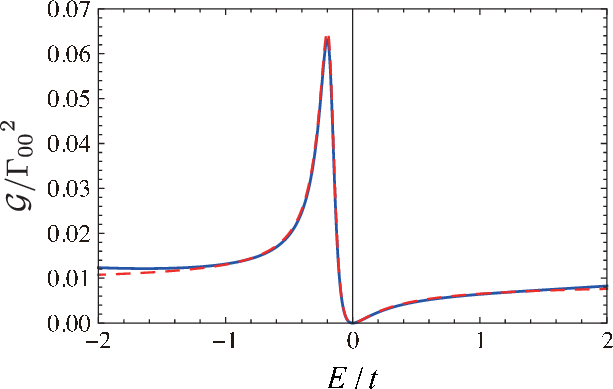}
\caption{(Color online) The function $\mathcal{G}(E)/{\Gamma_{00}}^2(E)$ (blue solid line)  and its fit by the Fano shape in eq.~(\ref{eq:FanoFit}) (red dashed line); 
the parameters of the quantum dot are the same as in Fig.~\ref{fig:2QD_eigen}(d). 
The fitting parameters are $A=0.009$, $E_{\rm r}^{\rm res}=0.1727$, $E_{\rm i}^{\rm res}=-0.06895$ and $q^{\mathcal{G}} = -2.505$.}
\label{fig:FanoFit}
\end{figure}

We will relate the Fano parameter controlling the asymmetry of the conductance $\mathcal{G}$ to the Fano parameter $q^{\rm pair}$ for the ``T''-shaped quantum dot considered in \S~\ref{subsec:T-shaped}. 
First note that in the present case the conductance $\mathcal{G}$ is proportional to $\Gamma_{00}^2 \left| G^R_{00}\right|^2$. 
When the energy is far from the band edges, the factor $\Gamma_{00}^2$ is approximately constant. 
The conductance is then essentially proportional to $\left| G^R_{00}\right|^2$. 
This quantity may be fit very accurately by a Fano shape, 
\begin{align}\label{eq:FanoFit}
  \left| G^R_{00}(E)\right|^2 \approx A \frac{\left({\tilde E} + q^{\mathcal{G}}\right)^2}{{\tilde E }^2 +1},
\end{align}
where $A$ and $q^{\mathcal{G}}$ are constants; 
see Fig.~\ref{fig:FanoFit}. 
The reason is that  the present case fits Fano's original consideration; 
there is only one isolated resonance interfering with a background.  

Looking at Fig.~\ref{fig:FanoFit},  we see that the conductance vanishes at $E=0$, which corresponds to the normalized energy ${\tilde E}=-E_{\rm r}^{\rm res}/E_{\rm i}^{\rm res}$.  
This means that the Fano parameter of the conductance is given by $q^{\mathcal{G}}=E_{\rm r}^{\rm res}/E_{\rm i}^{\rm res}$. 
On the other hand, when the conductance vanishes, $\Lambda$ must vanish as well; see eq.~(\ref{eq:conductance_J}). 
This implies that $\Lambda(E)=\Lambda^{\rm b + ab}(E)+\Lambda^{\rm pair}(E) =0$ for $E=0$, which with eq.~(\ref{LamPair}) gives
\begin{align}\label{eq83}
q^{\mathcal{G}} = q^{\rm pair} +  \frac{\cos\theta}{2{\tilde N}}
\frac{(E_{\rm r}^{\rm res})^2 + (E_{\rm i}^{\rm res})^2}{\left|E_i^{\rm res}\right |} \Lambda^{\rm b+ab}(0).
\end{align}
The Fano parameter of the conductance $q^{\mathcal{G}}$ may be found experimentally \cite{SAKKI2004}. 
Equation~\eqref{eq83} then shows that it is possible to connect the parameter $q^\mathcal{G}$ to the microscopic parameter $q^{\rm pair}$ that we have derived here.

We remark that  the function $\Lambda_{00}(E)^2$ contains the shape in eq.~(\ref{eq550}), which does not appear in the conductance $\mathcal{G}$.  
To understand why this is so, we invert the relation~\eqref{eq:C90} between $\Lambda_{00}$ and the Green's function to obtain
\begin{align}\label{eq:LGG}
 \Lambda_{00}(E)^2 = 4 \left| G^R_{00}(E)\right |^2 - \Gamma_{00}^2  \left| G^R_{00}(E)\right |^4
\end{align}
This relation shows that although in the present case $ \left| G^R_{00}(E)\right |^2$ is close to Fano's original shape, the function $\Lambda_{00}(E)^2$ contains the \textit{square} of Fano's original shape, $\left| G^R_{00}(E)\right |^4$, which ultimately gives the shape   in eq. (\ref{eq550}), corresponding to  the term $\Omega^{\rm pair}(E)$. 
In more complicated quantum dots than the present case, this term may produce the conductance shapes that are different from Fano's original shape. 

\subsection{Three-site quantum-dot system: $N=3$}\label{subsec:Three}
\begin{figure}
\includegraphics[width=0.4\textwidth]{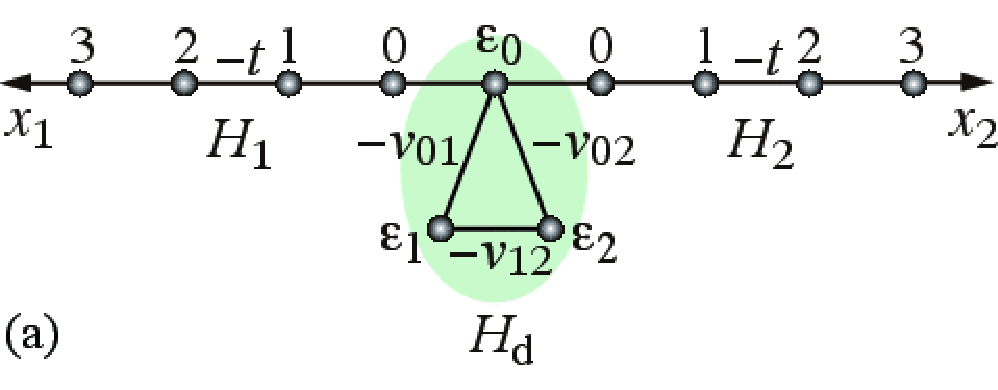}

\vspace*{\baselineskip}

\includegraphics[width=0.3\textwidth]{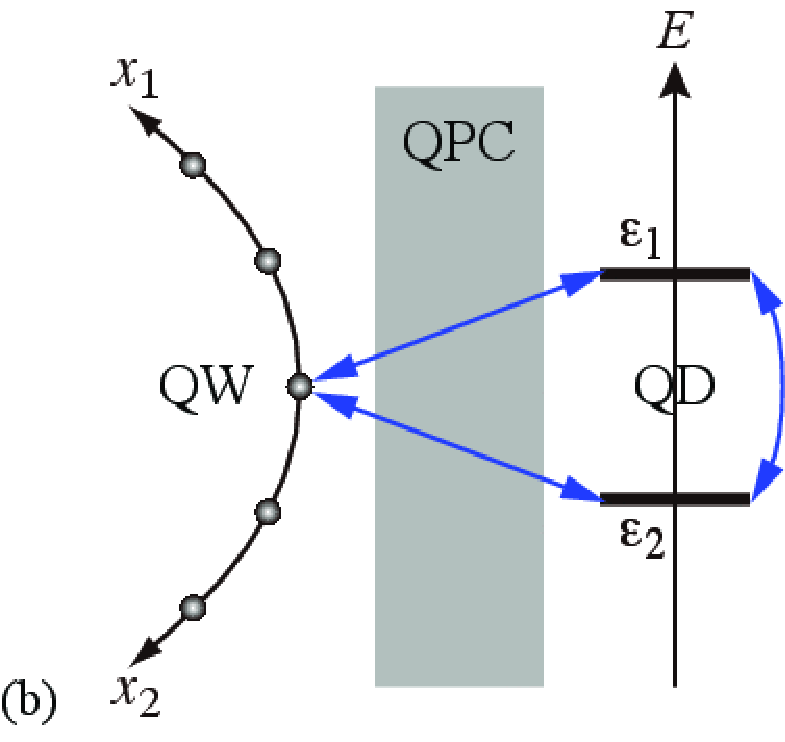}
\caption{(Color online) (a) The three-site quantum dot.
(b) The three-site dot may be realized when the quantum point contact (QPC) depicted in Fig.~\ref{fig:NQD} couples the quantum wire (QW) with two energy levels in the quantum dot (QD).}
\label{fig:3QD}
\end{figure}

Third, we discuss the conductance of the three-site quantum dot shown in Fig.~\ref{fig:3QD}(a).
The three-site quantum dot may correspond to the situation where the gate voltage of the quantum dot is adjusted so that two energy levels of the dot can couple to the quantum wire through a quantum point contact (Fig.~\ref{fig:3QD}(b)).

\begin{figure}
\includegraphics[width=0.4\textwidth]{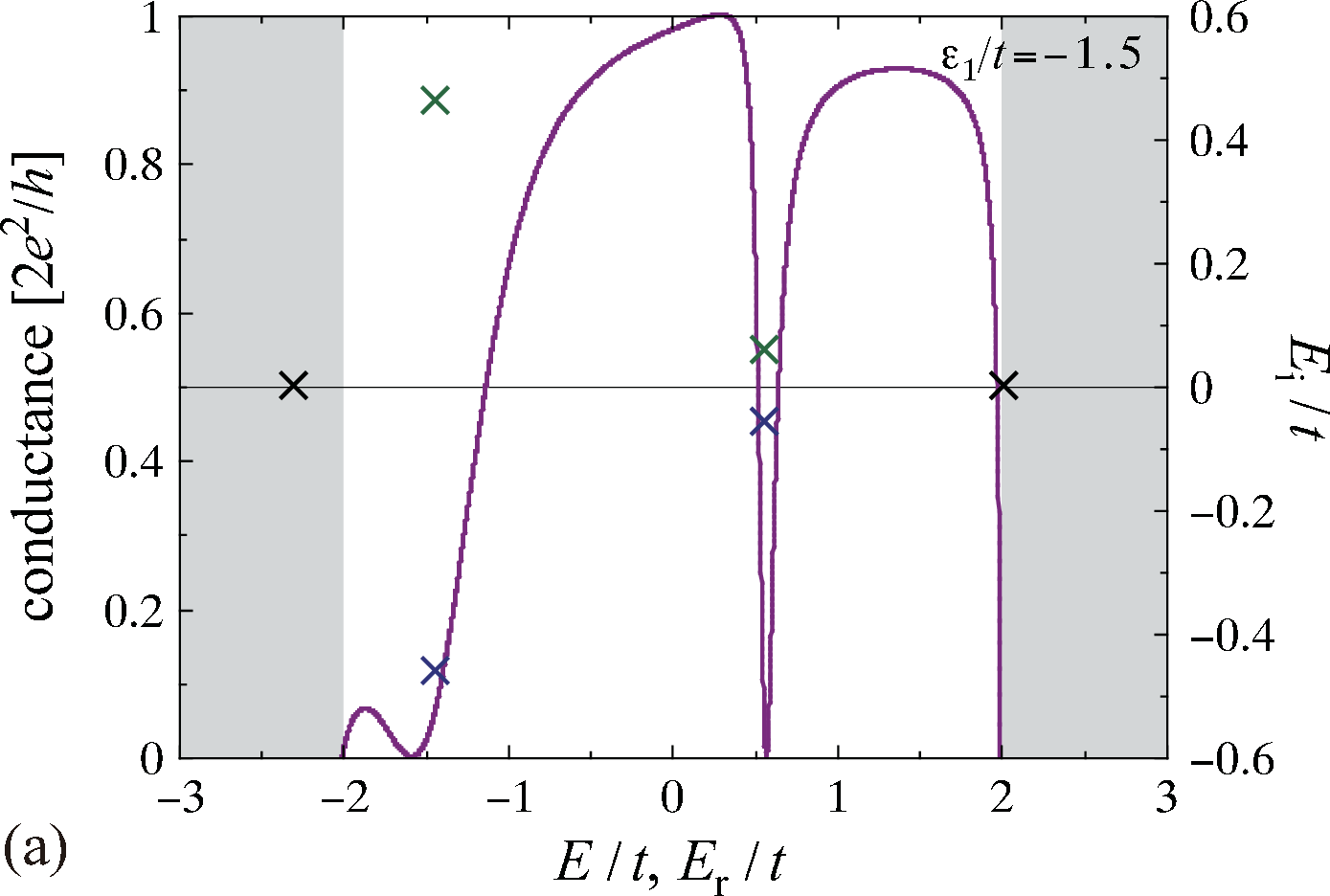}

\vspace*{\baselineskip}

\includegraphics[width=0.4\textwidth]{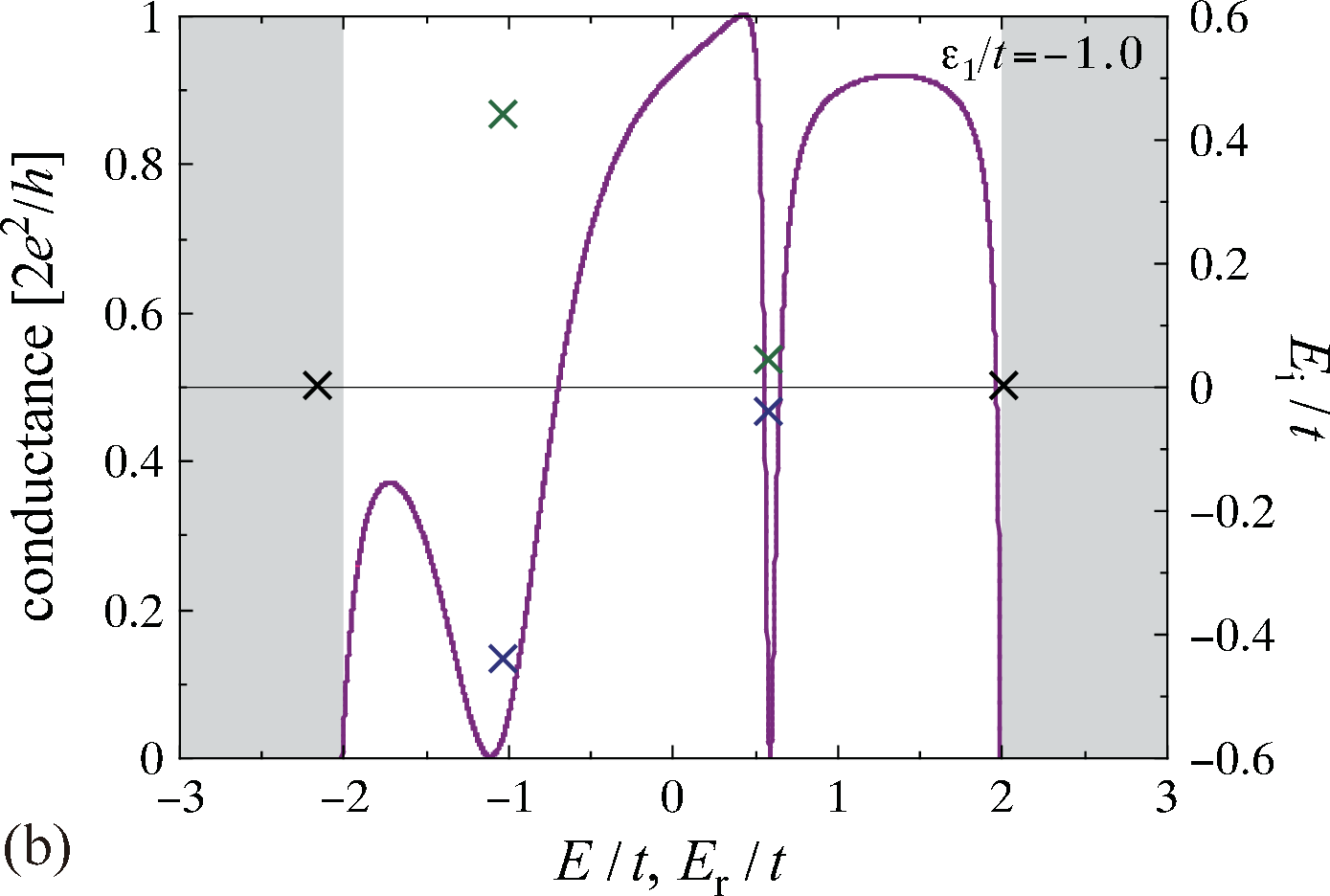}

\vspace*{\baselineskip}

\includegraphics[width=0.4\textwidth]{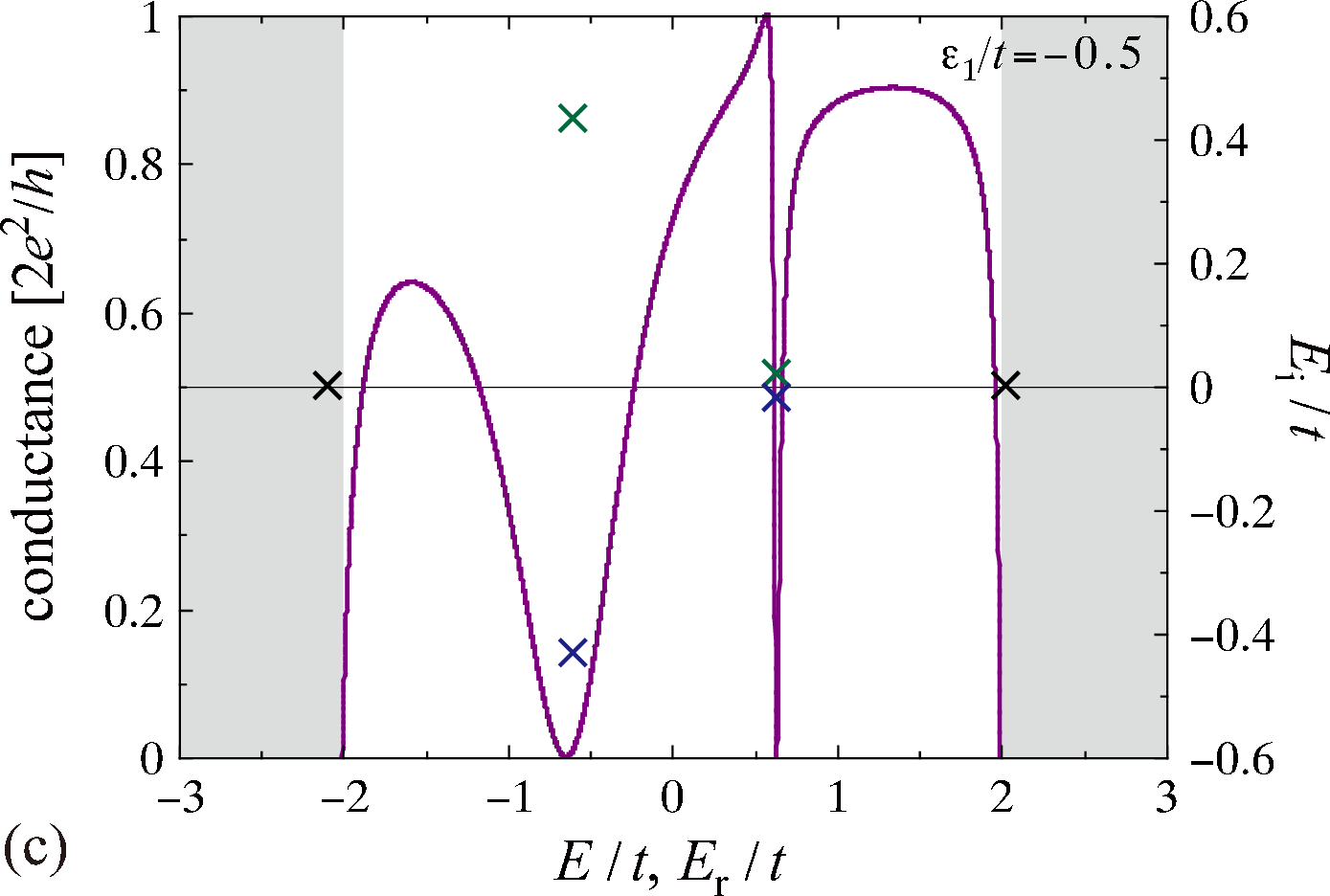}

\vspace*{\baselineskip}

\includegraphics[width=0.4\textwidth]{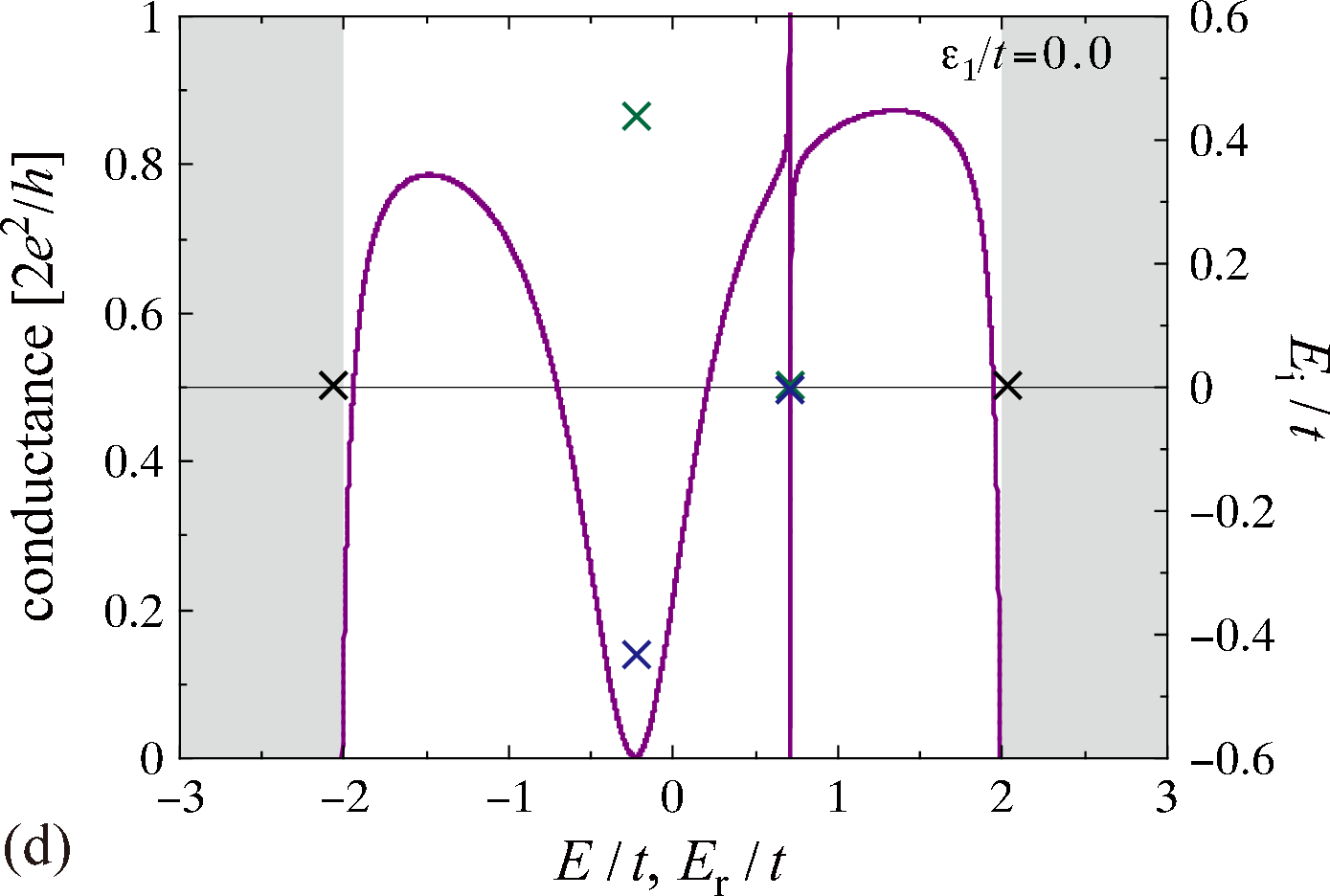}
\caption{(Color online) The conductance (curve for the left axis) for the three-site dot with (a) $\varepsilon_1/t=-1.5$, (b) $\varepsilon_1/t=-1.0$, (c) $\varepsilon_1/t=-0.5$ and (d) $\varepsilon_1/t=0$, plotted with all the discrete eigenvalues (crosses for the right axis).
We fixed $\varepsilon_0/t=0$, $\varepsilon_2/t=0.5$, $v_{01}/t=v_{10}/t=0.8$, $v_{02}/t=v_{20}/t=0.5$ and $v_{12}/t=v_{21}/t=0.4$.
}
\label{fig:N=3_ta=1_lead=2_e1_dependent} 
\end{figure}

The three-site system ($N=3$) has six discrete eigenvalues in total ($2N=6$), out of which are two resonant states for some parameter values.
This situation was not considered in Fano's argument~\cite{Fano61}.
We show in Fig.~\ref{fig:N=3_ta=1_lead=2_e1_dependent} the conductance, the eigenvalues of the two bound states, which are denoted by $E^\textrm{b}_1$ and $E^\textrm{b}_2$, as well as the eigenvalues of the two resonant-state pairs, which are denoted by $E^\textrm{res}_1$, $E^\textrm{ar}_1$, $E^\textrm{res}_2$ and $E^\textrm{ar}_2$, for $\varepsilon_1/t=-1.5$, $-1$, $-0.5$, $0$ with $\varepsilon_0/t=0$, $\varepsilon_2/t=0.5$, $v_{01}/t=v_{10}/t=0.8$, $v_{02}/t=v_{20}/t=0.5$ and $v_{12}/t=v_{21}/t=0.4$.
Upon increasing the parameter $\varepsilon_1$, the conductance dip that is generated by the resonant state on the left-hand side, $E^\textrm{res}_1$, approaches to the other conductance dip that is generated by the resonant state on the right-hand side, $E^\textrm{res}_2$.
Then the latter conductance peak develops strong asymmetry.

For the present system, we have yet another Fano parameter due to a crossing term between one resonant-state pair and the other resonant-state pair. 
The conductance formula~(\ref{eq:conductance_J}) contains the square of the sum over the discrete eigenvalues of the form
\begin{align}\label{eq60}
\Lambda_{00}(E)^2=\left(\Lambda^\textrm{b}(E)+\Lambda^\textrm{pair}_1(E)+\Lambda^\textrm{pair}_2(E)\right)^2,
\end{align}
where
\begin{align}
\Lambda^\textrm{b}(E)\equiv&\sum_{p=1,2}\frac{\langle d_0|\psi^\textrm{b}_p\rangle\langle\tilde{\psi}^\textrm{b}_p |d_0\rangle}{E-E^\textrm{b}_p},
\\ \label{eq:rho_cross_N=3}
\Lambda^\textrm{pair}_l(E)\equiv&\frac{\langle d_0|\psi^\textrm{res}_l\rangle\langle\tilde{\psi}^\textrm{res}_l |d_0\rangle}{E-E^\textrm{res}_l}
+\frac{\langle d_0|\psi^\textrm{ar}_l\rangle\langle\tilde{\psi}^\textrm{ar}_l |d_0\rangle}{E-E^\textrm{ar}_l}
\nonumber\\
&\qquad\qquad\qquad\mbox{for $l=1,2$.}
\end{align}

We compare in Fig.~\ref{fig:cross2345} the following quantities:
\begin{figure}
\includegraphics[width=0.45\textwidth]{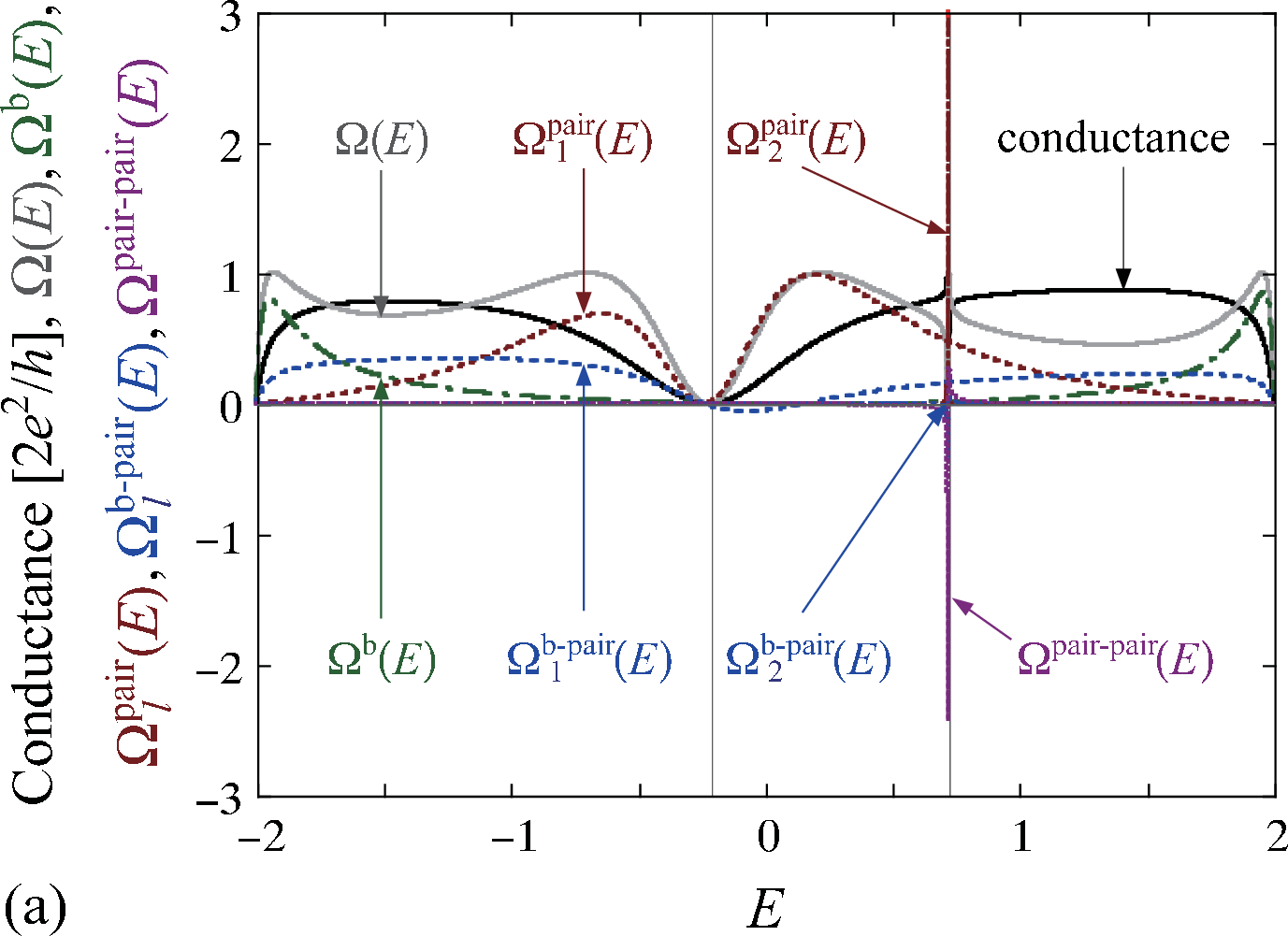}

\vspace*{\baselineskip}

\includegraphics[width=0.45\textwidth]{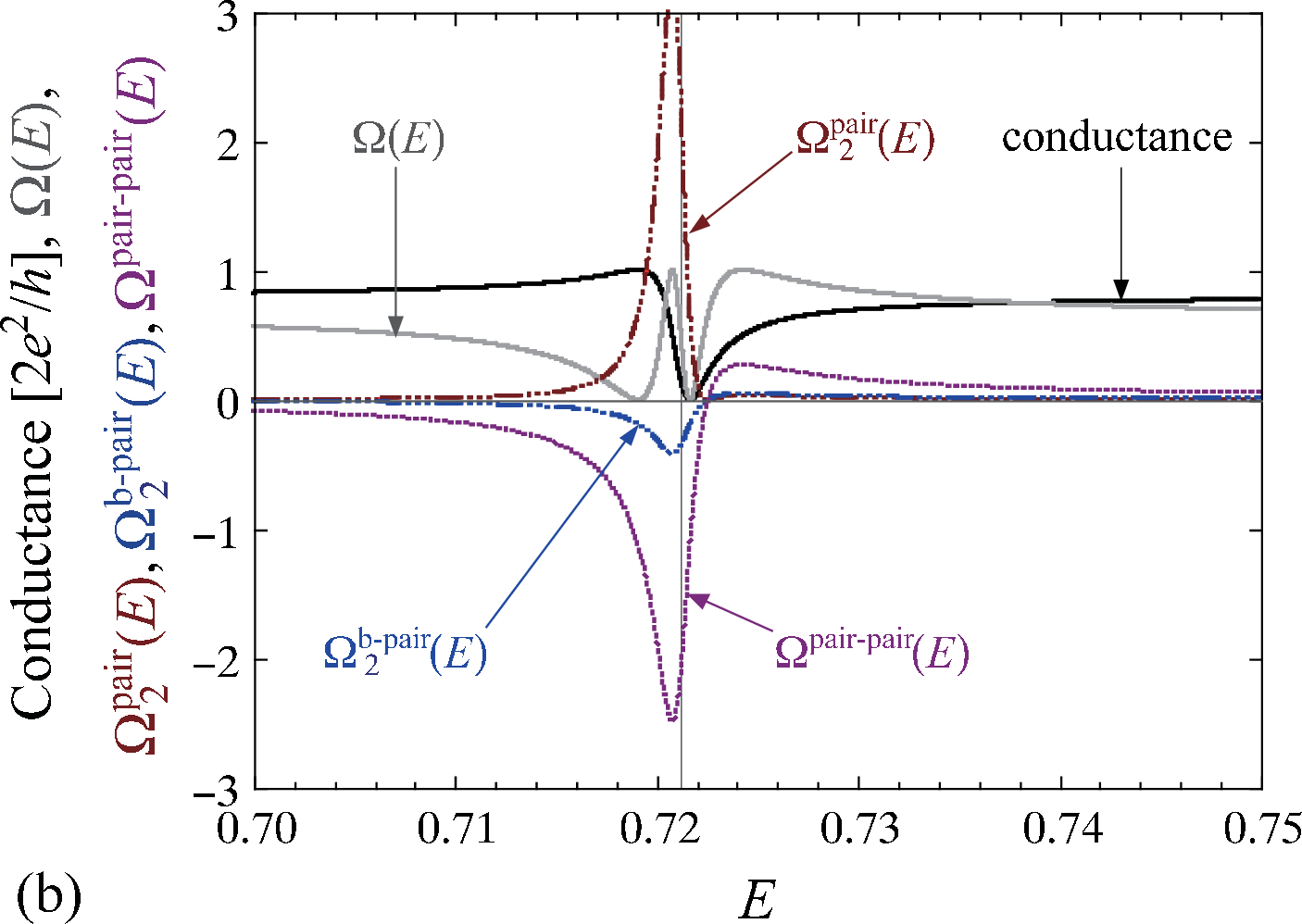}
\caption{(Color online) The quantities $\Omega(E)$ (gray curve), $\Omega^\textrm{b}(E)$ (green curve), $\Omega^\textrm{pair}_l(E)$ (broken and chained red curves), $\Omega^\textrm{b-pair}_l(E)$ (broken and chained blue curves) and $\Omega^\textrm{pair-pair}(E)$ (dotted purple curve), defined in eq.~(\ref{eq60})--(\ref{eq65}), plotted with the conductance (solid curve), eq.~(\ref{eq:conductance_G}), or eq.~(\ref{eq:conductance_J}).
The vertical gray lines indicate the real parts of the resonant eigenvalues $E=E^\textrm{res}_\textrm{r1}=-0.211544\ldots$ and $E=E^\textrm{res}_\textrm{r2}=0.721170\ldots$.
(b) shows the part of (a) around $E=E^\textrm{res}_\textrm{r2}$ with the plots of the conductance (solid curve), $\Omega(E)$ (gray curve), $\Omega^\textrm{pair}_2(E)$ (chained red curve), $\Omega^\textrm{b-pair}_2(E)$ (chained blue curve) and $\Omega^\textrm{pair-pair}(E)$ (dotted purple curve).
The system is the three-site dot.
We fixed $\varepsilon_0/t=0$, $\varepsilon_1/t=0$, $\varepsilon_2/t=0.5$, $v_{01}/t=v_{10}/t=0.8$, $v_{02}/t=v_{20}/t=0.5$ and $v_{12}/t=v_{21}/t=0.4$.}
\label{fig:cross2345} 
\end{figure}
\begin{align}
\Omega(E)\equiv&\left(\frac{\Gamma_{00}\Lambda_{00}}{2}\right)^2
\nonumber\\
=&\frac{{\Gamma_{00}}^2}{4}
\left(\Lambda^\textrm{b}(E)+\Lambda^\textrm{pair}_1(E)+\Lambda^\textrm{pair}_2(E)\right)^2,
\\ \label{eq62}
\Omega^\textrm{b}(E)\equiv&\frac{{\Gamma_{00}}^2}{4}
\Lambda^\textrm{b}(E)^2,
\\ \label{eq63}
\Omega^\textrm{pair}_l(E)\equiv&\frac{{\Gamma_{00}}^2}{4}
\Lambda^\textrm{pair}_l(E)^2
\quad\mbox{for $l=1,2$},
\\ \label{eq64}
\Omega^\textrm{b-pair}_l(E)\equiv&\frac{{\Gamma_{00}}^2}{2}
\Lambda^\textrm{b}(E)\Lambda^\textrm{pair}_l(E)
\quad\mbox{for $l=1,2$},
\\ \label{eq65}
\Omega^\textrm{pair-pair}(E)\equiv&\frac{{\Gamma_{00}}^2}{2}
\Lambda^\textrm{pair}_1(E)\Lambda^\textrm{pair}_2(E).
\end{align}
We can see that the following three terms are asymmetric: first, $\Omega^\textrm{pair}_2(E)$, which contains the crossing term between the resonant eigenstate $\psi^\textrm{res}_2$ and the anti-resonant eigenstate $\psi^\textrm{ar}_2$;
second, $\Omega^\textrm{b-pair}_2(E)$, which is the crossing term between the bound states $(\psi^\textrm{b}_1,\psi^\textrm{b}_2)$ and the resonant-state pair $(\psi^\textrm{res}_2,\psi^\textrm{ar}_2)$;
third, $\Omega^\textrm{pair-pair}(E)$, which is the crossing term between the two resonant-state pairs $(\psi^\textrm{res}_1,\psi^\textrm{ar}_1)$ and $(\psi^\textrm{res}_2,\psi^\textrm{ar}_2)$.

In order to derive the Fano parameters for the asymmetry of the three terms, we expand the terms~(\ref{eq63})--(\ref{eq65}) in the neighborhood of $E=E^\textrm{res}_{\textrm{r}2}$ by using the normalized energy
\begin{align}
\tilde{E}\equiv \frac{E-E_\textrm{r2}^\textrm{res}}{\left|E_\textrm{i2}^\textrm{res}\right|}.
\end{align} 
We can analyze the terms $\Omega^\textrm{pair}_2(E)$ and $\Omega^\textrm{b-pair}_2(E)$ in the same way as in \S~\ref{subsec:FanoG}.
We again use the expression
\begin{align}
\tilde{N}\mathrm{e}^{i\theta}\equiv\langle d_0|\psi^\textrm{res}_2\rangle\langle\tilde{\psi}^\textrm{res}_2|d_0\rangle.
\end{align}
Then the Fano parameter controlling the asymmetry of the term $\Omega^\textrm{pair}_2(E)$ is given by
\begin{equation}
q^\textrm{pair}_2=\tan\theta.
\end{equation}
Following the same logic as in eqs.~(\ref{eq57-1})--(\ref{eq63-1}), we obtain the Fano parameter that controls the asymmetry of the term $\Omega^\textrm{b-pair}_2(E)$ by solving
\begin{equation}
s\left(q^\textrm{b-pair}_2\right)^2-2(r-t)q^\textrm{b-pair}_2-s=0,
\end{equation}
where
\begin{align}
r&\equiv
\frac{\Lambda^\textrm{b}(E^\textrm{res}_{\textrm{r}2})}{\left|E^\textrm{res}_{\textrm{i}2}\right|}\sin\theta, \\
s&\equiv
\frac{\Lambda^\textrm{b}(E^\textrm{res}_{\textrm{r}2})}{\left|E^\textrm{res}_{\textrm{i}2}\right|}\cos\theta
+{\Lambda^\textrm{b}}'(E^\textrm{res}_{\textrm{r}2})\sin\theta, \label{eq:qb}\\
t&\equiv
{\Lambda^\textrm{b}}'(E^\textrm{res}_{\textrm{r}2})\cos \theta 
+\frac{1}{2}{\Lambda^\textrm{b}}''(E^\textrm{res}_{\textrm{r}2})\left|E^\textrm{res}_{\textrm{i}2}\right|\sin \theta.
\end{align}

Next, in order to discuss the quantity $\Omega^\textrm{pair-pair}(E)$, we use the expansion
\begin{align}
\Lambda^\textrm{pair}_{1}(E)
&\simeq
\Lambda^\textrm{pair}_1(E^\textrm{res}_{\textrm{r}2})+{\Lambda^\textrm{pair}_1}'(E^\textrm{res}_{\textrm{r}2})\left|E^\textrm{res}_{\textrm{i}2}\right|\tilde{E}
\nonumber\\
&+\frac{1}{2}{\Lambda^\textrm{pair}_1}''(E^\textrm{res}_{\textrm{r}2})\left|E^\textrm{res}_{\textrm{i}2}\right|^2\tilde{E}^2.
\end{align}
We then approximately have the crossing term between the two resonant-state pairs as
\begin{align}
\Omega^\textrm{pair-pair}(E)\equiv\frac{{\Gamma_{00}}^2}{4}
\Lambda^\textrm{pair}_1(E)\Lambda^\textrm{pair}_2(E)
\sim\frac{r' +s'\tilde{E} +t'\tilde{E}^2}{1+\tilde{E}^2}
\end{align}
with
\begin{align}
r'&\equiv
\frac{\Lambda^\textrm{pair}_1(E^\textrm{res}_{\textrm{r}2})}{\left|E^\textrm{res}_{\textrm{i}2}\right|}\sin\theta,
\\
s'&\equiv
\frac{\Lambda^\textrm{pair}_1(E^\textrm{res}_{\textrm{r}2})}{\left|E^\textrm{res}_{\textrm{i}2}\right|}\cos\theta
+{\Lambda^\textrm{pair}_1}'(E^\textrm{res}_{\textrm{r}2})\sin\theta, \label{eq:q_res}\\
t'&\equiv
{\Lambda^\textrm{pair}_1}'(E^\textrm{res}_{\textrm{r}2})\cos \theta
+\frac{1}{2}{\Lambda^\textrm{pair}_1}''(E^\textrm{res}_{\textrm{r}2})\left|E^\textrm{res}_{\textrm{i}2}\right|\sin \theta
\end{align}
We thus have yet another Fano parameter $q_2^\textrm{pair-pair}$ as the solution of
\begin{equation}
s'\left(q_2^\textrm{pair-pair}\right)^2-2(r'-t')q_2^\textrm{pair-pair}-s'=0.
\end{equation}

\begin{figure}
\includegraphics[width=0.45\textwidth]{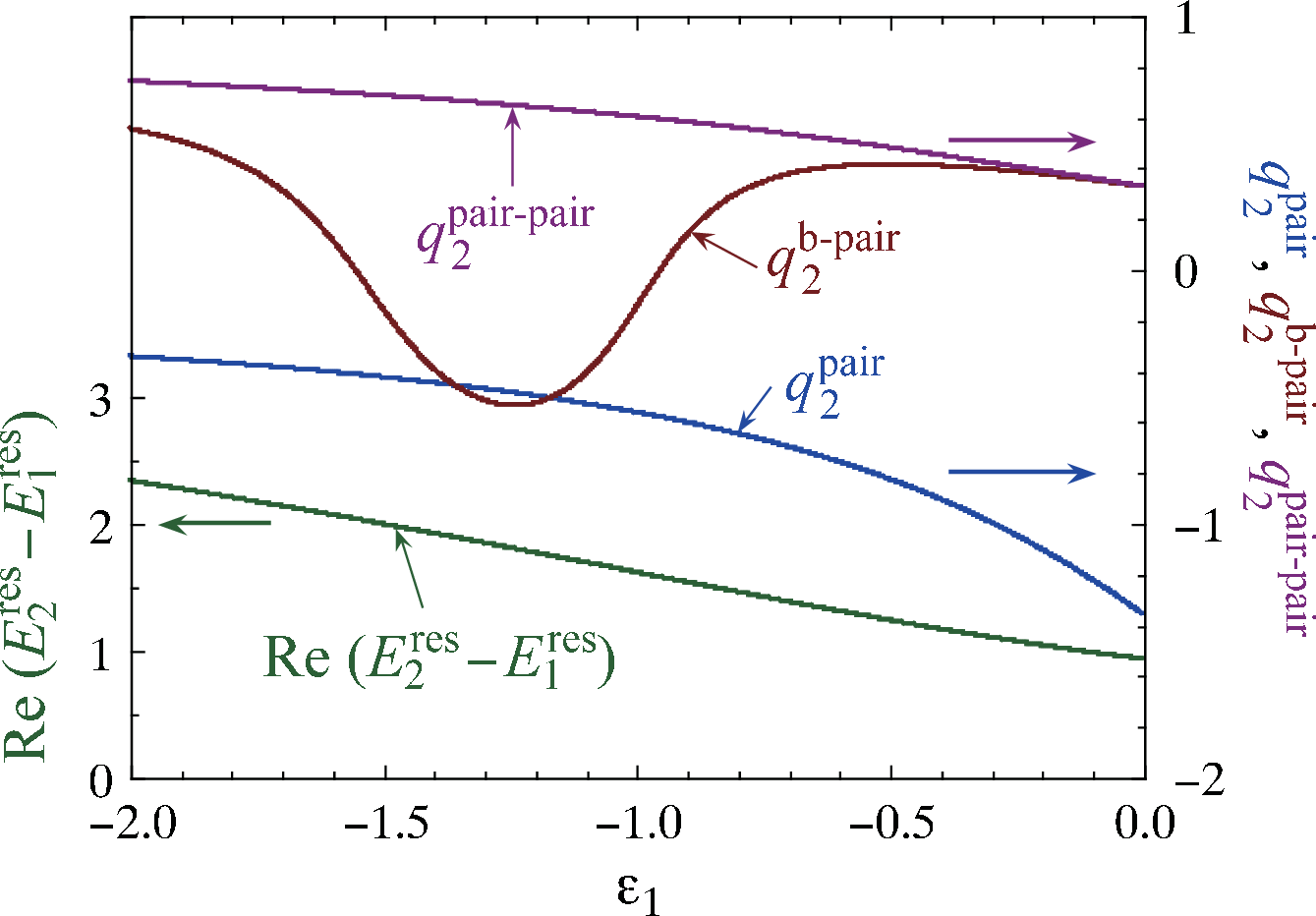}
\caption{(Color online) The Fano parameters $q^\textrm{pair}_2$ (blue curve), $q^\textrm{b-pair}_2$ (red curve) and $q_2^\textrm{pair-pair}$ (purple curve), plotted with the difference of the real parts of the two resonant eigenvalues, $E_\textrm{r2}^\textrm{res}-E_\textrm{r1}^\textrm{res}$.
Use the right axis for the Fano parameters and the left axis for the eigenvalue difference.
We fixed $\varepsilon_0/t=0$, $\varepsilon_2/t=0.5$, $v_{01}/t=v_{10}/t=0.8$, $v_{02}/t=v_{20}/t=0.5$ and $v_{12}/t=v_{21}/t=0.4$.}
\label{fig16}
\end{figure}
We show in Fig.~\ref{fig16} how the three Fano parameters $q^\textrm{pair}_2$,  $q^\textrm{b-pair}_2$ and $q_2^\textrm{pair-pair}$ depend on the system parameter $\varepsilon_1$.
In the particular case of Fig.~\ref{fig16}, the third Fano parameter $q_2^\textrm{pair-pair}$ is the greatest in most of the range.
This may be due to the following reason.
The first term of $s'$ for the parameter $q_2^\textrm{pair-pair}$ contains the Lorentzian
\begin{align}\label{eq:q_resE}
\Lambda^\textrm{pair}_1(E^\textrm{res}_{\textrm{r}2})
\sim\left[(E^\textrm{res}_{\textrm{r}1}-E^\textrm{res}_{\textrm{r}2})^{2}+{E^\textrm{res}_{\textrm{i}1}}^2\right]^{-1}.
\end{align}
Therefore, $s'$ grows fast as the resonant-state pair $E^\textrm{res}_{\textrm{r}1}$ approaches the resonant-state pair $E^\textrm{res}_{\textrm{r}2}$ up until $\left|E^\textrm{res}_{\textrm{r}1}-E^\textrm{res}_{\textrm{r}2}\right|\sim \left|E^\textrm{res}_{\textrm{i}1}\right|$.
This is in contrast to the first term of $s$ for the parameter $q^\textrm{b-pair}_2$, which contains
\begin{align}\label{eq:qbE}
\Lambda^\textrm{b}(E^\textrm{res}_{\textrm{r}2})\sim \left(E^\textrm{b}_p-E^\textrm{res}_{\textrm{r}2}\right)^{-1}
\end{align}
for $p=1,2$.
This is indeed demonstrated in Fig.~\ref{fig:N=3_ta=1_lead=2_e1_dependent}, where, as we increase $\varepsilon_1$, the asymmetry rapidly develops while the resonant-state pair $(E^\textrm{res}_1,E^\textrm{ar}_1)$ approaches $(E^\textrm{res}_2,E^\textrm{ar}_2)$.

Figure~\ref{fig:N=3_ta=1_lead=2_e1_dependent} then implies that the resonant-state pair with a large imaginary part strongly affects the resonant-state pair with a small imaginary part in the form of the asymmetry of the peak shape due to the latter.
This gives us an important moral that even a broad resonance peak can cause a drastic physical effect.

\subsection{Three-site quantum-dot system ($N=3$) with two contact points}
\label{subsec:two-contacts}

\begin{figure}
\includegraphics[width=0.42\textwidth]{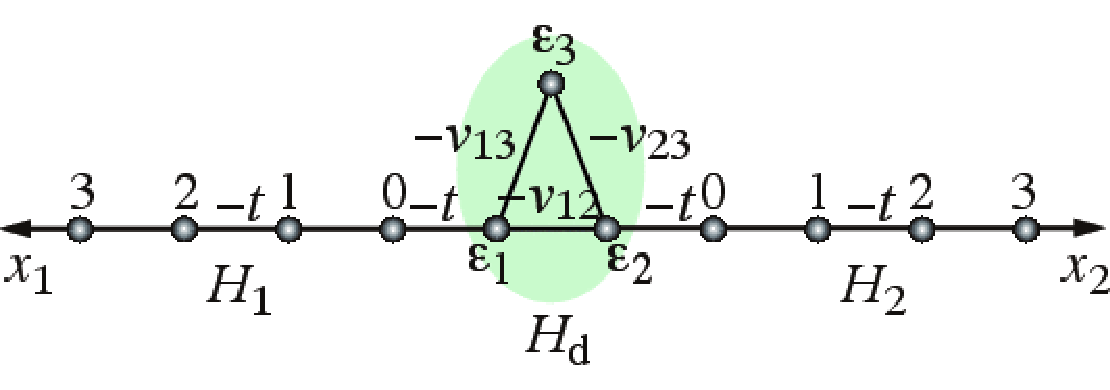}
\caption{(Color online) Three-site quantum-dot system with two contact points.}
\label{fig18n}
\end{figure}

Fourth, we discuss the conductance of the three-site quantum-dot system with two contact points as shown in Fig.~\ref{fig18n}.
We will consider a particular parameter set ($t_1/t=t_2/t=0.5$, $\varepsilon_3/t=-2$, $\varepsilon_1/t=\varepsilon_2/t=-0.5$, $v_{12}/t=v_{21}/t=0.5$, $v_{13}/t=v_{31}/t=1.5$ and $v_{23}/t=v_{32}/t=1.5$), for which an atypical situation occurs; 
that is, the Fano asymmetry does \textit{not} arise because one resonance has a specific symmetry and does not interfere with the other resonance.

This three-site system ($N=3$) also has six discrete eigenvalues ($2N=6$): one bound state, one anti-bound state and two resonance pairs in the case of the above parameter set.
(If we set $t_1/t=t_2/t=1$ in the present system with two contact points, two eigenvalues would tend to infinity, as is discussed in Appendix\ref{app:inf}.)
We show in Fig.~\ref{fig19n} the conductance as well as the two resonance pairs.
\begin{figure}
\includegraphics[width=0.4\textwidth]{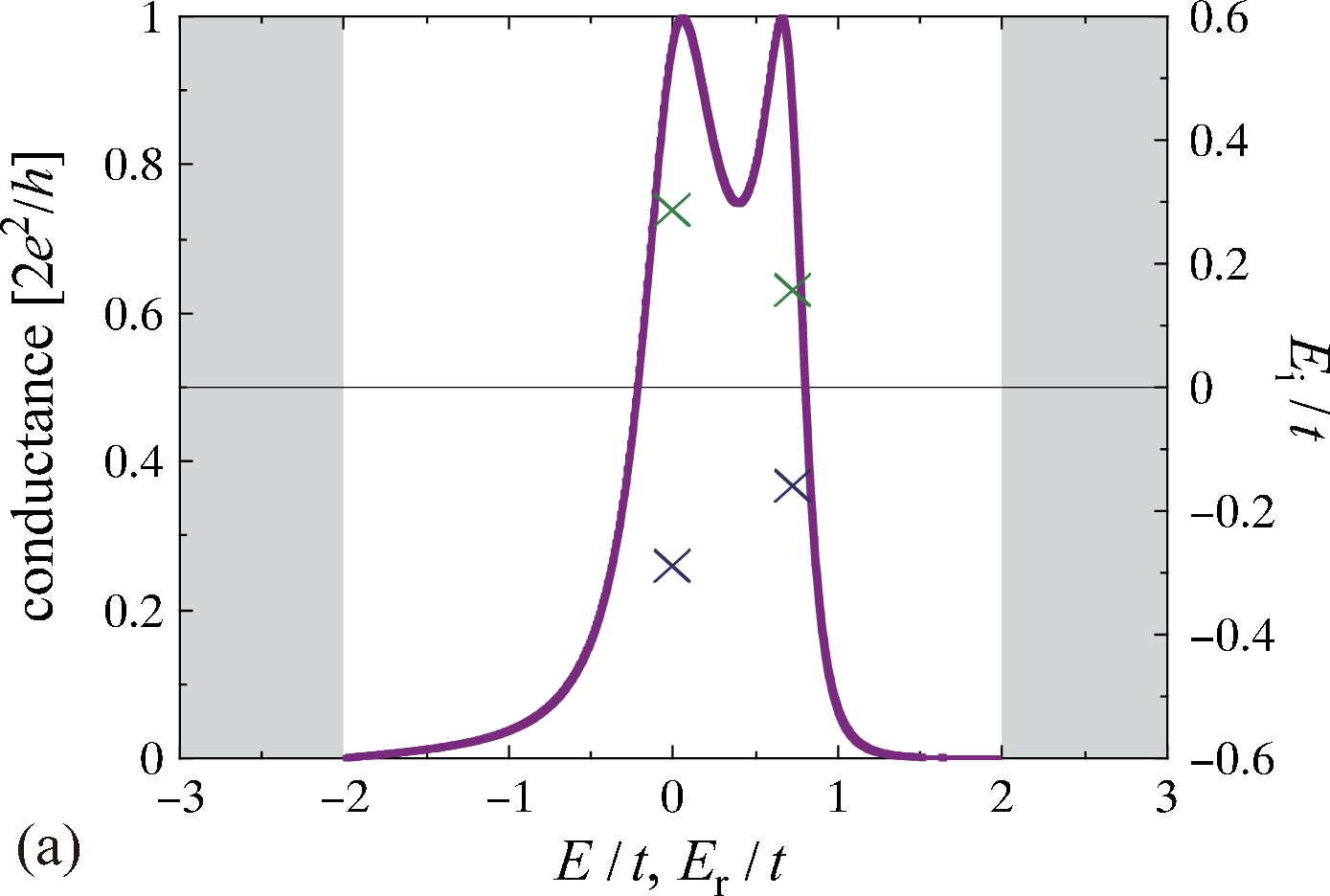}

\vspace*{\baselineskip}

\includegraphics[width=0.4\textwidth]{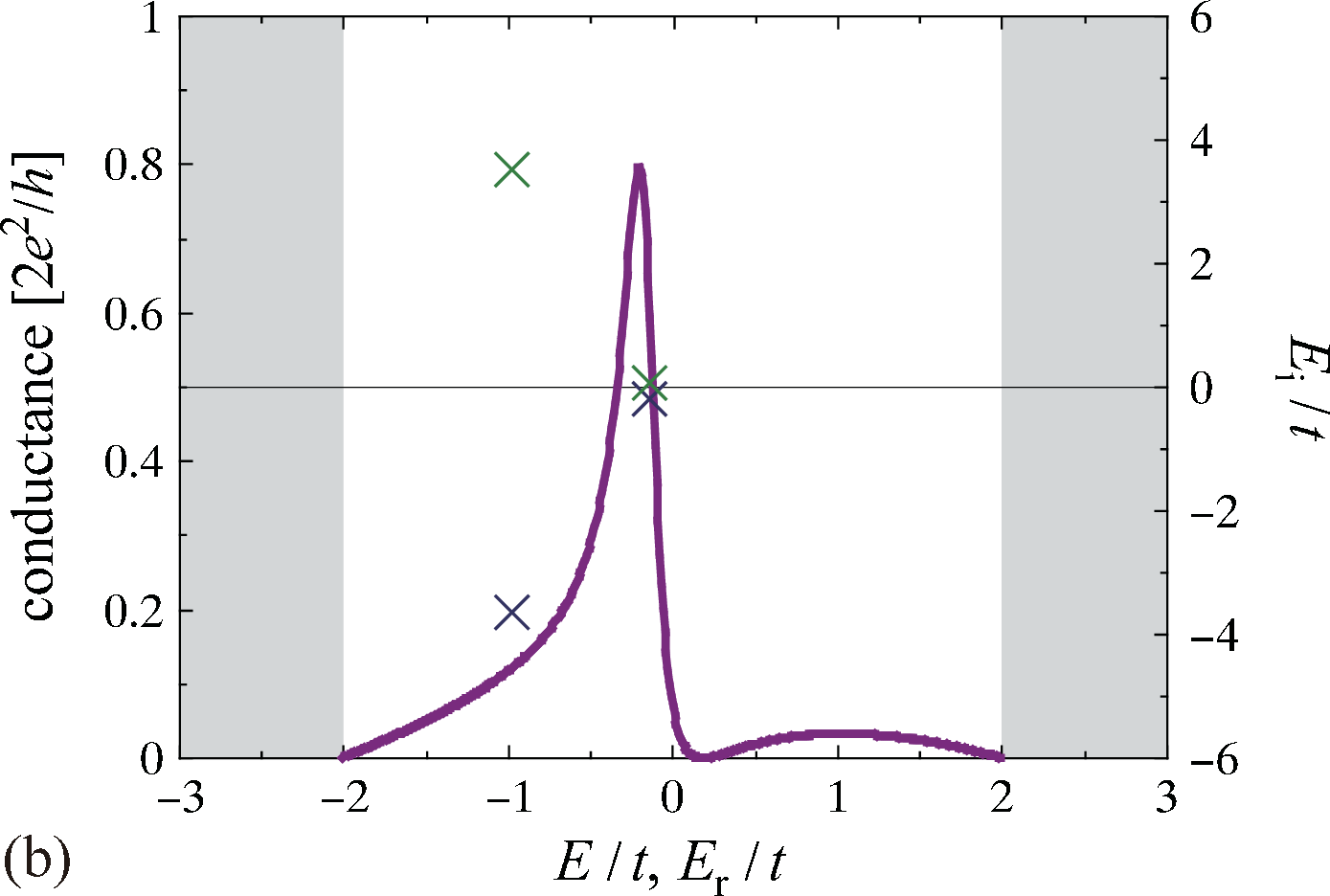}
\caption{(Color online) The conductance (curve for the left axis) for the three-site two-contact dot with: (a) $t_1/t=t_2/t=0.5$, $\varepsilon_1/t=\varepsilon_2/t=-0.5$, $\varepsilon_3/t=-2$, $v_{12}/t=v_{21}/t=0.5$, $v_{13}/t=v_{31}/t=1.5$ and $v_{23}/t=v_{32}/t=1.5$; (b) $t_1/t=t_2/t=0.97$, $\varepsilon_1/t=5$, $\varepsilon_2/t=\varepsilon_3/t=0$, $v_{12}/t=v_{21}/t=0.75$, $v_{13}/t=v_{31}/t=0.95$, and $v_{23}/t=v_{32}/t=0.15$.
The resonant and anti-resonant eigenvalues (crosses for the right axis) are also plotted. The bound and anti-bound states are located on the left and right sides out of this range for (a) and (b), respectively.
Note that the scale of the right axis is different between (a) and (b).}
\label{fig19n}
\end{figure}
We can see that these two resonance pairs do \textit{not} give not rise to any Fano asymmetry in the resonance peaks.

The reason in this particular case is that the wave functions for the resonant and anti-resonant states with $E^\textrm{res/ar}_1=0.0\mp i0.288675\ldots$ vanish on the site $d_3$, the top site of the triangle in Fig.~\ref{fig18n}: $\langle d_3|\psi_n\rangle=0$.
This is due to the symmetry $\varepsilon_1=\varepsilon_2$.
The energy eigenvalues of this resonance pair are pure imaginary, the real parts of the eigen-wave-numbers are $\pm\pi$ and the corresponding values of $z_n=\exp(i k_n)$ are pure imaginary.
The wave amplitude has also the symmetry $\langle d_1|\psi_n\rangle=-\langle d_2|\psi_n\rangle$ and are real.
If we define the the phase $\theta$ as in
\begin{align}
\tilde{N}\mathrm{e}^{i\theta}=\langle d_1|\psi_n^\textrm{res}\rangle\langle\tilde{\psi}_n^\textrm{res}|d_2\rangle,
\end{align}
the phase is $\pi$ and the resulting Fano parameter vanishes:
\begin{align}
q^\textrm{pair}=\tan\theta=0.
\end{align}
The moral of this case study is that a system with some symmetries may not exhibit the Fano asymmetry at all.

The above said, we note that the present system can also harbor asymmetric peaks as demonstrated in Fig.~\ref{fig19n}(b) for asymmetric couplings.
In this case, the Fano parameter for the pair of resonant and anti-resonant states with $E_2^\textrm{res/ar}=-0.154517\mp i0.120149$ is $q_2^\textrm{pair}=-1.66092$, which is rather large.
We can also see that the other pair of resonant and anti-resonant states with a relatively large imaginary part, $E_1^\textrm{res/ar}=-0.989392\mp i3.56912$, interferes with the pair $E_2^\textrm{res/ar}$, just as was discussed in \S~\ref{subsec:Three}.
The Fano parameter for the interference between these two pairs defined at the end of \S~\ref{subsec:Three} is $q_2^\textrm{pair-pair}=0.548253$ in the present case.

\subsection{Three-site quantum-dot system ($N=3$) with two contact points under an external magnetic field} 
\label{sec:mag}

In experiments,\cite{KAKI2002,KAKI2003,SAKKI2004} complex Fano parameters for the conductance have been obtained in the presence of a magnetic field. 
Here we will show that our formula indeed gives a complex Fano parameter $q^{\rm pair}$. 
As we saw in \S~\ref{subsec:FanoG}, the parameter $q^{\rm pair}$ is related to the Fano parameter of the conductance.

We will consider the effect of an external magnetic on the three-site quantum dot with a triangular shape, shown in Fig.~\ref{fig18n}.
When entering the dot at the site $1$, the electron can choose two paths, namely the straight path $1$--$2$ or the indirect path $1$--$3$--$2$, and thereby experiences the Aharonov-Bohm effect at the site $2$.
To model the effect we will add a complex phase (the Peierls phase~\cite{Peierls32}) $\varphi$ to the matrix elements $v_{12}$ and $v_{21}$ of the Hamiltonian that correspond to the $1$--$2$ path, as in $v_{12}\mathrm{e}^{i\varphi}$ and $v_{21}\mathrm{e}^{-i\varphi}$, while keeping the other matrix elements unchanged. 
This in effect will add a phase to the $1$--$2$ path relative to the $1$--$3$--$2$ path. 
Varying the phase of the $1$--$2$ path will produce varying conductance profiles.  

Recall that the conductance is proportional to the product $\Lambda_{12} \Lambda_{21}$. As we will show now, this product includes a Fano-like shape with a complex parameter $q$, whose phase depends on the strength of the magnetic field. 

We will focus on the components of $\Lambda_{12}$ and $\Lambda_{21}$ corresponding to a resonance/anti-resonance pair. For $\Lambda_{12}$ this component is given by
\begin{eqnarray}\label{eq5-110}
\Lambda_{12} ^{\rm pair} (\varphi)= \frac{\langle d_1| \psi^{\rm res}(\varphi)\rangle\langle {\tilde \psi}^{\rm res}(\varphi)| d_2\rangle}{E-E^{\rm res}(\varphi)} + \frac{\langle d_1| \psi^{\rm ar}(\varphi)\rangle\langle {\tilde \psi}^{\rm ar}(\varphi)| d_2\rangle}{E-E^{\rm ar}(\varphi)}.
\end{eqnarray}
For $\Lambda_{21}$ we simply exchange the indices $1$ and $2$. Let us introduce the notations
\begin{align}\label{eq5-111}
r_j (\varphi)&=  \langle d_j| \psi^{\rm res}(\varphi)\rangle
\end{align}
for simplicity.
The relations in Tab.~\ref{tabS1} then give
\begin{align}
\langle {\tilde \psi}^{\rm res}(\varphi)| d_j\rangle &=r_j(-\varphi),
\\
\langle d_j| \psi^{\rm ar}(\varphi)\rangle &=r_j(-\varphi)^\ast,
\\
\langle {\tilde \psi}^{\rm ar }(\varphi)| d_j\rangle &=r_j(\varphi)^\ast,
\\\label{eq5-115}
E^\mathrm{ar}(\varphi)&=\left(E^\mathrm{res}(\varphi)\right)^\ast.
\end{align}
Substituting eqs.~\eqref{eq5-111}--\eqref{eq5-115} into eq.~\eqref{eq5-110}, we have
\begin{eqnarray}
\Lambda_{12} ^{\rm pair}(\varphi) = 
\frac{r_1(\varphi) r_2(-\varphi)}{E-E^{\rm res}(\varphi)} 
+\frac{r_1(-\varphi)^\ast r_2(\varphi)^\ast}{E-E^{\rm res}(\varphi)^\ast}.
\end{eqnarray}
With the normalized energy~\eqref{eq57-1}, 
the expression above can be written as
\begin{eqnarray}
\Lambda_{12} ^{\rm pair}(\varphi) = \frac{\tilde{N}(\varphi)}{\left|E_{\rm i}^{\rm res}\right (\varphi)|} \frac{{\tilde E} + q^{\rm pair}(\varphi)}{1 + {\tilde E}^2},
\end{eqnarray}
where
\begin{eqnarray}
\tilde{N}(\varphi)= r_1(\varphi) r_2(-\varphi) + r_1(-\varphi)^\ast r_2(\varphi)^\ast
\end{eqnarray}
and
\begin{eqnarray}
q^{\rm pair} (\varphi)= -i \frac{r_1(\varphi) r_2(-\varphi) - r_1(-\varphi)^\ast r_2(\varphi)^\ast}%
{r_1(\varphi) r_2(-\varphi) + r_1(-\varphi)^\ast r_2(\varphi)^\ast}.
\end{eqnarray}
Without an external magnetic field $\varphi=0$, $\tilde{N}(\varphi)$ and $q^\mathrm{pair}(\varphi)$ are reduced to $\tilde{N}$ and $q^\mathrm{pair}$ defined in \S~\ref{subsec:T-shaped}.
With a magnetic field $\varphi\neq0$, however, they are both complex in general.
The pair contribution to the product $\Lambda_{12} \Lambda_{21}$ takes the modified Fano shape
\begin{eqnarray}
\Omega_{12}^{\rm pair}(\varphi) \propto 
\frac{|{\tilde E} + q^{\rm pair}(\varphi)|^2}{(1+{\tilde E}^2)^2}
\end{eqnarray}
with a complex asymmetry parameter.

\begin{figure}
\includegraphics[width=0.45\textwidth]{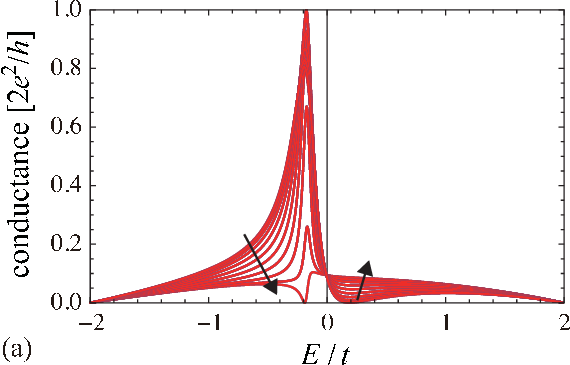}

\vspace*{\baselineskip}

\includegraphics[width=0.45\textwidth]{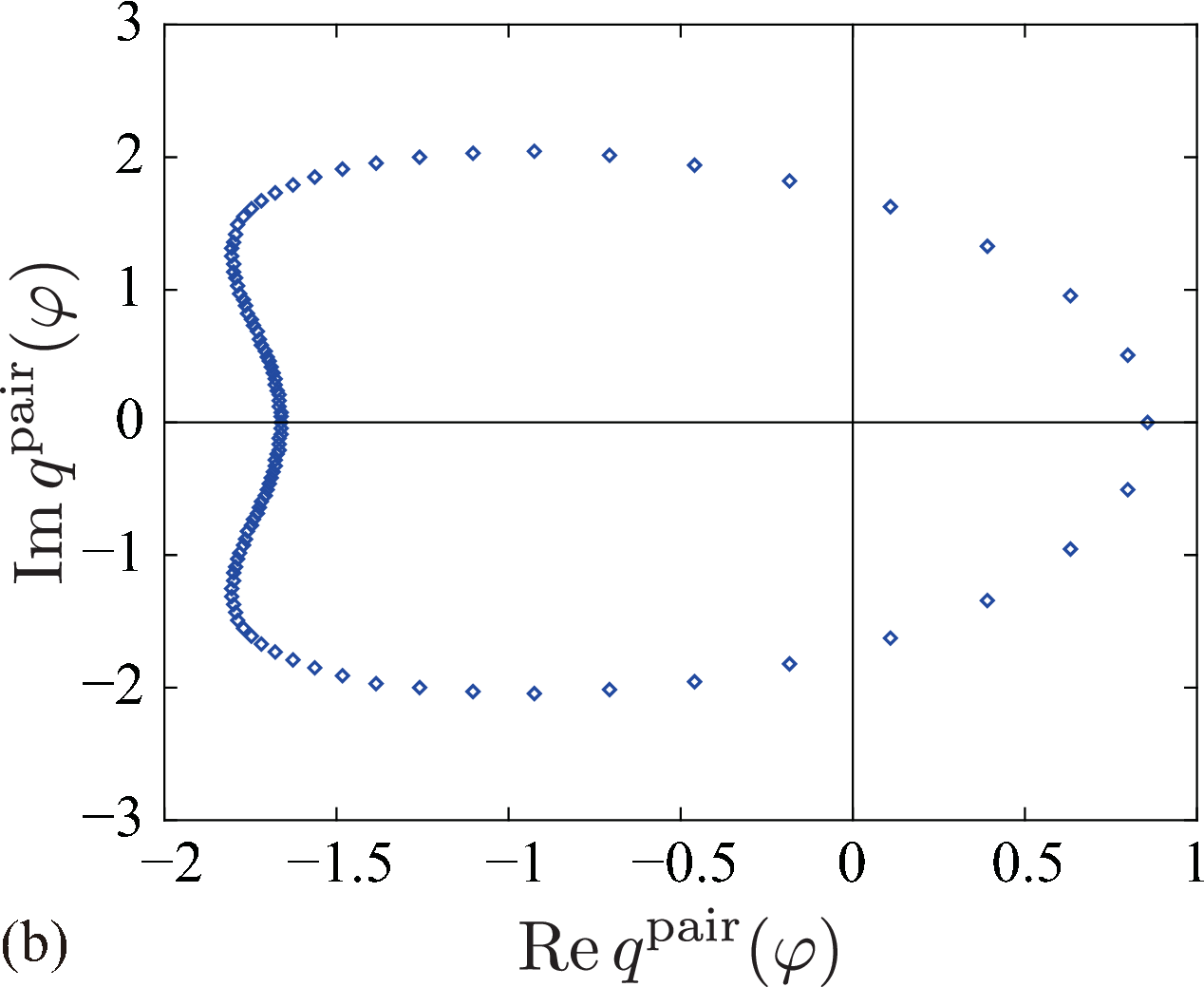}
\caption{(Color online) (a) Conductance profile for the quantum dot in Fig.~\ref{fig18n} with different strengths of the external magnetic field. 
The conductance changes in the direction of the arrows as the magnetic field $\varphi$ in $ v_{12}\mathrm{e}^{i\varphi}$ increases  with $\varphi=n\pi/10$ and $n=0,1\ldots, 19$.
All the  parameters other than $\varphi$ are the same as in Fig.~\ref{fig19n}(b).
(b) The complex Fano parameter $q^{\rm pair}(\varphi)$ of $\Omega_{12}^{\rm pair}$ for $\varphi=n\pi/50$ with   $n=0,1\cdots, 99$.
It starts on the negative real axis and circles around clockwise.}
\label{fig:ComplexQ}
\end{figure}

Monotonic increase of the magnetic field causes $q^{\rm pair}(\varphi)$ to trace a closed orbit on its complex plane, becoming real when $\varphi = n\pi$ with integer $n$. We show in Fig.~\ref{fig:ComplexQ} how the conductance of the triangular quantum dot changes as the Peierls phase $\varphi$ is varied as well as how the complex parameter $q^{\rm pair}(\varphi)$ changes. 

\subsection{The effect of the hopping energy $t_{\alpha}$ between the central dot and the leads}\label{subsec:effect of leads}

\begin{figure}
\includegraphics[width=0.4\textwidth]{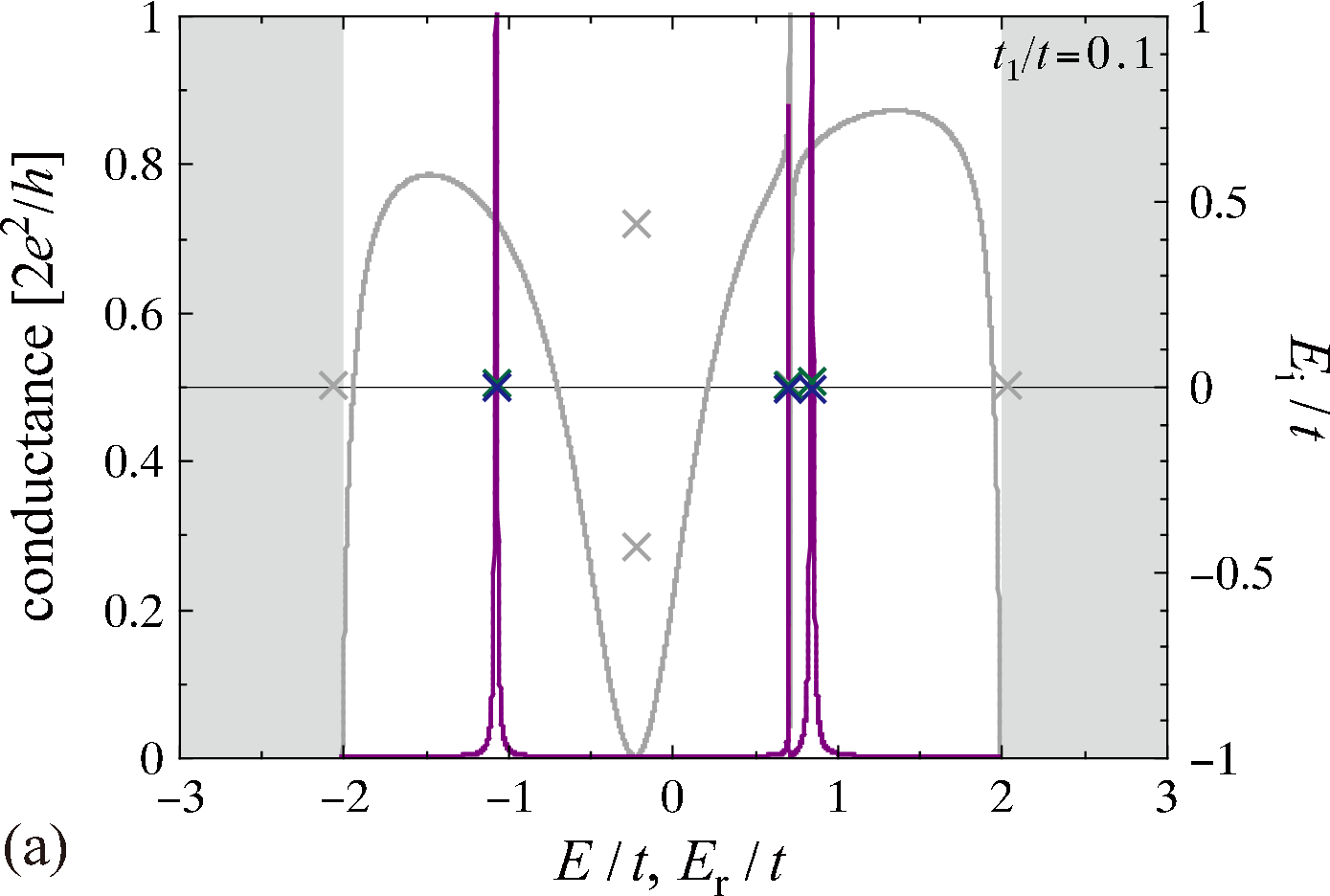}

\vspace*{\baselineskip}

\includegraphics[width=0.4\textwidth]{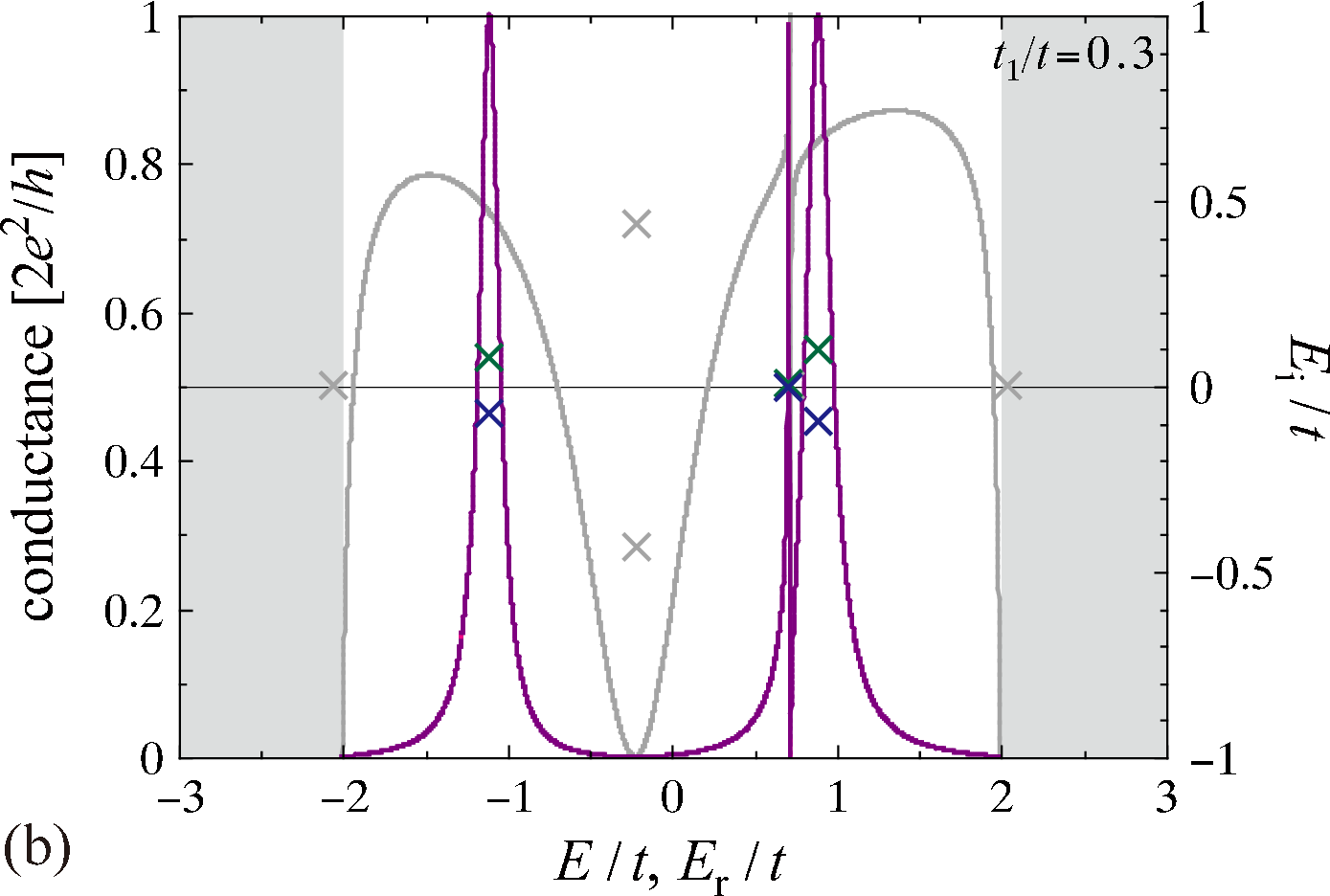}

\vspace*{\baselineskip}

\includegraphics[width=0.4\textwidth]{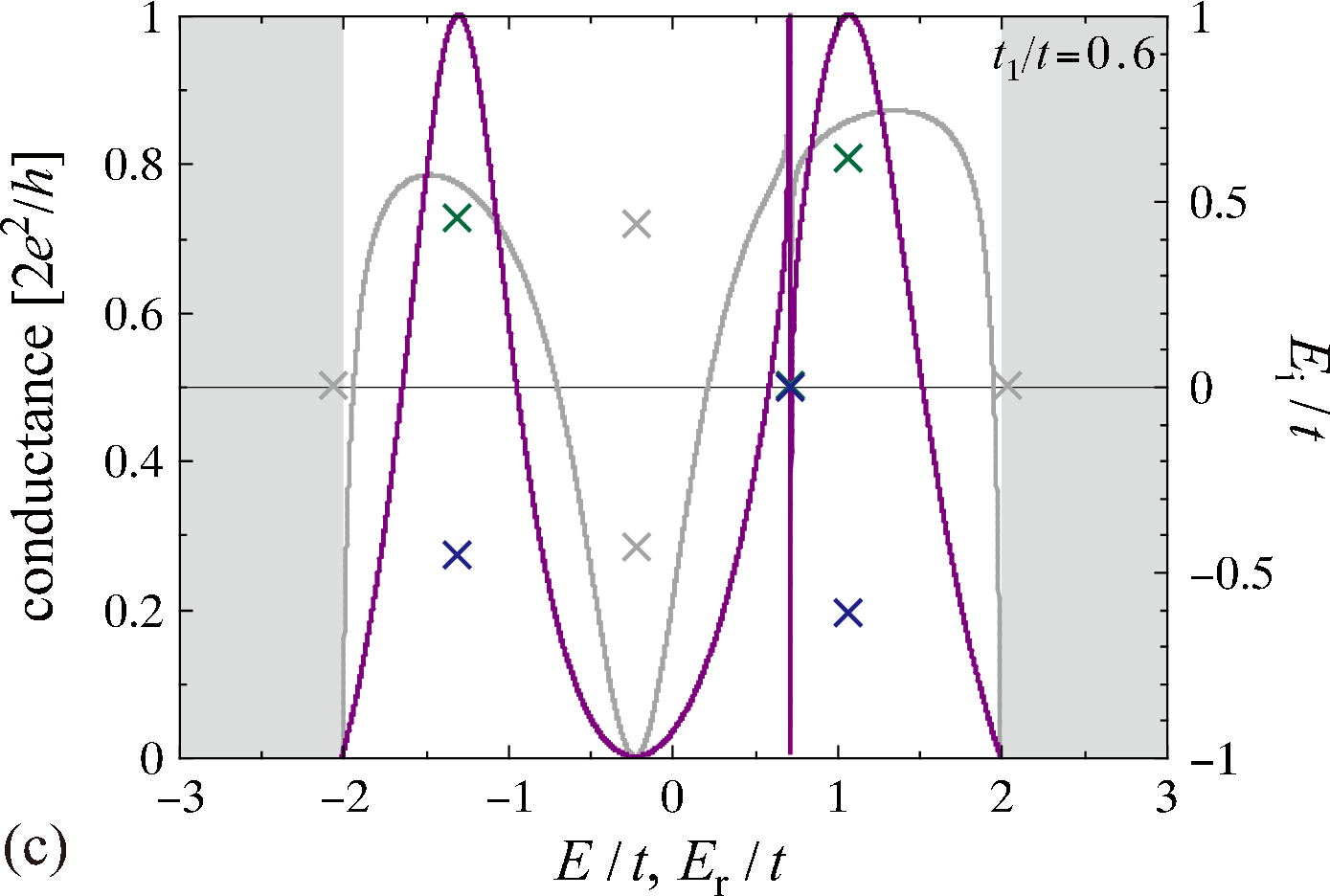}

\vspace*{\baselineskip}

\includegraphics[width=0.4\textwidth]{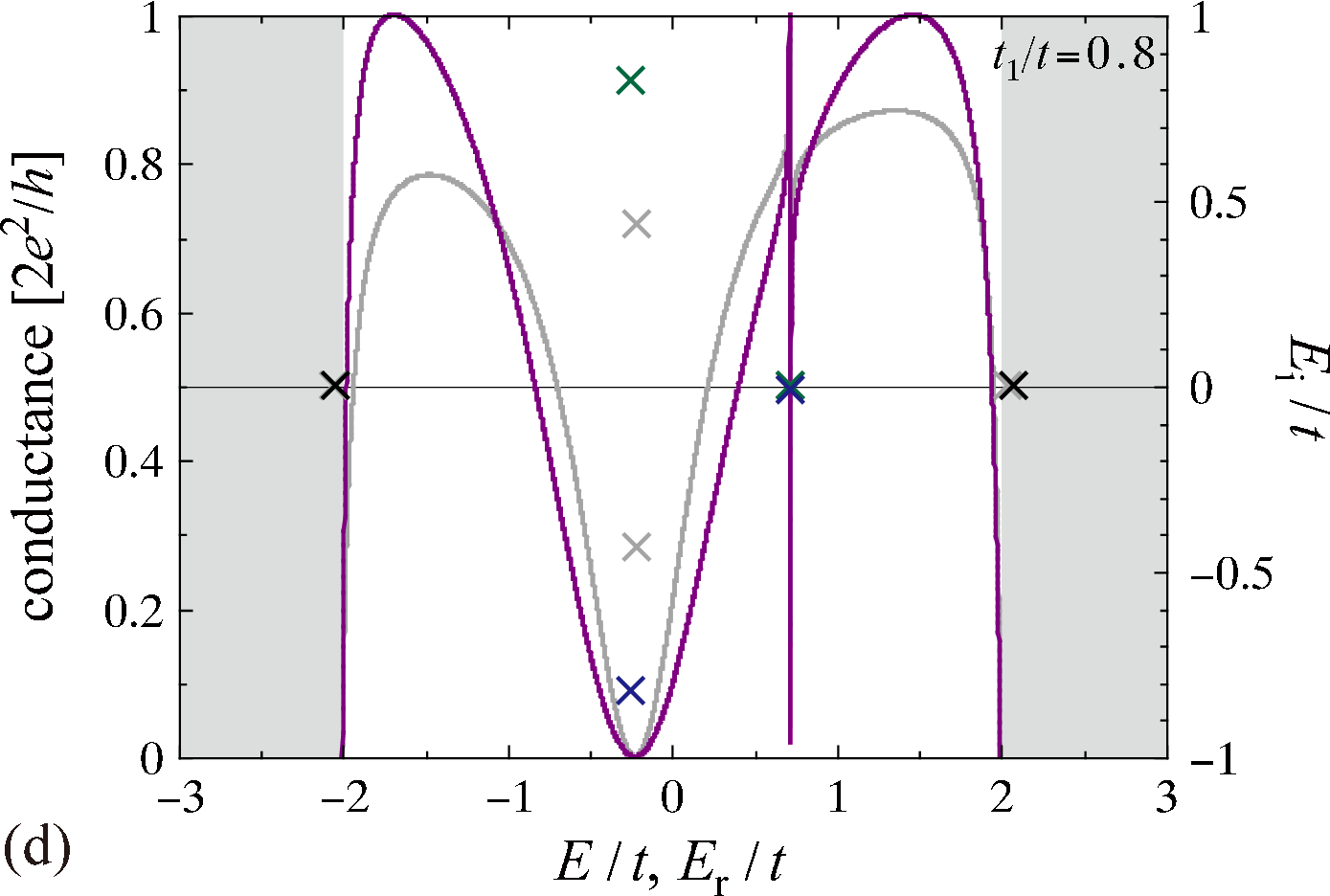}
\caption{(Color online) The conductance (curve for the left axis) for the three-site dot with (a) $t_1/t=t_2/t=0.1$, (b) $t_1/t=t_2/t=0.3$, (c) $t_1/t=t_2/t=0.6$ and (d) $t_1/t=t_2/t=0.8$, plotted with all the discrete eigenvalues (crosses for the right axis) 
The gray curves and the gray crosses indicate the conductance and the discrete eigenvalues for $t_1/t=t_2/t=1$, the same data as plotted in Fig.~\ref{fig:N=3_ta=1_lead=2_e1_dependent} .
We fixed $\varepsilon_0/t=0$, $\varepsilon_1/t=0$, $\varepsilon_2/t=0.5$, $v_{01}/t=v_{10}/t=0.8$, $v_{02}/t=v_{20}/t=0.5$ and $v_{12}/t=v_{21}/t=0.4$.}
\label{fig:N=3_lead=2_ta} 
\end{figure}
Finally, we briefly show the effect of the hopping energy $t_{\alpha}$ between the central dot and the lead $\alpha$.
We here use the case of the three-site dot with $t_1=t_2\neq t$,
$\varepsilon_0/t=0$, $\varepsilon_1/t=0$, $\varepsilon_2/t=0.5$, $v_{01}/t=v_{10}/t=0.8$, $v_{02}/t=v_{20}/t=0.5$ and $v_{12}/t=v_{21}/t=0.4$.
For $t_1=t_2<t/\sqrt{2}$, there are three resonant-state pairs and no bound states.
We have corresponding three sharp peaks in the weakly coupled case $t_1/t=t_2/t=0.1$ as in Fig.~\ref{fig:N=3_lead=2_ta} (a).
Upon increasing the hopping energy $t_1=t_2$, the second peak corresponding to the resonant-state pair with the least modulus of the imaginary part develops asymmetry.
At $t_1/t=t_2/t=1/\sqrt{2}$, the resonant and anti-resonant states of a resonant-state pair collide and become two anti-bound states, which leaves two resonant-state pairs.
For $t_1/t=t_2/t>1/\sqrt{2}$, the second peak continuously develop the asymmetry.
(The anti-bound states become bound states before $t_1=t_2=t$.)

\section{Conclusion}\label{sec:conclusion}
We carried out the spectrum analysis of the open quantum $N$-site (or $N$-level) dot with multiple leads. 
We obtained the simple conductance formula~(\ref{eq:conductance_G}) in terms of the matrices $\Lambda$ and $\Gamma$.
We then expanded the matrix $\Lambda$ purely in terms of discrete eigenstates, not including any background integrals.
To our knowledge, this is the first time the conductance is exactly given by the summation over all the simple poles.
(see ref.~\cite{endnote}).
We then showed that the Fano conductance arises from the crossing terms of three origins;
first between a pair of a resonant state and an anti-resonant state, second between a resonant-state pair and a bound state, and finally between two resonant-state pairs.
We also presented microscopic derivation of the Fano parameter.

The analysis in the present paper is applicable only to non-interacting systems.
It is an interesting and challenging problem to generalize the present approach to interacting systems~\cite{Presilla97}.
The Kondo effect, for example, has been observed in recent experiments on quantum dots and attracts much theoretical interest.
The present approach may be particularly useful in analyzing the interplay between the Fano resonance and Kondo resonance.
We are optimistic that the present argument can be generalized to such interacting systems~\cite{Nishino09,Imamura09,Nishino11}.

\begin{acknowledgments}
One of the present authors (N.H.) is grateful to illuminating discussions with Dr.~S.~Klaiman.
Another author (G.O.) thanks Institute of Industrial Science (Hatano Laboratory), University of Tokyo, and the JSPS Invitation Fellowship for Research in Japan for their hospitality and support.
This work is supported by Grant-in-Aid for Scientific Research  
No.~17340115 from the Ministry of Education, Culture, Sports, Science  
and Technology as well as by Core Research for Evolutional Science and  
Technology (CREST) of Japan Science and Technology Agency.
\end{acknowledgments}

\appendix

\section{State vectors of resonance and anti-resonance}
\label{app:states}

In the present Appendix, we review the relations among the ket and bra vectors of the resonant and anti-resonant states, which were briefly mentioned in \S~\ref{sec:resonant}.
We then show how the relations are modified when we introduce a magnetic field in order to utilize in \S~\ref{sec:mag}.

We consider the tight-binding model~\eqref{eq90}.
By the methods given in Appendices\ref{app:selfenergy}--\ref{app:Feshbach}, we can derive the effective Hamiltonian of a finite dimension:
\begin{align}\label{eqS20}
H_\mathrm{eff}(k)=H_\mathrm{d}+\Sigma(k),
\end{align}
where $\Sigma$ denotes the self-energy term with complex potentials only on the diagonal;
see eq.~\eqref{eqB20}.
The self-energy term generally depends on the energy $E$ itself as in eq.~\eqref{eqB31}, or on the wave number $k$ through the dispersion relation $E(k)$ as in eq.~\eqref{eqB165}.
We indicated the $k$ dependence in eq.~\eqref{eqS20} for convenience of the present Appendix, but indicate the $E$ dependence elsewhere.

\subsection{Effective Hamiltonian without a magnetic field}

Let us first consider the case where we do not have any magnetic fields,
or more specifically, the case where the dot Hamiltonian $H_\mathrm{d}$ observes the time-reversal symmetry as in eq.~\eqref{eqHd}:
\begin{align}\label{eqS30}
H_\textrm{d}\equiv&\sum_{i=1}^{N}\varepsilon_{i}|d_{i}\rangle\langle d_{i}|
\nonumber\\
&-\sum_{1\leqslant i<j\leqslant N}v_{ij}\left( |d_{i}\rangle\langle d_{j}|+|d_{j}\rangle\langle d_{i}|\right),
\end{align}
where all $v_{ij}\in\mathbb{R}$.
We then have the following symmetries:
\begin{align}\label{eqS40}
{H_\mathrm{d}}^\dag&={H_\mathrm{d}}^\ast={H_\mathrm{d}}^\mathrm{T}=H_\mathrm{d},
\\\label{eqS50}
\Sigma(k)^\dag&=\Sigma(k)^\ast=\Sigma(-k^\ast),
\\\label{eqS55}
\Sigma(k)^\mathrm{T}&=\Sigma(k),
\\\label{eqS60}
H_\mathrm{eff}(k)^\dag&=H_\mathrm{eff}(k)^\ast=H_\mathrm{eff}(-k^\ast),
\\
H_\mathrm{eff}(k)^\mathrm{T}&=H_\mathrm{eff}(k),
\end{align}
where the superscript T denotes the transpose.
The symmetries for $\Sigma(k)$ result from the fact that it has complex potentials proportional to $\mathrm{e}^{ik}$ on the diagonal; see Appendix\ref{app:selfenergy}, particularly eqs.~\eqref{eqB31} and~\eqref{eqB165}.
Note here that, if there is a resonant state at $k^\mathrm{res}=k$, the corresponding anti-resonant state is located at $k^\mathrm{ar}=-k^\ast$, because the transformation $k\rightarrow -k^\ast$ flips the real part of $k$.
We therefore have
\begin{align}
H_\mathrm{eff}(k^\mathrm{res})^\dag&=H_\mathrm{eff}(k^\mathrm{res})^\ast=H_\mathrm{eff}(k^\mathrm{ar}),
\\
H_\mathrm{eff}(k^\mathrm{res})^\mathrm{T}&=H_\mathrm{eff}(k^\mathrm{res}),
\\\label{eqS65}
H_\mathrm{eff}(k^\mathrm{ar})^\mathrm{T}&=H_\mathrm{eff}(k^\mathrm{ar}).
\end{align}

The transformation $k\rightarrow -k^\ast$ is the time-reversal transformation.
Equation~\eqref{eqS60}, therefore, implies that the effective Hamiltonian $H_\mathrm{eff}$ breaks the time-reversal symmetry.
Although the original Hamiltonian~\eqref{eq90} seemingly has the time-reversal symmetry, and indeed it does in the Hilbert space, each resonant and anti-resonant state, which resides outside the Hilbert space, breaks the time-reversal symmetry.
This breaking manifests itself in eq.~\eqref{eqS60}.

Below we forget the specifics of the Hamiltonian in eq.~\eqref{eqS30} and proceed on the basis of the symmetries~\eqref{eqS40}--\eqref{eqS65}.
We can show that the eigenvalues observe the following symmetry:
\begin{align}\label{eqS70}
E(k)^\ast=E(-k^\ast).
\end{align}
In other words, if $E(k)$ is the eigenenergy $E^\mathrm{res}$ of a resonant state at $k^\mathrm{res}=k$, the eigenenergy $E^\mathrm{ar}$ of the corresponding anti-resonant state at $k^\mathrm{ar}=-k^\ast$ is the complex conjugate $E(k)^\ast$,
\begin{align}
\left(E^\mathrm{res}\right)^\ast=E^\mathrm{ar},
\end{align}
which is indeed the case for the tight-binding model as well as for the standard Schr\"{o}dinger equation with the dispersion relation $E\propto k^2$.

The relation~\eqref{eqS70} is derived from the symmetry argument as follows.
Suppose that we have a resonant state $|\psi(k)\rangle$ with the wave number $k$ and the eigenenergy $E(k)$.
It should satisfy the eigenvalue equation
\begin{align}\label{eqS110}
H_\mathrm{eff}(k)|\psi(k)\rangle=E(k)|\psi(k)\rangle.
\end{align}
Then, the eigenvalue $E(k)$ is a solution of the secular equation
\begin{align}
\det\left(E(k)I-H_\mathrm{eff}(k)\right)=0,
\end{align}
where $I$ is the identity matrix.
Because $(\det A)^\ast=\det A^\ast$ for an arbitrary matrix $A$, we have
\begin{align}
\det\left(E(k)^\ast I-H_\mathrm{eff}(k)^\ast\right)&=0
\end{align}
which is followed by
\begin{align}
\det\left(E(k)^\ast I-H_\mathrm{eff}(-k^\ast)\right)&=0.
\end{align}
The latter equation is the secular equation for the Hamiltonian $H_\mathrm{eff}(-k^\ast)$, and hence the solution should be written as $E(-k^\ast)$.
This gives eq.~\eqref{eqS70}.

\subsection{Ket and bra vectors of resonant and anti-resonant states without a magnetic field}

Let us find the relation between the ket vectors of the resonant and anti-resonant states without magnetic fields.
By taking the complex conjugate of the eigenvalue equation~\eqref{eqS110}, we have
\begin{align}
H_\mathrm{eff}(k)^\ast|\psi(k)\rangle^\ast&=E(k)^\ast|\psi(k)\rangle^\ast,
\\
\mbox{or}\qquad
H_\mathrm{eff}(-k^\ast)|\psi(k)\rangle^\ast&=E(-k^\ast)|\psi(k)\rangle^\ast.
\end{align}
The later equation indicates that the vector defined by
\begin{align}\label{eqS150}
|\psi(-k^\ast)\rangle&:=|\psi(k)\rangle^\ast
\end{align}
is the ket vector $|\psi^\mathrm{ar}\rangle$ of the anti-resonant state at $k^\mathrm{ar}=-k^\ast$, which is paired with the resonant state $|\psi^\mathrm{res}\rangle$ at $k^\mathrm{res}=k$:
\begin{align}
|\psi^\mathrm{res}\rangle^\ast=|\psi^\mathrm{ar}\rangle.
\end{align}
(We here fixed the overall phase factor of $|\psi(-k^\ast)\rangle$ so that eq.~\eqref{eqS150} may hold.)

We go on to the derivation of the bra vectors of resonant and anti-resonant states without magnetic fields.
A seemingly strange consequence of the symmetry~\eqref{eqS65}  is the following.
The transpose of the eigenvalue equation~\eqref{eqS110} reads
\begin{align}\label{eqS210}
|\psi(k)\rangle^\mathrm{T} H_\mathrm{eff}(k)&=E(k)|\psi(k)\rangle^\mathrm{T}
\end{align}
because of the symmetry eq.~\eqref{eqS65}.
We can regard eq.~\eqref{eqS210} as a left-eigenvalue equation for the Hamiltonian $H_\mathrm{eff}(k)$.
Therefore, the bra vector corresponding to the ket vector $|\psi(k)\rangle$ is \textit{not} the complex conjugate but the simple transpose (after fixing the overall phase factor):
\begin{align}\label{eqS180}
\langle\tilde{\psi}(k)|:=|\psi(k)\rangle^\mathrm{T},
\end{align}
or
\begin{align}\label{eqS2300}
\langle\tilde{\psi}^\mathrm{res}|&=|\psi^\mathrm{res}\rangle^\mathrm{T},
\\\label{eqS2400}
\langle\tilde{\psi}^\mathrm{ar}|&=|\psi^\mathrm{ar}\rangle^\mathrm{T}.
\end{align}
We here used the notation $\tilde{\psi}$ for the bra vectors in order to stress that they are not the complex conjugate of the respective ket vectors.
An immediate consequence of eq.~\eqref{eqS180} is the norm
\begin{align}\label{eqS181}
\langle\tilde{\psi}(k)|\psi(k)\rangle=\int\psi(x;k)^2dx,
\end{align}
where $\psi(x;k)=\langle x|\psi(k)\rangle$.
Note that the norm is \textit{not} the square modulus, a simple square, and hence \textit{not} positive in general, but complex.
The unfamiliar results~\eqref{eqS180}--\eqref{eqS181} are originated from the non-Hermiticity of $H_\mathrm{eff}$, eq.~\eqref{eqS60}.
In other words, these results hold for a state that breaks the time-reversal symmetry.

On the other hand, by taking the Hermitian conjugate of the eigenvalue equation~\eqref{eqS110}, we have
\begin{align}
|\psi(k)\rangle^\dag H_\mathrm{eff}(k)^\dag&=E(k)^\ast|\psi(k)\rangle^\dag,
\\
\mbox{or}\qquad
|\psi(k)\rangle^\dag H_\mathrm{eff}(-k^\ast)&=E(-k^\ast)|\psi(k)\rangle^\dag.
\end{align}
The latter equation is the left-eigenvalue equation for the Hamiltonian $H_\mathrm{eff}(-k^\ast)$.
Therefore, the left-eigenvector defined by
\begin{align}\label{eqS170}
\langle\tilde{\psi}(-k^\ast)|&:=|\psi(k)\rangle^\dag
\end{align}
is the bra vector $\langle\tilde{\psi}^\mathrm{ar}|$ for the anti-resonant state at $k^\mathrm{ar}=-k^\ast$:
\begin{align}\label{eqS2900}
\langle\tilde{\psi}^\mathrm{ar}|&=|\psi^\mathrm{res}\rangle^\dag,
\\\label{eqS3000}
\langle\tilde{\psi}^\mathrm{res}|&=|\psi^\mathrm{ar}\rangle^\dag.
\end{align}
(This relation is also derived from eqs.~\eqref{eqS150} and~\eqref{eqS180}.)
Other relations for the corresponding anti-resonant states are
\begin{align}
\langle\tilde{\psi}(-k^\ast)|&=|\psi(-k^\ast)\rangle^\mathrm{T}
\\
&=\left(|\psi(-k^\ast)\rangle^\dag\right)^\ast
=\langle\tilde{\psi}(k)|^\ast
\\
\langle\tilde{\psi}(-k^\ast)|\psi(-k^\ast)\rangle&=\int\psi(x;-k^\ast)^2dx=\int\left(\psi(x;k)^2\right)^\ast dx
\\
&=\langle\tilde{\psi}(k)|\psi(k)\rangle^\ast,
\end{align}
or
\begin{align}
\langle\tilde{\psi}^\mathrm{ar}|&=\langle\tilde{\psi}^\mathrm{res}|^\ast,
\\
\langle\tilde{\psi}^\mathrm{ar}|\psi^\mathrm{ar}\rangle^\ast&=
\langle\tilde{\psi}^\mathrm{res}|\psi^\mathrm{res}\rangle.
\end{align}

\subsection{Effective Hamiltonian with a magnetic field}

Let us introduce a magnetic field to the dot Hamiltonian $H_\mathrm{d}$.
This can be done by the Peierls substitution\cite{Peierls32}:
\begin{align}\label{eqS280}
H_\textrm{d}(\varphi)&\equiv\sum_{i=1}^{N}\varepsilon_{i}|d_{i}\rangle\langle d_{i}|
\nonumber\\
&-\sum_{1\leqslant i<j\leqslant N}v_{ij}\left( \mathrm{e}^{i\varphi_{ij}}|d_{i}\rangle\langle d_{j}|+\mathrm{e}^{-i\varphi_{ij}}|d_{j}\rangle\langle d_{i}|\right),
\end{align}
where the phases $\{\varphi_{ij}\}$ are all real with $\varphi_{ji}=-\varphi_{ij}$; we simply expressed the argument of $H_\textrm{d}$ as $\varphi$ in order to indicate the dependence on the phases $\{\varphi_{ij}\}$.
On the other hand, we keep the lead Hamiltonian as well as the coupling Hamiltonian as they are.
Then the effective Hamiltonian~\eqref{eqS20} is generalized to
\begin{align}
H_\mathrm{eff}(k,\varphi)=H_\mathrm{d}(\varphi)+\Sigma(k).
\end{align}
Note that the self-energy part $\Sigma(k)$ does not depend on $\varphi$, because we do not apply the magnetic field to the leads.
We then have the following symmetries:
\begin{align}\label{eqS290}
H_\mathrm{d}(\varphi)^\dag&=H_\mathrm{d}(\varphi),
\\
H_\mathrm{d}(\varphi)^\ast&=H_\mathrm{d}(\varphi)^\mathrm{T}=H_\mathrm{d}(-\varphi),
\\
\Sigma(k)^\dag&=\Sigma(k)^\ast=\Sigma(-k^\ast),
\\
\Sigma(k)^\mathrm{T}&=\Sigma(k),
\\
H_\mathrm{eff}(k,\varphi)^\dag&=H_\mathrm{eff}(-k^\ast,\varphi),
\\
H_\mathrm{eff}(k,\varphi)^\ast&=H_\mathrm{eff}(-k^\ast,-\varphi),
\\
H_\mathrm{eff}(k,\varphi)^\mathrm{T}&=H_\mathrm{eff}(k,-\varphi),
\end{align}
and therefore
\begin{align}
H_\mathrm{eff}(k^\mathrm{res},\varphi)^\dag&=H_\mathrm{eff}(k^\mathrm{ar},\varphi),
\\
H_\mathrm{eff}(k^\mathrm{res},\varphi)^\ast&=H_\mathrm{eff}(k^\mathrm{ar},-\varphi),
\\
H_\mathrm{eff}(k^\mathrm{res},\varphi)^\mathrm{T}&=H_\mathrm{eff}(k^\mathrm{res},-\varphi),
\\\label{eqS350}
H_\mathrm{eff}(k^\mathrm{ar},\varphi)^\mathrm{T}&=H_\mathrm{eff}(k^\mathrm{ar},-\varphi).
\end{align}

We again forget the specifics of the Hamiltonian~\eqref{eqS280} and rely on the symmetries~\eqref{eqS290}--\eqref{eqS350} hereafter.
We can derive the following symmetries for the eigenvalues:
\begin{align}\label{eqS100}
E(k,\varphi)^\ast &=E(-k^\ast,\varphi)=E(-k^\ast,-\varphi),
\\\label{eqS101}
E(k,\varphi)&=E(k,-\varphi),
\end{align}
or
\begin{align}
E^\mathrm{res}(\varphi)&=E^\mathrm{res}(-\varphi),
\\
E^\mathrm{ar}(\varphi)&=E^\mathrm{ar}(-\varphi),
\\
E^\mathrm{res}(\varphi)^\ast&=E^\mathrm{ar}(\varphi).
\end{align}
The derivation is given as follows.
A resonant state $|\psi(k,\varphi)\rangle$ with the eigenstate $E(k,\varphi)$ should now satisfy the eigenvalue equation
\begin{align}\label{eqS200}
H_\mathrm{eff}(k,\varphi)|\psi(k,\varphi)\rangle=E(k,\varphi)|\psi(k,\varphi)\rangle.
\end{align}
Therefore, the eigenvalue $E(k,\varphi)$ is a solution of the secular equation
\begin{align}
\det\left(E(k,\varphi)I-H_\mathrm{eff}(k,\varphi)\right)=0.
\end{align}
Because $(\det A)^\ast=\det A^\dag=\det A^\ast$ and $\det A=\det A^\mathrm{T}$ for an arbitrary matrix $A$, we have
\begin{align}
\det\left(E(k,\varphi)^\ast I-H_\mathrm{eff}(k,\varphi)^\dag\right)&=0,
\\
\det\left(E(k,\varphi)^\ast I-H_\mathrm{eff}(k,\varphi)^\ast\right)&=0
\\
\mbox{and}\qquad
\det\left(E(k,\varphi) I-H_\mathrm{eff}(k,\varphi)^\mathrm{T}\right)&=0,
\end{align}
or
\begin{align}\label{eqS430}
\det\left(E(k,\varphi)^\ast I-H_\mathrm{eff}(-k^\ast,\varphi)\right)&=0,
\\\label{eqS440}
\det\left(E(k,\varphi)^\ast I-H_\mathrm{eff}(-k^\ast,-\varphi)\right)&=0
\\\label{eqS450}
\mbox{and}\qquad
\det\left(E(k,\varphi) I-H_\mathrm{eff}(k,-\varphi)\right)&=0.
\end{align}
Equations~\eqref{eqS430}--\eqref{eqS450} are the secular equation for the Hamiltonians $H_\mathrm{eff}(-k^\ast,\varphi)$, $H_\mathrm{eff}(-k^\ast,-\varphi)$ and $H_\mathrm{eff}(k,-\varphi)$, respectively.
Hence the solutions should be written as  $E(-k^\ast,\varphi)$, $E(-k^\ast,-\varphi)$ and $E(k,-\varphi)$, respectively, which gives the symmetries~\eqref{eqS100}--\eqref{eqS101}.

\subsection{Ket and bra vectors of resonant and anti-resonant states with a magnetic field}

Next, we consider the ket vectors under a magnetic field.
The complex conjugate of the eigenvalue equation~\eqref{eqS200} reads
\begin{align}
H_\mathrm{eff}(k,\varphi)^\ast|\psi(k,\varphi)\rangle^\ast
&=E(k,\varphi)^\ast|\psi(k,\varphi)\rangle^\ast
\\
\mbox{or}\qquad
H_\mathrm{eff}(-k^\ast,-\varphi)|\psi(k,\varphi)\rangle^\ast
&=E(-k^\ast,-\varphi)|\psi(k,\varphi)\rangle^\ast.
\end{align}
The latter equation indicates that the vector defined by
\begin{align}\label{eqS230}
|\psi(-k^\ast,-\varphi)\rangle&:=|\psi(k,\varphi)\rangle^\ast
\end{align}
is the ket vector $|\psi^\mathrm{ar}(-\varphi)\rangle$ of a state at $k^\mathrm{ar}=-k^\ast$ under the reversed magnetic field $-\varphi$.
We can recast eq.~\eqref{eqS230} into the forms
\begin{align}\label{eqS240}
|\psi(-k^\ast,\varphi)\rangle&=|\psi(k,-\varphi)\rangle^\ast,
\end{align}
which is the ket vector of the anti-resonant state under the original (not reversed) magnetic field $\varphi$:
\begin{align}
|\psi^\mathrm{ar}(\varphi)\rangle&=|\psi^\mathrm{res}(-\varphi)\rangle^\ast.
\end{align}

We then consider the bra vectors under a magnetic field.
The transpose of the eigenvalue equation~\eqref{eqS200} reads
\begin{align}
|\psi(k,\varphi)\rangle^\mathrm{T} H_\mathrm{eff}(k,\varphi)^\mathrm{T}
&=E(k,\varphi)|\psi(k,\varphi)\rangle^\mathrm{T},
\\
\mbox{or}\qquad
|\psi(k,\varphi)\rangle^\mathrm{T} H_\mathrm{eff}(k,-\varphi)
&=E(k,\varphi)|\psi(k,\varphi)\rangle^\mathrm{T}.
\end{align}
The latter equation is the left-eigenvalue equation for the Hamiltonian $H_\mathrm{eff}(k,-\varphi)$. 
Therefore, the vector defined by
\begin{align}
\langle\tilde{\psi}(k,-\varphi)|&:=|\psi(k,\varphi)\rangle^\mathrm{T}
\end{align}
is the bra vector $\langle\tilde{\psi}^\mathrm{res}(-\varphi)|$ of the resonant state at $k^\mathrm{res}=k$ under the reversed magnetic field $-\varphi$:
\begin{align}
\langle\tilde{\psi}^\mathrm{res}(-\varphi)|&=|\psi^\mathrm{res}(\varphi)\rangle^\mathrm{T},
\\
\langle\tilde{\psi}^\mathrm{ar}(-\varphi)|&=|\psi^\mathrm{ar}(\varphi)\rangle^\mathrm{T}.
\end{align}
This is followed by the norm
\begin{align}
\langle\tilde{\psi}(k,\varphi)|\psi(k,\varphi)\rangle=\langle\tilde{\psi}(k,-\varphi)|\psi(k,-\varphi)\rangle
=\int\psi(x;k,\varphi)\psi(x;k,-\varphi)dx,
\end{align}
where $\psi(x;k,\varphi)=\langle x|\psi(k,\varphi)\rangle$.

On the other hand, the complex conjugate of the eigenvalue equation~\eqref{eqS200} reads
\begin{align}
|\psi(k,\varphi)\rangle^\dag H_\mathrm{eff}(k,\varphi)^\dag
&=E(k,\varphi)^\ast|\psi(k,\varphi)\rangle^\dag
\\
\mbox{or}\qquad
|\psi(k,\varphi)\rangle^\dag H_\mathrm{eff}(-k^\ast,\varphi)
&=E(-k^\ast,\varphi)|\psi(k,\varphi)\rangle^\dag.
\end{align}
The latter equation is the left-eigenvalue equation for the Hamiltonian $H_\mathrm{eff}(-k^\ast,\varphi)$.
Therefore, the left-eigenvector defined by
\begin{align}\label{eqS320}
\langle\tilde{\psi}(-k^\ast,\varphi)|&:=|\psi(k,\varphi)\rangle^\dag
\end{align}
is the bra vector $\langle\tilde{\psi}^\mathrm{ar}(\varphi)|$ of the anti-resonant state at $k^\mathrm{ar}=-k^\ast$ under the original magnetic field $\varphi$:
\begin{align}
\langle\tilde{\psi}^\mathrm{ar}(\varphi)|=|\psi^\mathrm{res}(\varphi)\rangle^\dag.
\end{align}

Other possible relations are summarized in Table~\ref{tabS1}.
\begin{table}
\caption{Summary of the relations among the resonant and anti-resonant states under a magnetic field.}
\label{tabS1}
\centering
\begin{tabular}{rllll}
\hline
$E^\mathrm{res}(\varphi)=$ & \multicolumn{1}{c}{---} & $E^\mathrm{res}(-\varphi)$, & $E^\mathrm{ar}(\varphi)^\ast$, & $E^\mathrm{ar}(-\varphi)^\ast$, \\
\hline
$E^\mathrm{res}(-\varphi)=$ & $E^\mathrm{res}(\varphi)$, & \multicolumn{1}{c}{---} & $E^\mathrm{ar}(\varphi)^\ast$, & $E^\mathrm{ar}(-\varphi)^\ast$, \\
\hline
$E^\mathrm{ar}(\varphi)=$ & $E^\mathrm{res}(\varphi)^\ast$, & $E^\mathrm{res}(-\varphi)^\ast$, & \multicolumn{1}{c}{---} & $E^\mathrm{ar}(-\varphi)$, \\
\hline
$E^\mathrm{ar}(-\varphi)=$ & $E^\mathrm{res}(\varphi)^\ast$, & $E^\mathrm{res}(-\varphi)^\ast$, & $E^\mathrm{ar}(\varphi)$, & \multicolumn{1}{c}{---} \\
\hline
\hline
$|\psi^\mathrm{res}(\varphi)\rangle=$ & \multicolumn{1}{c}{---} & $|\psi^\mathrm{ar}(-\varphi)\rangle^\ast$, & $\langle\tilde{\psi}^\mathrm{res}(-\varphi)|^\mathrm{T}$, & $\langle\tilde{\psi}^\mathrm{ar}(\varphi)|^\dag$, \\
\hline
$|\psi^\mathrm{ar}(-\varphi)\rangle=$ & $|\psi^\mathrm{res}(\varphi)\rangle^\ast$, & \multicolumn{1}{c}{---} & $\langle\tilde{\psi}^\mathrm{res}(-\varphi)|^\dag$, & $\langle\tilde{\psi}^\mathrm{ar}(\varphi)|^\mathrm{T}$, \\
\hline
$\langle\tilde{\psi}^\mathrm{res}(-\varphi)|=$ & $|\psi^\mathrm{res}(\varphi)\rangle^\mathrm{T}$, & $|\psi^\mathrm{ar}(-\varphi)\rangle^\dag$, & \multicolumn{1}{c}{---} & $\langle\tilde{\psi}^\mathrm{ar}(\varphi)|^\ast$, \\
\hline
$\langle\tilde{\psi}^\mathrm{ar}(\varphi)|=$ & $|\psi^\mathrm{res}(\varphi)\rangle^\dag$, & $|\psi^\mathrm{ar}(-\varphi)\rangle^\mathrm{T}$, & $\langle\tilde{\psi}^\mathrm{res}(-\varphi)|^\ast$, & \multicolumn{1}{c}{---} \\
\hline
\hline
$|\psi^\mathrm{res}(-\varphi)\rangle=$& \multicolumn{1}{c}{---} & $|\psi^\mathrm{ar}(\varphi)\rangle^\ast$, & $\langle\tilde{\psi}^\mathrm{res}(\varphi)|^\mathrm{T}$, & $\langle\tilde{\psi}^\mathrm{ar}(-\varphi)|^\dag$, \\
\hline
$|\psi^\mathrm{ar}(\varphi)\rangle=$  & $|\psi^\mathrm{res}(-\varphi)\rangle^\ast$,  & \multicolumn{1}{c}{---} & $\langle\tilde{\psi}^\mathrm{res}(\varphi)|^\dag$, & $\langle\tilde{\psi}^\mathrm{ar}(-\varphi)|^\mathrm{T}$, \\
\hline
$\langle\tilde{\psi}^\mathrm{res}(\varphi)|=$& $|\psi^\mathrm{res}(-\varphi)\rangle^\mathrm{T}$, & $|\tilde{\psi}^\mathrm{ar}(\varphi)\rangle^\dag$, & \multicolumn{1}{c}{---} & $\langle\tilde{\psi}^\mathrm{ar}(-\varphi)|^\ast$, \\
\hline
$\langle\tilde{\psi}^\mathrm{ar}(-\varphi)|=$& $|\psi^\mathrm{res}(-\varphi)\rangle^\dag$, & $|\tilde{\psi}^\mathrm{ar}(\varphi)\rangle^\mathrm{T}$, & $\langle\tilde{\psi}^\mathrm{res}(\varphi)|^\ast$, & \multicolumn{1}{c}{---} \\
\hline
\end{tabular}
\end{table}
We obtain from this Table the norm for the anti-resonant state as
\begin{align}
\langle\tilde{\psi}(-k^\ast,\varphi)|\psi(-k^\ast,\varphi)\rangle
&=\langle\tilde{\psi}(-k^\ast,-\varphi)|\psi(-k^\ast,-\varphi)\rangle
\nonumber\\
&=\int\psi(x;-k^\ast,\varphi)\psi(x;-k^\ast,-\varphi)dx
\nonumber\\
&=\langle\tilde{\psi}(k,\varphi)|\psi(k,\varphi)\rangle^\ast,
\end{align}
or
\begin{align}
\langle\tilde{\psi}^\mathrm{res}(\varphi)|\psi^\mathrm{res}(\varphi)\rangle
&=\langle\tilde{\psi}^\mathrm{ar}(\varphi)|\psi^\mathrm{res}(\varphi)\rangle^\ast.
\end{align}
The relations in Table~\ref{tabS1} reduce to eqs.~\eqref{eqS150},~\eqref{eqS180} and~\eqref{eqS170} for $\varphi=0$.

\subsection{Relations for the bound and anti-bound states}

Finally, we briefly mention the relations for the bound and anti-bound states, for which $k$ is pure imaginary and hence $k^\mathrm{b/ab}=k=-k^\ast$.
Equation~\eqref{eqS100} then dictates that the eigenenergy $E^\mathrm{b/ab}$ must be real, which is indeed the case for the bound and anti-bound states.
Table~\ref{tabS1} reduces to Table~\ref{tabS2};
\begin{table}
\caption{Summary of the relations among the bound and anti-bound states under a magnetic field.}
\label{tabS2}
\centering
\begin{tabular}{rllll}
\hline
$|\psi^\mathrm{b/ab}(\varphi)\rangle=$ & \multicolumn{1}{c}{---} & $|\psi^\mathrm{b/ab}(-\varphi)\rangle^\ast$, & $\langle\tilde{\psi}^\mathrm{b/ab}(-\varphi)|^\mathrm{T}$, & $\langle\tilde{\psi}^\mathrm{b/ab}(\varphi)|^\dag$, \\
\hline
$|\psi^\mathrm{b/ab}(-\varphi)\rangle=$ & $|\psi^\mathrm{b/ab}(\varphi)\rangle^\ast$, & \multicolumn{1}{c}{---} & $\langle\tilde{\psi}^\mathrm{b/ab}(-\varphi)|^\dag$, & $\langle\tilde{\psi}^\mathrm{b/ab}(\varphi)|^\mathrm{T}$, \\
\hline
$\langle\psi^\mathrm{b/ab}(-\varphi)|=$ & $|\psi^\mathrm{b/ab}(\varphi)\rangle^\mathrm{T}$, & $|\psi^\mathrm{b/ab}(-\varphi)\rangle^\dag$, & \multicolumn{1}{c}{---} & $\langle\tilde{\psi}^\mathrm{b/ab}(\varphi)|^\ast$, \\
\hline
$\langle\psi^\mathrm{b/ab}(\varphi)|=$ & $|\psi^\mathrm{b/ab}(\varphi)\rangle^\dag$, & $|\psi^\mathrm{b/ab}(-\varphi)\rangle^\mathrm{T}$, & $\langle\tilde{\psi}^\mathrm{b/ab}(-\varphi)|^\ast$, & \multicolumn{1}{c}{---} \\
\hline
\end{tabular}
\end{table}
in particular, eq.~\eqref{eqS150} shows that, if there is no magnetic field, the wave function $|\psi^\mathrm{b/ab}\rangle$ can be put to be real, which is also a well-known fact for bound states.

\section{The Green's functions in the central dot}
\label{app:selfenergy}
In the present Appendix, we describe the calculation of the Green's function $G^\textrm{R}_{ij}(E)$ for the states in the central dot, $\{|d_i\rangle\}$.
The fact that we can reduce the calculation to the inversion of a finite-dimensional matrix is fully utilized in \S~\ref{sec:conductance}.
The calculation uses the self-energy of the semi-infinite leads~\cite{Datta95,SH2008,Livshits57,Feshbach58,Feshbach62,Albeverio96,Fyodorov97,Dittes00,Pichugin01,Sadreev03,Okolowicz03,Shapiro06,Shapiro08,Rotter09}.
Using the expression of the Green's function, we give in Appendix\ref{app:calceig} an equation that gives the resonant states.

The basic statement is the fact
\begin{equation}\label{eqB10}
G_{ij}^\textrm{R}(E)\equiv\left\langle d_i\right|
\frac{1}{E-H+i\delta}\left| d_j \right\rangle
=\left\langle d_i\right|
\frac{1}{E-H^\textrm{R}_\textrm{eff}(E)}\left| d_j \right\rangle,
\end{equation}
where the thus-defined effective Hamiltonian $H^\textrm{R}_\textrm{eff}$ has degrees of freedom only on the central dot.
Below, we will review the derivation of the following form:
\begin{equation}\label{eqB20}
H^\textrm{R}_\textrm{eff}(E)=H_\textrm{d}+
\sum_{\alpha=1,2}
\Sigma^\textrm{R}_\alpha(E),
\end{equation}
where
\begin{align}\label{eqB30}
H_\textrm{d}&\equiv\sum_{i=1}^{N}\varepsilon_{i}|d_{i}\rangle\langle d_{i}|
\nonumber\\
&-\sum_{1\leq i < j \leq N} v_{ij}\left(\left|d_i\right\rangle\left\langle d_j\right|+\left|d_j\right\rangle\left\langle d_i\right|\right),
\\ \label{eqB31}
\Sigma^\textrm{R}_\alpha(E)&\equiv
\left(\frac{t_\alpha}{t}\right)^2\frac{E-i\sqrt{4t^2-E^2}}{2}|d_\alpha\rangle\langle d_\alpha|.
\end{align}
Therefore, we can calculate the Green's function $G_{ij}^\textrm{R}$ by inverting an $N$-by-$N$ matrix~(\ref{eqB20}).
The second term on the right-hand side of eq.~\eqref{eqB20} is often called the self-energy of the leads.

For the advanced Green's function, we can similarly derive
\begin{equation}\label{eqB32}
G_{ij}^\textrm{A}(E)\equiv\left\langle d_i\right|
\frac{1}{E-H-i\delta}\left| d_j \right\rangle
=\left\langle d_i\right|
\frac{1}{E-H^\textrm{A}_\textrm{eff}(E)}\left| d_j \right\rangle
\end{equation}
with
\begin{align}
H^\textrm{A}_\textrm{eff}(E)&\equiv H_\textrm{d}+
\sum_{\alpha=1,2}
\Sigma^\textrm{A}_\alpha(E)
\\
\Sigma^\textrm{A}_\alpha(E)&\equiv
\left(\frac{t_\alpha}{t}\right)^2\frac{E+i\sqrt{4t^2-E^2}}{2}|d_\alpha\rangle\langle d_\alpha|.
\end{align}
Then we have
\begin{align}
\left(G^\textrm{R}\right)^{-1}&-\left(G^\textrm{A}\right)^{-1}
=H^\textrm{A}_\textrm{eff}-H^\textrm{R}_\textrm{eff}
=\sum_{\alpha=1,2}
\left(\Sigma^\textrm{A}_\alpha-\Sigma^\textrm{R}_\alpha\right)
\nonumber\\
&=\sum_{\alpha=1,2}\left(\frac{t_\alpha}{t}\right)^2i\sqrt{4t^2-E^2}
|d_\alpha\rangle\langle d_\alpha|.
\end{align}
This gives eq.~\eqref{eq:Gamma} with eq.~\eqref{eq28}.

There are several ways of deriving eq.~(\ref{eqB10}).
We present a method using the Feshbach formalism in Appendix\ref{app:Feshbach}.
Another way that we describe here is to use the resolvent expansion
\begin{align}\label{eqB40}
&\frac{1}{E-H+i\delta}
=\frac{1}{E-H_0+i\delta}
\nonumber\\
&+\frac{1}{E-H_0+i\delta}H_1
\frac{1}{E-H_0+i\delta}
\nonumber\\
&+\frac{1}{E-H_0+i\delta}H_1
\frac{1}{E-H_0+i\delta}H_1
\frac{1}{E-H_0+i\delta}
&+\cdots,
\end{align}
where
\begin{align}\label{eqB50}
H_0 \equiv & H_\textrm{d}+\sum_{\alpha=1,2} H_\alpha
\nonumber\\
=&\sum_{i=1}^{N}\varepsilon \left| d_i \right\rangle \left\langle d_i \right|
\nonumber\\
&-\sum_{1\leq i<j\leq N}v_{ij}\left(\left|d_i\right\rangle\left\langle d_j \right|
+\left|d_j\right\rangle\left\langle d_i \right| \right)
\nonumber\\
&-t\sum_{\alpha=1,2}\sum_{x_{\alpha}=0}^{\infty}\left(|x_{\alpha}+1\rangle\langle x_{\alpha}|+|x_{\alpha}\rangle\langle x_{\alpha}+1|\right),
\\ \label{eqB51}
H_1 \equiv & \sum_{\alpha=1,2} H_{\textrm{d},\alpha}
\nonumber\\
=&-\sum_{\alpha=1,2} t_\alpha \left(\left|x_\alpha=0\right\rangle\left\langle d_\alpha \right|
+\left|d_\alpha\right\rangle\left\langle x_\alpha=0\right|\right).
\end{align}
In calculating $G_{ij}^\textrm{R}(E)$ defined in eq.~(\ref{eqB10}), we should note the following.
Let $\mathcal{H}_\textrm{d}$ denote the Hilbert space spanned by the states on the central dot, $\{|d_i\rangle\}$,
and $\mathcal{H}_\textrm{lead}$ denote the Hilbert space spanned by the states on the leads, $\{|x_\alpha\rangle\}$.
Then we have
\begin{align}\label{eqB60}
\frac{1}{E-H_0+i\delta}|d_i\rangle=\frac{1}{E-H_\textrm{d}+i\delta}|d_i\rangle \in& \mathcal{H}_\textrm{d},
\\ \label{eqB61}
\frac{1}{E-H_0+i\delta}|x_\alpha\rangle=\frac{1}{E-H_\alpha+i\delta}|x_\alpha\rangle \in& \mathcal{H}_\textrm{lead},
\\ \label{eqB62}
H_1|d_i\rangle =-\sum_{\alpha=1,2} \delta_{i\alpha}t_\alpha |x_\alpha=0\rangle \in& \mathcal{H}_\textrm{lead},
\\ \label{eqB63}
H_1|x_\alpha\rangle = -\delta_{x_\alpha 0}t_\alpha |d_\alpha\rangle \in& \mathcal{H}_\textrm{d}.
\end{align}
That is, the operator $\left(E-H_0+i\delta\right)^{-1}$, when applied to a state either in $\mathcal{H}_\textrm{d}$ or $\mathcal{H}_\textrm{lead}$, does not change its Hilbert space, whereas the operator $H_1$ switches it.
Therefore, all terms of odd orders of $H_1$ in the resolvent expansion of $G_{ij}^\textrm{R}$ vanish.
All terms of even orders of $H_1$ (except the zeroth order) have powers of the summation over $\alpha$ of the following factor:
\begin{align}\label{eqB70}
&|d_\alpha\rangle
\langle d_\alpha | H_1 | x_\alpha=0 \rangle 
\nonumber\\
&\qquad\times \langle x_\alpha=0 |\frac{1}{E-H_0+i\delta}|x_\alpha=0\rangle
\nonumber\\
&\qquad\times \langle x_\alpha=0 | H_1 |d_\alpha\rangle\langle d_\alpha |
\nonumber\\
&=\left({t_\alpha}^2 \langle x_\alpha=0 |\frac{1}{E-H_\alpha+i\delta}|x_\alpha=0\rangle\right)
 |d_\alpha\rangle\langle d_\alpha |.
\end{align}
We will show below that the above operator is equal to $\Sigma^\textrm{R}_\alpha(E)$ defined in eq.~(\ref{eqB31}).
We therefore have 
\begin{align}\label{eqB80}
&G_{ij}^\textrm{R}(E)
\nonumber\\
&=\langle d_i | \frac{1}{E-H_\textrm{d}+i\delta} | d_j \rangle
\nonumber\\
&\quad+\langle d_i | \frac{1}{E-H_\textrm{d}+i\delta}
\left(\sum_{\alpha=1,2}\Sigma^\textrm{R}_\alpha(E)\right)
 \frac{1}{E-H_\textrm{d}+i\delta} |d_j \rangle 
\nonumber\\
&\quad+\cdots,
\end{align}
which can be summarized as
\begin{align}\label{eqB90}
&G_{ij}^\textrm{R}(E)
=\langle d_i | 
\frac{1}{\displaystyle E-H_\textrm{d}-\left(\sum_\alpha 
\Sigma^\textrm{R}_\alpha(E)
\right)+i\delta}
|d_j\rangle.
\end{align}
This is almost the same as eq.~\eqref{eqB10}.
The infinitesimal $+i\delta$ in the denominator becomes unnecessary because $\Sigma^\textrm{R}_\alpha$ already has an explicitly negative imaginary part, as can be seen in eq.~\eqref{eqB31}.

The remaining task is to show that the operator in eq.~(\ref{eqB70}) is indeed equal to $\Sigma^\textrm{R}_\alpha(E)$ defined in eq.~(\ref{eqB31}).
For the purpose, we calculate $\langle x_\alpha=0 |(E-H_0+i\delta)^{-1}|x_\alpha=0\rangle$ in eq.~(\ref{eqB70}), or
\begin{equation}\label{eqB100}
G_\textrm{lead}^\textrm{R}(E;0)\equiv
\langle x=0 | \frac{1}{E-H_\textrm{lead}(0)+i\delta}|x=0\rangle,
\end{equation}
where
\begin{equation}\label{eqB110}
H_\textrm{lead}(X)=-t\sum_{x=X}^\infty\left(|x+1\rangle\langle x|+|x\rangle\langle x+1|\right).
\end{equation}
We then use the resolvent expansion
\begin{align}\label{eqB120}
&\frac{1}{E-H_\textrm{lead}(0)+i\delta}
=\frac{1}{E-H_\textrm{lead}(1)+i\delta}
\nonumber\\
&+\frac{1}{E-H_\textrm{lead}(1)+i\delta}
\nonumber\\
&\quad\times(-t)\left(|1\rangle\langle 0|+|0\rangle\langle1|\right)
\frac{1}{E-H_\textrm{lead}(1)+i\delta}
\nonumber\\
&+\cdots.
\end{align}
Similar reasoning as the one described in eqs.~(\ref{eqB40})--(\ref{eqB90}) leads us to
\begin{equation}\label{eqB130}
G_\textrm{lead}^\textrm{R}(E;0)=\frac{1}{E-t^2G_\textrm{lead}^\textrm{R}(E;1)+i\delta}
\end{equation}
with
\begin{equation}\label{eqB140}
G_\textrm{lead}^\textrm{R}(E;1)=\langle x=1 | \frac{1}{E-H_\textrm{lead}(1)+i\delta} | x=1 \rangle.
\end{equation}
Thanks to the translational invariance, we should have $G_\textrm{lead}^\textrm{R}(E;0)=G_\textrm{lead}^\textrm{R}(E;1)$.
Then, eq.~(\ref{eqB130}) reduces to a quadratic equation
\begin{equation}\label{eqB150}
t^2\left(G_\textrm{lead}^\textrm{R}\right)^2-EG_\textrm{lead}^\textrm{R}+1=0,
\end{equation}
which is followed by
\begin{equation}\label{eqB160}
G_\textrm{lead}^\textrm{R}(E;0)=\frac{E-i\sqrt{4t^2-E^2}}{2t^2}
\qquad\mbox{for $-2t\leq E \leq 2t$},
\end{equation}
where we fixed the sign in front of the square root so that the imaginary part may be negative.
Thus the the operator in eq.~(\ref{eqB70}) was indeed shown to be equal to $\Sigma^\textrm{R}_\alpha(E)$ defined in eq.~(\ref{eqB31}).

To summarize the above, the retarded Green's function is expressed in the form on the right-hand side of eq.~(\ref{eqB10}) with the definitions in eqs.~\eqref{eqB20}--(\ref{eqB31}).
The Green's functions 
are therefore obtained by inverting the $N$-by-$N$ non-Hermitian matrix $\langle d_i| (E-H^\textrm{R/A}_\textrm{eff}(E)) | d_j \rangle$ for a fixed value of $E$.

Incidentally, the factor $(E\mp i\sqrt{4t^2-E^2})/2$ in $\Sigma^\textrm{R/A}_\alpha$ can be rewritten as
\begin{align}\label{eqB165}
\frac{E\mp i\sqrt{4t^2-E^2}}{2}=-t\mathrm{e}^{\pm ik}
\end{align}
if we use the dispersion relation of the tight-binding leads $E=-2t\cos k$.
In fact, there is a much easier but non-standard way of deriving the self-energy of the leads, eq.~\eqref{eqB31}, directly in the form~\eqref{eqB165};
see ref.~\cite{SH2008}.

Next, we show that the inversion problem of the above non-Hermitian matrix can be reduced to the inversion problem of the Hermitian matrix $\langle d_i| (E-H_\textrm{d}) | d_j \rangle$.
Going back to eq.~\eqref{eqB80}, we rewrite the resolvent expansion in the matrix form
\begin{align}\label{eqA27}
G^\textrm{R}=G^\textrm{d}+G^\textrm{d}\Sigma^\textrm{R} G^\textrm{d}+G^\textrm{d}\Sigma^\textrm{R} G^\textrm{d}\Sigma^\textrm{R} G^\textrm{d}+\cdots,
\end{align}
where
\begin{align}
G^\textrm{d}&\equiv\left(E-H_\textrm{d}\right)^{-1},
\\
\Sigma^\textrm{R}&\equiv\sum_{\alpha=1,2}\Sigma^\textrm{R}_\alpha.
\end{align}
By multiplying $\Sigma^\textrm{R}$ from the left once, we have
\begin{align}\label{eqA300}
&\Sigma^\textrm{R}G^\textrm{R}
\nonumber\\
&=\Sigma^\textrm{R} G^\textrm{d}+
\Sigma^\textrm{R} G^\textrm{d}\Sigma^\textrm{R} G^\textrm{d}+
\Sigma^\textrm{R} G^\textrm{d}\Sigma^\textrm{R} G^\textrm{d}\Sigma^\textrm{R} G^\textrm{d}+\cdots
\end{align}

It is important to notice here that the self-energy of the leads in eq.~\eqref{eqB20} has only diagonal elements at the two contact sites;
all other elements are zero.
Equation~\eqref{eqA300}, therefore, is an equation essentially in the two-dimensional space spanned by the contact-site states $|d_1\rangle$ and $|d_2\rangle$.
In the following, let $\check{A}$ denote a two-by-two matrix constructed from an $N$-by-$N$ matrix $A$ as
\begin{align}\label{eqA310}
\check{A}=\begin{pmatrix}
A_{11} & A_{12} \\
A_{21} & A_{22}
\end{pmatrix}.
\end{align}
Indeed, we will need only the elements of $\check{G}^\textrm{R}$ in Appendix\ref{app:calceig}.
Then we have
\begin{align}
&\check{\Sigma}^\textrm{R}\check{G}^\textrm{R}
\nonumber\\
&=\check{\Sigma}^\textrm{R} \check{G}^\textrm{d}+
\check{\Sigma}^\textrm{R} \check{G}^\textrm{d}\check{\Sigma}^\textrm{R} \check{G}^\textrm{d}+
\check{\Sigma}^\textrm{R} \check{G}^\textrm{d}\check{\Sigma}^\textrm{R} \check{G}^\textrm{d}\check{\Sigma}^\textrm{R} \check{G}^\textrm{d}+\cdots
\nonumber\\
&=\frac{\check{\Sigma}^\textrm{R}\check{G}^\textrm{d}}{\check{I}-\check{\Sigma}^\textrm{R}\check{G}^\textrm{d}}
\nonumber\\
&=\frac{\check{I}}{\left(\check{G}^\textrm{d}\right)^{-1}\left(\check{\Sigma}^\textrm{R}\right)^{-1}-\check{I}},
\end{align}
where $\check{I}$ is the two-by-two identity matrix and
\begin{align}
\check{\Sigma}^\textrm{R}=\begin{pmatrix}
\Sigma^\textrm{R}_1 & 0 \\
0 & \Sigma^\textrm{R}_2
\end{pmatrix}.
\end{align}
We thus arrive at
\begin{align}\label{eqA34}
\check{G}^\textrm{R}=\left[\left(\check{G}^\textrm{d}\right)^{-1}-\check{\Sigma}^\textrm{R}\right]^{-1}.
\end{align}
The calculation of this matrix involves the calculation of $G^\textrm{d}$, or the inversion of the $N$-by-$N$ \textit{Hermitian} matrix $E-H_\textrm{d}$.
The other two matrix inversions are done in the two-dimensional space.

\section{Derivation of the effective Hamiltonian using the Feshbach formalism}
\label{app:Feshbach}

We here show another way of deriving the effective Hamiltonian~\eqref{eqB20}, namely the Feshbach formalism,~\cite{Feshbach58,Feshbach62,Rotter09} which was first developed for nuclear physics.
Let us introduce for the present model~\eqref{eq:Hamiltonian_original} the following projection operators:
\begin{align}\label{eqCC10}
P&\equiv\sum_{i=1}^N|d_i\rangle\langle d_i|,
\\\label{eqCC20}
Q&\equiv1-P=\sum_{\alpha=1,2}\sum_{x_\alpha=0}^\infty|x_\alpha\rangle\langle x_\alpha|.
\end{align}
We operate these projection operators on the time-independent Schr\"{o}dinger equation for the total Hamiltonian
\begin{align}
H|\psi_n\rangle=E_n|\psi_n\rangle
\end{align}
and derive the equation for the projected component $P|\psi_n\rangle$.
We will show that the result is
\begin{align}\label{eqCC40}
H_\textrm{eff}(E_n)\left(P|\psi_n\rangle\right)&=E_n\left(P|\psi_n\rangle\right)
\end{align}
with $H_\textrm{eff}$ given in eq.~\eqref{eqB20}.

We first have
\begin{align}\label{eqCC50}
PHP|\psi_n\rangle+PHQ|\psi_n\rangle&=E_nP|\psi_n\rangle,
\\\label{eqCC60}
QHP|\psi_n\rangle+QHQ|\psi_n\rangle&=E_nQ|\psi_n\rangle.
\end{align}
We formally solve eq.~\eqref{eqCC60} with respect to $Q|\psi_n\rangle$ in the form
\begin{align}
Q|\psi_n\rangle&=\frac{1}{E_n-QHQ}QHP|\psi_n\rangle
\end{align}
and substitute it into eq.~\eqref{eqCC50}, obtaining
\begin{align}\label{eqCC80}
PHP|\psi_n\rangle+PHQ\frac{1}{E_n-QHQ}QHP|\psi_n\rangle&=E_nP|\psi_n\rangle.
\end{align}
We can cast eq.~\eqref{eqCC80} into the form~\eqref{eqCC40} with the effective Hamiltonian given by
\begin{align}\label{eqCC90}
H_\mathrm{eff}(E)&=PHP+PHQ\frac{1}{E-QHQ}QHP.
\end{align}

Let us here note that for the present Hamiltonian~\eqref{eq:Hamiltonian_original} with the projection operators~\eqref{eqCC10}--\eqref{eqCC20}, we have
\begin{align}
PHP&=H_\mathrm{d},
\\
PHQ+QHP&=\sum_{\alpha=1,2}H_{\mathrm{d},\alpha},
\\
QHQ&=\sum_{\alpha=1,2}H_\alpha.
\end{align}
Therefore, the term $(E-QHQ)^{-1}$ in the expression~\eqref{eqCC90} is indeed the Green's function~\eqref{eqB100}, where the convergence factor $+i\delta$ is added to give the retarded one.
Equation~\eqref{eqCC90} thereby results in the effective Hamiltonian~\eqref{eqB20}.

\section{Calculation of discrete eigenvalues}
\label{app:calceig}

We show in the present Appendix how we can calculate all resonant states for the system~(\ref{eq:Hamiltonian_original}).
As is evident in the Fisher-Lee relation~(\ref{eq:conductance}), the conductance of the present system has poles in the complex energy plane wherever the Green's function $G^\textrm{R}(E)$ has poles.
Since the matrix $G^\textrm{R}$ is the inversion of the matrix $E-H^\textrm{R}_\textrm{eff}(E)$ as is shown in Appendix\ref{app:selfenergy}, all poles $E_n$ (or all discrete eigenstates including the resonant states) can be calculated by solving the equation
\begin{equation}\label{eqB171}
\det(E-H^\textrm{R}_\textrm{eff}(E))=0
\end{equation}
and the corresponding eigenvector by solving
\begin{align}\label{eqB200}
H^\textrm{R}_\textrm{eff}(E_n)|\psi_n\rangle=E_n|\psi_n\rangle.
\end{align}
Although this seems a usual eigenvalue problem, we should note that the Hamiltonian $H^\textrm{R}_\textrm{eff}$ itself is energy-dependent, and therefore it is not a standard eigenvalue problem.
In fact, the number of the eigenvalues is \textit{not} equal to the dimensionality $N$ of the Hamiltonian $H^\textrm{R}_\textrm{eff}$.

Let us count the number of the solutions of the resonance equation~\eqref{eqB171}.
It is convenient to use the variable
\begin{align}
z=\mathrm{e}^{ik}.
\end{align}
In eq.~\eqref{eqB171}, we have
\begin{align}
E=-t\left(z+\frac{1}{z}\right).
\end{align}
The energy dependence of $H^\textrm{R}_\textrm{eff}(E)$ comes from $\Sigma^\textrm{R}_\alpha$, which contains $-tz$ as was shown in eq.~\eqref{eqB165}.
Therefore, we can cast eq.~\eqref{eqB171} into a $2N$th-order polynomial in $z$.
We thereby conclude that the system generally has $2N$ discrete eigenstates in total.

In the cases where the inversion of the Hermitian matrix $E-H_\textrm{d}$ can be carried out easily, it may be more convenient for finding the discrete eigenvalues to use the expression~\eqref{eqA34}, from which the resonance equation is given by
\begin{align}
\det\left[\left(\check{G}^\textrm{d}\right)^{-1}-\check{\Sigma}^\textrm{R}\right]=0.
\end{align}
Here the matrix whose determinant is to be calculated is a two-by-two matrix.
Particularly when the two leads are attached to one site $0$, the resonance equation reduces to
\begin{align}
G^\textrm{d}_{00}(E)=-\frac{t\mathrm{e}^{-ik}}{t_1^2+t_2^2}.
\end{align}

The corresponding eigenvector inside the dot is obtained by solving eq.~\eqref{eqB200}.
As is shown in eq.~\eqref{eq6}, the eigenvector in the lead is given by
\begin{align}\label{eqB700}
\langle x_\alpha | \psi_n\rangle=\frac{t_\alpha}{t}{z_n}^{x_\alpha}\langle d_\alpha | \psi_n\rangle,
\end{align}
where $z_n=\exp(ik_n)$ is related to the eigenenergy as
\begin{align}\label{eqB8000}
E_n=-t\left(z_n+\frac{1}{z_n}\right)
\end{align}
because of eq.~\eqref{eq:dispersion_resonant};
see ref.~\citen{SH2008} for the derivation of eq.~\eqref{eqB700}.

\section{Solution of the matrix Riccati equation}
\label{app:sign}

In the present Appendix, we will solve eq.~\eqref{eq:RiccatiGR} and derive the formula~\eqref{eq:conductance_G}.
In eq.~\eqref{eq:RiccatiGR}, we restrict ourselves to the two-dimensional space spanned by $|d_1\rangle$ and $|d_2\rangle$.
This is possible because the matrix $\Gamma$ has diagonal elements only in this two-dimensional space, as can be seen in eqs.~\eqref{eq:GA-GR_gamma2} and~\eqref{eq28}. 
In the present Appendix, we let $G^\textrm{R}$, $\Gamma$ and $\Lambda$ all denote two-by-two matrices for simplicity.
(From the viewpoint of the notation given in eq.~\eqref{eqA310}, it would be proper to express them as $\check{G}^\textrm{R}$, $\check{\Gamma}$ and $\check{\Lambda}$, but we avoid to use them for brevity of the notations.)

Then the matrix equation to be solved is
\begin{align}
G^\textrm{R}\Gamma G^\textrm{R}+2iG^\textrm{R}-G^\textrm{R}\Gamma\Lambda-i\Lambda=0.
\end{align}
By multiplying $\Gamma$ from the left and rearranging the terms, we have
\begin{align}\label{eq:Xi-Theta}
\Xi^2-\Xi (\Theta-2iI)-i\Theta=0,
\end{align}
where
\begin{align}
\Xi&\equiv \Gamma G^\textrm{R},
\\
\Theta&\equiv \Gamma \Lambda,
\end{align}
and $I$ denotes the two-by-two identity matrix.

We here show that $[\Xi,\Theta]=0$.
Since
\begin{align}
\Theta=\Gamma(G^\textrm{R}+G^\textrm{A})=\Xi+\Gamma G^\textrm{A},
\end{align}
what we should show is $[\Gamma G^\textrm{R},\Gamma G^\textrm{A}]=0$.
Because of eq.~\eqref{eq:Gamma}, we have
\begin{align}\label{eq:C60}
i\Gamma G^\textrm{R}=I-\left(G^\textrm{A}\right)^{-1}G^\textrm{R},
\\ \label{eq:C70}
i\Gamma G^\textrm{A}=\left(G^\textrm{R}\right)^{-1}G^\textrm{A}-I.
\end{align}
After these expressions, it is straightforward to see $[\Gamma G^\textrm{R},\Gamma G^\textrm{A}]=0$.

Because $\Xi$ and $\Theta$ commute with each other, we can solve eq.~\eqref{eq:Xi-Theta} just as a usual quadratic equation to obtain
\begin{align}
\Xi&=\frac{\Theta-2iI\pm\sqrt{(\Theta-2iI)^2+4i\Theta}}{2}
\nonumber\\
&=\frac{\Theta}{2}-iI\pm\sqrt{\frac{\Theta^2}{4}-I}.
\end{align}
Then we obtain the Green's function
\begin{align}\label{eq:C90}
G^\textrm{R}=\frac{\Lambda}{2}-i\Gamma^{-1}\left(I\pm\sqrt{I-\frac{\Theta^2}{4}}\right).
\end{align}
We here have flipped the sign in the square root and extracted the imaginary number, because then we have
\begin{align}\label{eq:C100}
G^\textrm{A}&=\left(G^\textrm{R}\right)^\ast
\nonumber\\
&=\frac{\Lambda}{2}+i\Gamma^{-1}\left(I\pm\sqrt{I-\frac{\Theta^2}{4}}\right)
\end{align}
and the two Green's functions give the consistent result
\begin{align}
G^\textrm{R}+G^\textrm{A}=\Lambda.
\end{align}

The next step is to simplify the expression of the matrix square root.
For two-by-two matrices, we have the Cayley-Hamilton equality:
\begin{align}
\Theta^2-T\Theta+DI=0,
\end{align}
where $T=\mathop{\textrm{Tr}}\Theta=\mathop{\textrm{Tr}}\Gamma\Lambda$ and $D=\det\Theta=\det\Gamma\Lambda$.
This implies that any functions of the matrix $\Theta$ that can be expanded in the Taylor series is reduced to a linear function $\alpha\Theta+\beta$.
In the present case, let us express
\begin{align}\label{eqC130}
\sqrt{I-\frac{\Theta^2}{4}}=\alpha\Theta+\beta I,
\end{align}
or
\begin{align}
\Xi=\left(\frac{1}{2}\mp i\alpha\right)\Theta-i\left(1\pm\beta\right)I.
\end{align}
and look for the coefficients $\alpha$ and $\beta$.
Once we obtain the coefficients, the Fisher-Lee relation~\eqref{eq:conductance} gives the conductance as
\begin{align}\label{eqD15}
\mathcal{G}_{12}(E)&=\frac{2e^2}{h}\Xi_{12}\Xi_{21}^\ast
\nonumber\\
&=\frac{2e^2}{h}\Theta_{12}\Theta_{21}\left(\frac{1}{2}-i\alpha\right)\left(\frac{1}{2}+i\alpha\right)
\nonumber\\
&=\frac{2e^2}{h}\Gamma_{11}\Lambda_{12}\Gamma_{22}\Lambda_{21}\left(\frac{1}{4}+\alpha^2\right).
\end{align}

The coefficients $\alpha$ and $\beta$ in eq.~\eqref{eqC130} are given in terms of the two eigenvalues of the matrix $\Theta$, which will be denoted by $\theta_1$ and $\theta_2$ hereafter.
We then have
\begin{align}\label{eq:C160}
\begin{cases}
\displaystyle \sqrt{1-\frac{{\theta_1}^2}{4}}=\alpha\theta_1+\beta,\\
\displaystyle \pm\sqrt{1-\frac{{\theta_2}^2}{4}}=\alpha\theta_2+\beta,
\end{cases}
\end{align}
where the multiple sign in front of the second line actually indicates the \textit{relative} sign of the square roots on the left-hand sides.
If we flip the signs of the square roots at the same time, the signs of $\alpha$ and $\beta$ flip, which does not affect the final result~\eqref{eqD15}.
The solution is given in the form
\begin{align}
\begin{pmatrix}
\alpha \\
\beta
\end{pmatrix}
=\frac{1}{\theta_1-\theta_2}
\begin{pmatrix}
1 & -1 \\
-\theta_2 & \theta_1
\end{pmatrix}
\begin{pmatrix}
\sqrt{1-\frac{{\theta_1}^2}{4}} \\
\pm\sqrt{1-\frac{{\theta_2}^2}{4}}
\end{pmatrix},
\end{align}
which is followed by
\begin{align}\label{eqD18}
\alpha^2&=\frac{1}{\left(\theta_1-\theta_2\right)^2}
\left(\sqrt{1-\frac{{\theta_1}^2}{4}}\pm\sqrt{1-\frac{{\theta_2}^2}{4}}\right)^2
\nonumber\\
&=\frac{1}{\left(\theta_1-\theta_2\right)^2}
\left(2-\frac{{\theta_1}^2+{\theta_2}^2}{4}\phantom{\sqrt{\frac{{\theta_1}^2{\theta_2}^2}{16}}}\right.
\nonumber\\
&\phantom{=\frac{1}{\left(\theta_1-\theta_2\right)^2}}
\left.\pm2\sqrt{1-\frac{{\theta_1}^2+{\theta_2}^2}{4}+\frac{{\theta_1}^2{\theta_2}^2}{16}}\right).
\end{align}

Because the two eigenvalues $\theta_1$ and $\theta_2$ are the solutions of the quadratic equation
\begin{align}
\theta^2-T\theta+D=0,
\end{align}
they satisfy the equalities
\begin{align}
\theta_1+\theta_2&=T,
\\
\theta_1\theta_2&=D,
\\
{\theta_1}^2+{\theta_2}^2&=T^2-2D,
\\
\left(\theta_1-\theta_2\right)^2&=T^2-4D.
\end{align}
Using these equalities in eq.~\eqref{eqD18}, we have
\begin{align}
\alpha^2&=\frac{1}{T^2-4D}\left(2-\frac{T^2-2D}{4}\right.
\nonumber\\
&\phantom{=\frac{1}{T^2-4D}}
\left.\pm\frac{1}{2}\sqrt{16-4T^2+8D+D^2}\right)
\nonumber\\
&=-\frac{1}{4}+\frac{1}{2\left(T^2-4D\right)}\left(4-D\pm\sqrt{\left(D+4\right)^2-4T^2}\right).
\end{align}
Combining this with eq.~\eqref{eqD15}, we arrive at the formula~\eqref{eq:conductance_G}.

Let us finally present a way of determining the sign of the multiple sign.
From eqs.~\eqref{eq:C90} and~\eqref{eq:C100}, we have
\begin{align}
\Gamma G^\textrm{R}-\Gamma G^\textrm{A}=-2iI\mp 2i\sqrt{I-\frac{\Theta^2}{4}}.
\end{align}
As we discussed below eq.~\eqref{eq:C160}, the sign of the square-root operator in fact means the relative sign of the two eigenvalues.
We can therefore know the appropriate sign from the sign of
\begin{align}
\det\sqrt{I-\frac{\Theta^2}{4}}=\det\left[I+\frac{i}{2}\left(\Gamma G^\textrm{A}-\Gamma G^\textrm{R}\right)\right].
\end{align}
By using eqs.~\eqref{eq:C60} and~\eqref{eq:C70}, we can also write the above quantity as
\begin{align}
\det\sqrt{I-\frac{\Theta^2}{4}}&=\det\left[\frac{1}{2}\left(G^\textrm{R}\right)^{-1}G^\textrm{A}
+\frac{1}{2}\left(G^\textrm{A}\right)^{-1}G^\textrm{R}\right]
\nonumber\\
&=\det\left[\mathop{\textrm{Re}}\left(G^\textrm{R}\right)^{-1}G^\textrm{A}\right].
\end{align}
We remind the readers that all matrix calculations in the present Appendix should be done as two-by-two matrices.

\section{Friedrichs solution of the system~(\ref{eq:Hamiltonian_original})}
\label{app:Friedrichs}
In the present Appendix, we solve the Lippmann-Schwinger equation for the present system~(\ref{eq:Hamiltonian_original}) to obtain the Friedrichs solution~\cite{Friedrichs1948} of the scattering states that appears in eq.~\eqref{eq450} in \S~\ref{sec:spectrum}.
The Lippmann-Schwinger equation may be written down as
\begin{equation}\label{eqA10}
\left|\psi_{k,\alpha}\right\rangle
=\left|k,\alpha\right\rangle
+\frac{1}{E_k-H_0+i\delta}H_1\left|\psi_{k,\alpha}\right\rangle,
\end{equation}
where
\begin{align}\label{eqA20}
H_0 \equiv & H_\textrm{d}+\sum_\alpha H_\alpha
\nonumber\\
=&\sum_{i=1}^{N}\varepsilon \left| d_i \right\rangle \left\langle d_i \right|
\nonumber\\
&-\sum_{1\leq i<j\leq N}v_{ij}\left(\left|d_i\right\rangle\left\langle d_j \right|
+\left|d_j\right\rangle\left\langle d_i \right| \right)
\nonumber\\
&+\sum_\alpha\int_{-\pi}^\pi \frac{dk}{2\pi}E_k\left|k,\alpha\right\rangle\left\langle k,\alpha\right|,
\\ \label{eqA21}
H_1 \equiv & \sum_\alpha H_{\textrm{d},\alpha}
\nonumber\\
=&-\sum_\alpha t_\alpha \int_{-\pi}^\pi\frac{dk}{2\pi} \sqrt{2}\sin k \left(\left|k,\alpha\right\rangle\left\langle d_\alpha \right|
+\left|d_\alpha\right\rangle\left\langle k,\alpha\right|\right),
\end{align}
the state $\left|k,\alpha\right\rangle$ is an eigenstate of $H_0$ (more specifically, of $H_\alpha$) with the eigenvalue $E_k=-2t\cos k$, and $\delta$ is a positive infinitesimal ensuring that the solution is an outgoing wave. For semi-infinite leads the states $\left|k,\alpha\right\rangle$ are normalized as $
\langle x,\alpha' \left|k,\alpha\right\rangle =\sqrt{2} \sin [k(x+1)]  \delta_{\alpha',\alpha}$.

The formal solution of the Lippmann-Schwinger equation~(\ref{eqA10}) is given in the form
\begin{align}\label{eqA30}
\left|\psi_{k,\alpha}\right\rangle
=&\left|k,\alpha\right\rangle
+\frac{1}{E_k-H+i\delta}H_1\left|k,\alpha\right\rangle
\nonumber\\
=&\left|k,\alpha\right\rangle
-\frac{t_\alpha \sqrt{2}\sin k}{E_k-H+i\delta}\left|d_\alpha\right\rangle.
\end{align}
Using the resolution of unity
\begin{equation}\label{eqA40}
1=\sum_{i=1}^{N}\left|d_i\right\rangle \left\langle d_i\right|
+\sum_\beta\int_{-\pi}^\pi\frac{dq}{2\pi}\left|q,\beta\right\rangle\left\langle q,\beta\right|,
\end{equation}
we then have
\begin{align}\label{eqA50}
\left|\psi_{k,\alpha}\right\rangle
=&\left|k,\alpha\right\rangle
-t_\alpha \sqrt{2}\sin k \left(\sum_{i=1}^{N}G^\textrm{R}_{i\alpha}(E_k)\left|d_i\right\rangle
\right.
\nonumber\\
&
+\sum_\beta\int_{-\pi}^\pi\frac{dq}{2\pi}\left\langle q,\beta\right|\frac{1}{E_k-H+i\delta}\left|d_\alpha\right\rangle
\left|q,\beta\right\rangle\Biggr),
\end{align}
where
\begin{equation}\label{eqA55}
G^\textrm{R}_{ij}(E_k)\equiv\left\langle d_i\right|\frac{1}{E_k-H+i\delta}\left|d_j\right\rangle.
\end{equation}
In order to transform the final term on the right-hand side of eq.~(\ref{eqA50}), we calculate the following:
\begin{align}\label{eqA60}
&\frac{1}{E_k-H-i\delta}\left|q,\beta\right\rangle
\nonumber\\
&=\left(1+\frac{1}{E_k-H-i\delta}H_1\right)\frac{1}{E_k-H_0-i\delta}\left|q,\beta\right\rangle
\nonumber\\
&=\frac{1}{E_k-E_q-i\delta}\left(\left|q,\beta\right\rangle-\frac{t_\beta\sqrt{2}\sin q}{E_k-H-i\delta}\left|d_\beta\right\rangle\right).
\end{align}
We thereby have
\begin{align}\label{eqA70}
\left\langle q,\beta\right|\frac{1}{E_k-H+i\delta}\left|d_\alpha\right\rangle
=-\frac{t_\beta \sqrt{2}\sin q G^\textrm{R}_{\beta\alpha}(E_k)}{E_k-E_q+i\delta}.
\end{align}
We therefore arrive at
\begin{align}\label{eqA5000}
|\psi_{k,\alpha}\rangle=& |k,\alpha\rangle -t_{\alpha} \sqrt{2}\sin k
\left(\sum_{i=1}^{N}G^\textrm{R}_{i\alpha}(E_k)|d_{i}\rangle\right. 
\nonumber \\
&\left.-\sum_{\beta}t_\beta G^\textrm{R}_{\beta\alpha}(E_k) \int_{-\pi}^{\pi}\frac{dq}{2\pi}\frac{\sqrt{2}\sin q |q,\beta\rangle}{E_k-E_q+i\delta}\right).
\end{align}
We describe in Appendix\ref{app:selfenergy} how we can calculate the Green's function $G^\textrm{R}_{ij}$.

We have calculated so far the right-eigenvector of the Hamiltonian $H$.
Since the Hamiltonian has semi-infinite leads and its effective Hamiltonian $H^\textrm{R}_\textrm{eff}$ is non-Hermitian, the left-eigenvector of the Hamiltonian $H$ is \textit{not} Hermitian conjugate to the corresponding right-eigenvector.
Starting from the Lippmann-Schwinger equation for the left-eigenvector
\begin{align}
\langle\tilde{\psi}_{k,\alpha}|
=\langle k,\alpha|+\langle\tilde{\psi}_{k,\alpha}|H_1\frac{1}{E-H_0+i\delta},
\end{align}
we have the final form
\begin{align}\label{eqA5200}
\langle\tilde{\psi}_{k,\alpha}|=& \langle k,\alpha| -\sqrt{2}t_{\alpha}\sin k
\left(\sum_{i=1}^{N}G^\textrm{R}_{i\alpha}(E_k)\langle d_{i}|\right. 
\nonumber \\
&\left.-\sum_{\beta}t_\beta G^\textrm{R}_{\beta\alpha}(E_k) \int_{-\pi}^{\pi}\frac{dq}{2\pi}\frac{\sqrt{2}\sin q\langle q,\beta|}{E_k-E_q+i\delta}\right).
\end{align}
Since the vector $|k,\alpha\rangle$ is a plane wave and the vector $|d_i\rangle$ is a site state, we can choose their phases such that
\begin{align}
\langle k,\alpha|&=|k,\alpha\rangle^\textrm{T},
\\
\langle d_i|&=|d_i\rangle^\textrm{T}.
\end{align}
Then we observe
\begin{align}
\langle\tilde{\psi}_{k,\alpha}|=|\psi_{k,\alpha}\rangle^\textrm{T}
\quad\left(\neq|\psi_{k,\alpha}\rangle^\dagger\right).
\end{align}
In fact, if $|\psi_{k,\alpha}\rangle$ is the right-eigenvector of a resonant state, the vector $|\psi_{k,\alpha}\rangle^\dagger$ is the left-eigenvector of the corresponding anti-resonant state, because $|\psi_{k,\alpha}\rangle^\ast$ is the right-eigenvector of the anti-resonant state; see eq.~\eqref{eq:eigenfunction_anti-resonant}, eqs.~\eqref{eqS2300}--\eqref{eqS2400} and eqs.~\eqref{eqS2900}--\eqref{eqS3000}.

\section{Proof of eq.~(\ref{eq:GA+GR_parallel})}
\label{appendix:A}
In the present Appendix, we prove eq.~(\ref{eq:GA+GR_parallel}).
Using the expressions~(\ref{eqA5000}) and~\eqref{eqA5200} of the scattering state, we have
\begin{align}\label{eqC10}
\langle d_i | \psi_k \rangle \langle\tilde{\psi}_k | d_j \rangle
=&\sum_\alpha 
\langle d_i | \psi_{k,\alpha} \rangle \langle\tilde{\psi}_{k,\alpha} | d_j \rangle
\nonumber\\
=&\sum_\alpha{t_\alpha}^2G_{i\alpha}^\textrm{R}(E_k)G_{\alpha j}^\textrm{A}(E_k).
\end{align}
We therefore have
\begin{align}\label{eqC20}
&\int\frac{dk}{2\pi}\frac{\langle d_i | \psi_k \rangle \langle\tilde{\psi}_k | d_j \rangle}{E-E_k}
\nonumber\\
=&\sum_\alpha {t_\alpha}^2
\int\frac{dk}{2\pi}\frac{1}{E-E_k}
\nonumber\\
&\quad\times
\langle d_i|\frac{1}{\displaystyle E_k-H_\textrm{d}-\sum_\alpha ({t_\alpha}^2/t) \mathrm{e}^{ik}|d_\alpha\rangle\langle d_\alpha|} | d_\alpha\rangle
\nonumber\\
&\quad\times
\langle d_\alpha|\frac{1}{\displaystyle E_k-H_\textrm{d}-\sum_\alpha ({t_\alpha}^2/t) \mathrm{e}^{-ik}|d_\alpha\rangle\langle d_\alpha|} | d_j\rangle,
\end{align}
where $E_k=-t\left(\mathrm{e}^{ik}+\mathrm{e}^{-ik}\right)$, and we used eqs.~(\ref{eqB10}) and~\eqref{eqB32} for the Green's functions with the expression~(\ref{eqB165}) for the effective potential.

On the paths $C_\parallel^\textrm{R}(\kappa_0)$ and $C_\parallel^\textrm{A}(\kappa_0)$, we let $k=k_\textrm{r}\pm i \kappa_0$ and integrate with respect to $k_\textrm{r}$.
For $k=k_\textrm{r}+ i \kappa_0$, the element $\mathrm{e}^{-ik}$ grows to infinity in the limit $\kappa_0\to\infty$ in the three denominators on the right-hand side of eq.~(\ref{eqC20}).
Conversely, for $k=k_\textrm{r}- i \kappa_0$, the element $\mathrm{e}^{ik}$ grows to infinity in the limit $\kappa_0\to\infty$  in the three denominators on the right-hand side of eq.~(\ref{eqC20}). 
Therefore the integral~(\ref{eqC20}) vanishes on the paths $C_\parallel^\textrm{R}(\kappa_0)$ and $C_\parallel^\textrm{A}(\kappa_0)$ in the limit $\kappa_0\to\infty$.
Thus eq.~(\ref{eq:GA+GR_parallel}) is proved for the system~(\ref{eq:Hamiltonian_original}).

\section{The case $t_1=t_2=t$ with infinite eigenvalues}
\label{app:inf}

In the present Appendix, we will show the following fact mentioned near the end of \S~\ref{sec:spectrum}:
when the couplings between the quantum dot and the leads are equal, and are equal to the hopping energy of the leads, i.e., $t_1=t_2=t$, the effective Hamiltonian has two infinite eigenvalues; the contribution of these eigenvalues to the function $\Lambda$ in eq. (\ref{eq:expansionGRGA}) is a finite constant and is equal to
\begin{eqnarray} \label{Lambda_inf}
\Lambda_{\infty}(E)=  - {\check{H}_\textrm{d}}^{-1}
 \end{eqnarray}
for any finite energy $E$, where $\check{H}_\textrm{d}$ is the ``contact'' Hamiltonian, the part of the quantum-dot Hamiltonian that involves the sites in contact with the leads, spanned by the contact sites $|d_1\rangle$ and $|d_2\rangle$.
Note, however, that the above does \textit{not} apply to the case where the two leads are attached to one site $0$.
 
As we showed in Appendix\ref{app:selfenergy}, and particularly in eqs.~\eqref{eqB20}--\eqref{eqB31},
the matrix $E_n - H^\textrm{R}_{\rm{eff}}(E_n)$ is an $N$-by-$N$ matrix in the quantum-dot subspace consisting of $N$ sites, $\{d_i\}$:
\begin{eqnarray}\label{eqF20}
 E_n - H^\textrm{R}_{\rm{eff}}(E_n) 
=  \left( \begin{array}{cccc}
         E_n - \varepsilon_{1} + \frac{t_1^2}{t} z_n & v_{12} & v_{13} &\cdots \\ 
         v_{21} & E_n -\varepsilon_2 +  \frac{t_2^2}{t} z_n &  v_{23}  & \cdots\\
         v_{31} & v_{32} & E_n -\varepsilon_{3} & \cdots\\
         \vdots & \vdots & \vdots & \ddots 
        \end{array} \right),
\end{eqnarray}
\noindent
where
\begin{align}
z_n&=\mathrm{e}^{ik_n},\\
E_n&=-t\left(z_n+\frac{1}{z_n}\right). \label{EqF4_}
\end{align}
Note that we used the expression~\eqref{eqB165} here.
Since we set $t_1 = t_2$, we will introduce the single parameter
\begin{align}
\gamma =  t- \frac{{t_1}^2}{t} = t- \frac{{t_2}^2}{t}.
\end{align}
Then the matrix~\eqref{eqF20} becomes
\begin{eqnarray}\label{eqF40}
E_n - H^\textrm{R}_{\rm{eff}}(E_n) 
=  \left( \begin{array}{cccc}
         - z_n \gamma - \varepsilon_{1} -\frac{t}{z_n} & v_{12} & v_{13} & \cdots \\ 
         v_{21} &- z_n \gamma - \varepsilon_{2} -\frac{t}{z_n}&  v_{23}  & \cdots\\
         v_{31} & v_{32} & - t(z_n +\frac{1}{z_n})-\varepsilon_{3} &\cdots\\
         \vdots & \vdots & \vdots & \ddots
        \end{array} \right)
\end{eqnarray}
\noindent
because of eq.~\eqref{EqF4_}.
We will consider the limit $t_1=t_2\to t$, or $\gamma \to 0$.
Hereafter we will try to find values of $z_n$ that tend to infinity as $\gamma \to 0$ in such a way that the product $z_n \gamma$ remains finite. 
In the limit $|z_n|\to\infty$ we drop the terms $t/z_n$ in eq.~\eqref{eqF40} and have
\begin{align}\label{eqF50}
&\lim_{|z_n|\to\infty}\left( E_n - H^\textrm{R}_{\rm{eff}}(E_n)\right)
\nonumber\\
&\quad=  \left( \begin{array}{cccc}
         - z_n \gamma - \varepsilon_{1}  & v_{12} & v_{13} &\cdots \\ 
         v_{21} &- z_n \gamma - \varepsilon_{2} &  v_{23}  & \cdots\\
         v_{31} & v_{32} & - t z_n -\varepsilon_{3} &\cdots\\
         \vdots & \vdots & \vdots & \ddots
        \end{array} \right).
\end{align}
 
We will call $\check{C}$ (for ``contact'' matrix) the two-by-two upper-left matrix within the matrix~\eqref{eqF50}. Thus
\begin{eqnarray} \label{HC}
\check{C} &=&  \left( \begin{array}{cc}
         - z_n \gamma - \varepsilon_{1}  & v_{12}  \\ 
         v_{21} &- z_n \gamma - \varepsilon_{2} 
               \end{array} \right) \nonumber\\
    &=& -z_n\gamma \check{I}  - \check{H}_\textrm{d}
\end{eqnarray}
where $\check{I}$ is the two-by-two identity matrix and
\begin{eqnarray}
\check{H}_\textrm{d} &=&  \left( \begin{array}{cc}
          \varepsilon_{1}  & -v_{12}  \\ 
         -v_{21} &\varepsilon_{2} 
               \end{array} \right).
 \end{eqnarray}
(We here used the notation~\eqref{eqA310}.)
In the limit $|z_n| \to \infty$, we then have
\begin{eqnarray}\label{eqF80}
\lim_{|z_n|\to\infty}\det \left(E_n - H^\textrm{R}_{\rm{eff}}(E_n)\right) = \left(-tz_n\right)^{N-2}\det\check{C},
\end{eqnarray}
because in the subspace of $|d_3\rangle$ to $|d_N\rangle$ the diagonal elements $-tz_n$ dominate.
Since the determinant~\eqref{eqF80} must vanish as in eq.~\eqref{eqB171}, this implies that the two eigenvalues $z_n$ tending to infinity must be the solutions of the equation $\det\check{C} = 0$ and the corresponding eigenvectors must have the form
\begin{eqnarray}\label{eqF110}
|\psi_n\rangle =  \left( \begin{array}{c}
         \langle d_1|\psi_{n}\rangle  \\ 
         \langle d_2|\psi_{n}\rangle  \\
         0 \\
         0 \\
         \vdots
          \end{array} \right) 
\end{eqnarray}
with
\begin{align}
&\check{C} \left( \begin{array}{c}
         \langle d_1|\psi_{n}\rangle  \\ 
         \langle d_2|\psi_{n}\rangle  \\
         \end{array} \right) 
\nonumber\\
&\quad=  \left(-z_n\gamma \check{I}  -\check{H}_\textrm{d}\right) \left( \begin{array}{c}
         \langle d_1|\psi_{n}\rangle  \\ 
         \langle d_2|\psi_{n}\rangle  \\
          \end{array} \right)=0.
\end{align}
This shows that $-z_n \gamma$ must be the (real) eigenvalues of the contact Hamiltonian $\check{H}_\textrm{d}$.
Denoting the eigenvalues of $\check{H}_\textrm{d}$ by $\zeta_n$, we have $z_n = -\zeta_n/\gamma$ in the limit $\gamma\to 0$.
 
In the limit $\gamma\to 0$ (or $|z_n|\to\infty $) we have
\begin{align}
E-E_n=E+t\left(z_n+\frac{1}{z_n}\right) \stackrel{|z_n|\to\infty}{\longrightarrow} tz_n= -t\frac{\zeta_n }{\gamma}.
\end{align}
Hence the contribution to $\Lambda(E)$ in eq.~\eqref{eq:expansionGRGA} from the infinite eigenvalues $E_n$ reduces to
\begin{eqnarray} \label{Lambda2}
\Lambda_{\infty}(E) =  -\sum_{n=1,2} \frac{ |\psi_n\rangle\langle{\tilde\psi}_n|}{t\zeta_n/\gamma}.
 \end{eqnarray}
Let us here notice that the eigenstates $|\psi_n\rangle$ include the normalization constant
\begin{align} \label{Nn}
{\cal N}_n &= \sum_{i=1}^N \left|\langle d_i|\psi_n'\rangle\right|^2 
\nonumber\\
&+\frac{{z_n}^2}{1-{z_n}^2}
\left(\frac{t_1}{t}\right)^2
\sum_{\alpha=1,2}\left|\langle d_\alpha|\psi_n'\rangle\right|^2,
\end{align}
where $|\psi_n'\rangle$ is the non-normalized eigenstate of the Hamiltonian so that  $|\psi_n\rangle =  {\cal N}_n^{-1/2} |\psi_n'\rangle$. The second  term in \eqref{Nn} comes from the summation of the square modulus of eq.~\eqref{eqB700} over $x_\alpha$.
For the eigenstates with the infinite eigenvalues, eq.~(\ref{Nn}) reduces to
\begin{align} \label{Nn2}
\lim_{|z_n|\to\infty}{\cal N}_n&=
\left[1-\left(\frac{t_1}{t}\right)^2\right]\sum_{\alpha=1,2}\left|\langle d_\alpha|\psi_n'\rangle\right|^2
\nonumber\\
&= \frac{\gamma}{t} \sum_{\alpha=1,2} \left|\langle d_\alpha|\psi_n'\rangle\right|^2,
\end{align}
where we used eq.~\eqref{eqF110}.
This is the normalization constant for the eigenvectors of the total Hamiltonian $H$.
Introducing the eigenvectors 
\begin{eqnarray} \label{phi}
|\phi_n\rangle = \left(\sum_{\alpha=1,2} \left|\langle d_\alpha|\psi_n'\rangle\right|^2\right)^{-1/2} |\psi_n'\rangle,
\end{eqnarray}
which are normalized for the two-by-two contact Hamiltonian $\check{H}_\textrm{d}$, we have
\begin{eqnarray} \label{Lambda3}
\Lambda_{\infty}(E) = - \sum_n\frac{ |\phi_n\rangle\langle {\tilde \phi_n} |}{\zeta_n} = -{\check{H}_\textrm{d}}^{-1}.
\end{eqnarray}
This completes the proof.

\begin{figure}
\includegraphics[width=0.45\columnwidth]{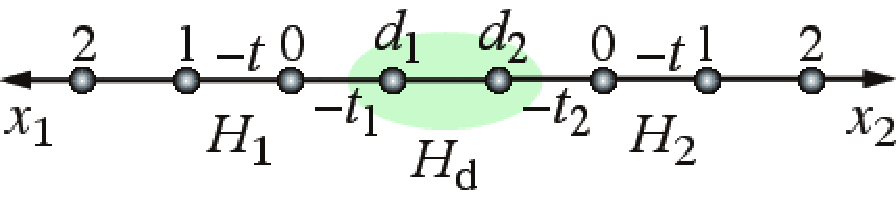}
\caption{(Color online) A system with the dot Hamiltonian of two sites.}
\label{fig:appF}
\end{figure}

Two comments are in order.
First, for a simple case shown in Fig.~\ref{fig:appF}, we can explain why the eigenvalues must tend to infinity.
When $t_1\neq t$ in Fig.~\ref{fig:appF}, the dot Hamiltonian $H_\textrm{d}$ consists of two sites ($N=2$) and hence the system must have $2N=4$ pieces of discrete eigenstates.
As $t_1\to t$, the site $d_1$ becomes a part of the lead and therefore the dot Hamiltonian $H_\textrm{d}$ now consists of only one site;
the system now must have only two pieces of discrete eigenstates.
Two eigenstates thereby must vanish when their corresponding eigenvalues go to infinity.

Second, the two eigenvalues that tend to infinity must correspond to anti-bound states because of the following reason.
As is shown above, the values of $z_n=-\zeta_n/\gamma$, and hence $E_n$, are both real.
Since it is impossible for the bound-state eigenenergy to tend to infinity just as $t\to t_1=t_2$, the only possibility is that they are both anti-bound states.


\begin{thebibliography}{100}

\bibitem{TB1993}
E.~Tekman and P.~F.~Bagwell:
{Phys.~Rev.~B} \textbf{48} (1993) 2553.

\bibitem{Datta95}
S.~Datta:
Electronic Transport in Mesoscopic Systems
(Cambridge University Press, Cambridge, 1995).

\bibitem{RW2001}
E.~R.~Racec and U.~Wulf:
{Phys.~Rev.~B} \textbf{64} (2001) 115318.

\bibitem{CWB2001}
A.~A.~Clerk, X.~Waintal, and P.~W.~Brouwer:
{Phys.~Rev.~Lett.} \textbf{86} (2001) 4636.

\bibitem{LLZ2005}
H.~Lu, R.~L{\"{u}}, and B.~F.~Zhu:
{Phys.~Rev.~B} \textbf{71} (2005) 235320.

\bibitem{Chakrabarti2006}
A.~Chakrabarti:
{Phys.~Rev.~B} \textbf{74} (2006) 205315.

\bibitem{JHS2007}
Y.~S.~Joe, E.~R.~Hedin, and A.~M.~Satanin:
{Phys.~Rev.~B} \textbf{76} (2007) 085419.

\bibitem{FN2008}
S.~Fujimoto and Y.~Natsume:
{J.~Phys.~Soc.~Jpn.} \textbf{77} (2008) 024712.

\bibitem{KAKI2002}
K.~Kobayashi, H.~Aikawa, S.~Katsumoto, and Y.~Iye:
{Phys.~Rev.~Lett.} \textbf{88} (2002) 256806.

\bibitem{KAKI2003}
K.~Kobayashi, H.~Aikawa, S.~Katsumoto, and Y.~Iye:
{Phys.~Rev.~B} \textbf{68} (2003) 235304.

\bibitem{SAKKI2004}
K.~Kobayashi, H.~Aikawa, A.~Sano, S.~Katsumoto, and Y.~Iye:
{Phys.~Rev.~B} \textbf{70} (2004) 035319.

\bibitem{SAKKI2005}
M.~Sato, H.~Aikawa, K.~Kobayashi, S.~Katsumoto, and Y.~Iye:
{Phys.~Rev.~Lett.} \textbf{95} (2005) 066801.

\bibitem{GGHKSMM2000}
J.~G{\"{o}}res, D.~Goldhaber-Gordon, S.~Heemeyer, M.~A.~Kastner, H.~Shtrikman,
  D.~Mahalu, and U.~Meirav:
{Phys.~Rev.~B} \textbf{62} (2000) 2188.

\bibitem{ZGGKK2001}
I.~G.~Zacharia, D.~Goldhaber-Gordon, G.~Granger, M.~A.~Kastner, Yu.~B.~Khavin,
  H.~Shtrikman, D.~Mahalu, and U.~Meirav:
{Phys.~Rev.~B} \textbf{64} (2001) 155311.

\bibitem{KKLPSKKYK2003}
J.~Kim, J.~R.~Kim, J.~O.~Lee, J.~W.~Park, H.~M.~So, N.~Kim, K.~Kang, K.~H.~Yoo,
  and J.~J.~Kim:
{Phys.~Rev.~Lett.} \textbf{90} (2003) 166403.

\bibitem{BS2004}
B.~Babi{\'{c}} and C.~Sch{\"{o}}nenberger:
{Phys.~Rev.~B} \textbf{70} (2004) 195408.

\bibitem{Brisker08}
D.~Brisker, I.~Cherkes, C.~Gnodtke, D.~Jarukanont, S.~Klaiman, W.~Koch,
  S.~Weissman, R.~Volkovich, M.~C.~Toroker, and U.~Peskin:
{Mol.~Phys.} \textbf{106} (2008) 281.

\bibitem{SH2005}
K.~Sasada and N.~Hatano:
{Physica E} \textbf{29} (2005) 609.

\bibitem{HSNP2007}
N.~Hatano, K.~Sasada, H.~Nakamura, and T.~Petrosky:
{Prog.~Theor.~Phys.} \textbf{119} (2008) 187.

\bibitem{Gamow28}
G.~Gamow:
{Z.~Phys.} \textbf{51} (1928) 204.

\bibitem{Siegert1939}
A.~J.~F.~Siegert:
{Phys.~Rev.} \textbf{56} (1939) 750.

\bibitem{Peierls59}
R.~E.~Peierls:
{Proc.~Roy.~Soc.~London A} \textbf{253} (1959) 16.

\bibitem{leCouteur60}
K.~J.~le~Couteur:
{Proc.~Roy.~Soc.~London A} \textbf{256} (1960) 115.

\bibitem{Zeldovich60}
Ya.~B.~Zel'dovich:
{Zh.~Eksp.~Teor.~Fiz.} \textbf{39} (1960) 776
[Sov.~Phys.~JETP {\textbf{12}} (1961) 542].

\bibitem{Hokkyo65}
N.~Hokkyo:
{Prog.~Theor.~Phys.} \textbf{33} (1965) 1116.

\bibitem{Romo68}
W.~J.~Romo:
{Nucl.~Phys.~A} \textbf{116} (1968) 617.

\bibitem{Berggren70}
T.~Berggren:
{Phys.~Lett.} \textbf{33B} (1970) 547.

\bibitem{Gyarmati71}
B.~Gyarmati and T.~Vertse:
{Nucl.~Phys.~A} \textbf{160} (1971) 523.

\bibitem{Romo80}
W.~J.~Romo:
{J.~Math.~Phys.} \textbf{21} (1980) 311.

\bibitem{Berggren82}
T.~Berggren:
{Nucl.~Phys.~A} \textbf{389} (1982) 261.

\bibitem{Berggren96}
T.~Berggren:
{Phys.~Lett.~B} \textbf{373} (1996) 1.

\bibitem{Madrid05}
R.~de~la Madrid, G.~Garc{\'{i}}a-Calder{\'{o}}n, and J.~G.~Muga:
{Czech.~J.~Phys.} \textbf{55} (2005) 1141.

\bibitem{Eckart30}
C.~Eckart:
{Phys.~Rev.} \textbf{35} (1930) 1303.

\bibitem{Bethe36}
H.~A.~Bethe and R.~F.~Bacher:
{Rev.~Mod.~Phys.} \textbf{8} (1936) 82.

\bibitem{Hulthen42a}
L.~Hulth{\'{e}}n:
{Arkiv Matematik Astronomi Fysik} \textbf{28A} (1942) 5.

\bibitem{Hulthen42b}
L.~Hulth{\'{e}}n:
{Arkiv Mat.~Astr.~Fysik} \textbf{29B} (1942) 1.

\bibitem{Jost51}
R.~Jost and A.~Pais:
{Phys.~Rev.} \textbf{82} (1951) 840.

\bibitem{Vogt54}
E.~Vogt and G.~H.~Wannier:
{Phys.~Rev.} \textbf{95} (1954) 1190.

\bibitem{Wigner55}
E.~P.~Wigner:
{Phys.~Rev.} \textbf{98} (1955) 145.

\bibitem{Corinaldesi56}
E.~Corinaldesi:
{Nucl.~Phys.} \textbf{2} (1956) 420.

\bibitem{Nussenzveig59}
H.~M.~Nussenzveig:
{Nucl.~Phys.} \textbf{11} (1959) 499.

\bibitem{Fivel60}
D.~I.~Fivel and A.~Klein:
{J.~Math.~Phys.} \textbf{1} (1960) 274.

\bibitem{Humblet61}
J.~Humblet and L.~Rosenfeld:
{Nucl.~Phys.} \textbf{26} (1961) 529.

\bibitem{Rosenfeld61}
L.~Rosenfeld:
{Nucl.~Phys.} \textbf{26} (1961) 579.

\bibitem{Humblet62}
J.~Humblet:
{Nucl.~Phys.} \textbf{31} (1962) 544.

\bibitem{Humblet64-1}
J.~Humblet:
{Nucl.~Phys.} \textbf{50} (1964) 1.

\bibitem{Jeukenne64}
J.~P.~Jeukenne:
{Nucl.~Phys.} \textbf{58} (1964) 1.

\bibitem{Humblet64-2}
J.~Humblet:
{Nucl.~Phys.} \textbf{57} (1964) 386.

\bibitem{Mahaux65}
C.~Mahaux:
{Nucl.~Phys.} \textbf{68} (1965) 481.

\bibitem{Rosenfeld65}
L.~Rosenfeld:
{Nucl.~Phys.} \textbf{70} (1965) 1.

\bibitem{Bhattacharjie62}
A.~Bhattacharjie and E.~C.~G.~Sudarshan:
{Il Nuovo Cimento} \textbf{25} (1962) 864.

\bibitem{Wojtczak63}
L.~Wojtczak:
{Nucl.~Phys.} \textbf{48} (1963) 325.

\bibitem{Spector64}
R.~M.~Spector:
{J.~Math.~Phys.} \textbf{5} (1964) 1185.

\bibitem{Bose64}
A.~K.~Bose:
{Il Nuovo Cimento} \textbf{32} (1964) 679.

\bibitem{Aly65}
H.~H.~Aly and R.~M.~Spector:
{Il Nuovo Cimento} \textbf{38} (1965) 149.

\bibitem{McVoy67}
K.~W.~McVoy, L.~Heller, and M.~Bolsterli:
{Rev.~Mod.~Phys.} \textbf{39} (1967) 245.

\bibitem{Bahethi71}
O.~P.~Bahethi and M.~G.~Fuda:
{J.~Math.~Phys.} \textbf{12} (1971) 2076.

\bibitem{Fuda71}
M.~G.~Fuda:
{J.~Math.~Phys.} \textbf{12} (1971) 1163.

\bibitem{Bawin74}
M.~Bawin and J.~P.~Lavine:
{Il Nuovo Cimento} \textbf{23A} (1974) 311.

\bibitem{Doolen78}
G.~D.~Doolen:
{Int.~J.~Quant.~Chem.} \textbf{14} (1978) 523.

\bibitem{Narnhofer81}
H.~Narnhofer and W.~Thirring:
{Phys.~Rev.~A} \textbf{23} (1981) 1688.

\bibitem{Rittby82}
M.~Rittby, N.~Elander, and E.~Br{\"{a}}ndas:
{Mol.~Phys.} \textbf{45} (1982) 553.

\bibitem{Alhassid85}
Y.~Alhassid, F.~Iachello, and R.~D.~Levine:
{Phys.~Rev.~Lett.} \textbf{54} (1985) 1746.

\bibitem{Massmann85}
H.~Massmann:
{Am.~J.~Phys.} \textbf{53} (1985) 679.

\bibitem{Colbert86}
D.~T.~Colbert, R.~Mayrhofer, and P.~R.~Certain:
{Phys.~Rev.~A} \textbf{33} (1986) 3560.

\bibitem{Benjamin86}
I.~Benjamin and R.~D.~Levine:
{Phys.~Rev.~A} \textbf{33} (1986) 2833.

\bibitem{Amrein87}
W.~O.~Amrein and M.~B.~Cibils:
{Helv.~Phys.~Acta} \textbf{60} (1987) 481.

\bibitem{Bohm89}
A.~Bohm, M.~Gadella, and G.~B.~Mainland:
{Am.~J.~Phys.} \textbf{57} (1989) 1103.

\bibitem{Ginocchio94}
J.~N.~Ginocchio:
{Ann.~Phys.} \textbf{152} (1984) 203.

\bibitem{Rakityansky96}
S.~A.~Rakityansky, S.~A.~Sofianos, and K.~Amos:
{Il Nuovo Cimento} \textbf{111B} (1996) 363.

\bibitem{Homma97}
M.~Homma, T.~Myo, and K.~Kat{\={o}}:
{Prog.~Theor.~Phys.} \textbf{97} (1997) 561.

\bibitem{Masui99}
H.~Masui, S.~Aoyama, T.~Myo, and K.~Kat{\={o}}:
{Prog.~Theor.~Phys.} \textbf{102} (1999) 1119.

\bibitem{Barkay01}
H.~Barkay and N.~Moiseyev:
{Phys.~Rev.~A} \textbf{64} (2001) 044702.

\bibitem{Carvalho02}
C.~A.~A.~de~Carvalho and H.~M.~Nussenzveig:
{Phys.~Rep.} \textbf{364} (2002) 83.

\bibitem{Razavy03}
M.~Razavy:
Quantum Theory of Tunneling
(World Scientific, Singapore, 2003).

\bibitem{Ahmed04}
Z.~Ahmed and S.~R.~Jain:
{J.~Phys.~A.~Math.~Gen.} \textbf{37} (2004) 867.

\bibitem{Kelkar04}
N.~G.~Kelkar, M.~Nowakowski, K.~P.~Khemchandani, and S.~R.~Jain:
{Nucl.~Phys.~A} \textbf{730} (2004) 121.

\bibitem{Jain05}
S.~R.~Jain:
{Phys.~Lett.~A} \textbf{335} (2005) 83.

\bibitem{Amrein06}
W.~O.~Amrein and K.~B.~Sinha:
{J.~Phys.~A.~Math.~Gen.} \textbf{39} (2006) 9231.

\bibitem{Moiseyev08}
N.~Moiseyev, M.~{\v{S}}indelka, and L.~S.~Cederbaum:
{J.~Phys.~B.~At.~Mol.~Opt.~Phys.} \textbf{41} (2008) 221001.

\bibitem{Rotter09}
I.~Rotter:
{J.~Phys.~A.~Math.~Theor.} \textbf{42} (2009) 153001.

\bibitem{Fano61}
U.~Fano:
{Phys.~Rev.} \textbf{124} (1961) 1866.

\bibitem{Presilla97}
C.~Presilla and J.~Sj{\"{o}}strand:
{Phys.~Rev.~B} \textbf{55} (1997) 9310.

\bibitem{Nishino09}
A.~Nishino, T.~Imamura, and N.~Hatano:
Phys.~Rev.~Lett. \textbf{102} (2009) 146803.

\bibitem{Imamura09}
T.~Imamura, A.~Nishino, and N.~Hatano:
Phys.~Rev.~B \textbf{80} (2009) 245323.

\bibitem{Nishino11}
A.~Nishino, T.~Imamura, and N.~Hatano:
Phys.~Rev.~B \textbf{83} (2011) 035306.

\bibitem{Magunov03}
A.~I.~Magunov, I.~Rotter, and S.~I.~Strakhova:
{J.~Phys.~B.~At.~Mol.~Opt.~Phys.} \textbf{36} (2003) L401.

\bibitem{Berggren68}
T.~Berggren:
{Nucl.~Phys.} \textbf{A109} (1968) 265.

\bibitem{Newton1961}
R.~G.~Newton:
Scattering Theory of Waves and Particles
(Springer-Verlag, New York, 1982) 2nd ed.

\bibitem{endnote}
After submitting an earlier version of the present paper, we became aware of works where the
  transmission amplitude of systems in continuum space is given as an expansion
  with respect to resonant
  eigen-wave-numbers without back-ground integrals~\cite{Tolstikhin01,Ostrovsky05,Klaiman10}.
  This approach was recently extended to the present discretized system by authors including one of the present author (N.H.)~\cite{Klaiman11}.

\bibitem{Tolstikhin01}
O.~I.~Tolstikhi, V.~N.~Ostrovsky, and H.~Nakamura:
Phys.~Rev.~A \textbf{63} (2001) 042707.

\bibitem{Ostrovsky05}
V.~N.~Ostrovsky and N.~Elander:
Phys.~Rev.~A \textbf{71} (2005) 052707.

\bibitem{Klaiman10}
S.~Klaiman and N.~Moiseyev:
J.~Phys.~B \textbf{43} (2010) 185205.

\bibitem{Klaiman11}
S.~Klaiman and N.~Hatano:
J.~Chem.~Phys. \textbf{134} (2011) 154111.

\bibitem{Kim01}
T.-S.~Kim and S.~Hershfield:
{Phys.~Rev.~B} \textbf{63} (2001) 245326.

\bibitem{Kikoin01}
K.~Kikoin and Y.~Avishai:
{Phys.~Rev.~Lett.} \textbf{86} (2001) 2090.

\bibitem{Kang01}
K.~Kang, S.~Y.~Cho, J.-J.~Kim, and S.-C.~Shin:
{Phys.~Rev.~B} \textbf{63} (2001) 113304.

\bibitem{Affleck01}
I.~Affleck and P.~Simon:
{Phys.~Rev.~Lett.} \textbf{86} (2001) 2854.

\bibitem{Simon01}
P.~Simon and I.~Affleck:
{Phys.~Rev.~B} \textbf{64} (2001) 085308.

\bibitem{Affleck07}
I.~Affleck and E.~S.~S{\o}rensen:
{Phys.~Rev.~B} \textbf{75} (2007) 165316.

\bibitem{Torio02}
M.~E.~Torio, K.~Hallberg, A.~H.~Ceccatto, and C.~R.~Proetto:
{Phys.~Rev.~B} \textbf{65} (2002) 085302.

\bibitem{Orellana03a}
P.~A.~Orellana, F.~Dom{\'{\i}}nguez-Adame, I.~G{\'{o}}mez, and
  M.~L.~{Ladr{\'{o}}n} de~Guevara:
{Phys.~Rev.~B} \textbf{67} (2003) 085321.

\bibitem{Orellana03b}
P.~A.~Orellana, M.~L.~{Ladr{\'{o}}n} de~Guevara, M.~Pacheco, and
  A.~Latg{\'{e}}:
{Phys.~Rev.~B} \textbf{68} (2003) 195321.

\bibitem{Rodriguez03}
A.~Rodr{\'{\i}}guez, F.~Dom{\'{\i}}nguez-Adame, I.~G{\'{o}}mez, and P.~A.~Orellana:
{Phys.~Lett.~A} \textbf{320} (2003) 242.

\bibitem{MSU2004}
I.~Maruyama, N.~Shibata, and K.~Ueda:
{J.~Phys.~Soc.~Jpn.} \textbf{73} (2004) 3239.

\bibitem{Aligia04}
A.~A.~Aligia and L.~A.~Salguero:
{Phys.~Rev.~B} \textbf{70} (2004) 075307.

\bibitem{Tanaka05}
Y.~Tanaka and N.~Kawakami:
{Phys.~Rev.~B} \textbf{72} (2005) 085304.

\bibitem{Lara05}
J.~M.~Y{\'a}{\~{n}}ez G.~A.~Lara, P.~Orellana and E.~V.~Anda:
{Solid State Comm.} \textbf{136} (2005) 323.

\bibitem{Franco06}
R.~Franco, M.~S.~Figueira, and E.~V.~Anda:
{Phys.~Rev.~B} \textbf{73} (2006) 195305.

\bibitem{Wang06}
R.~Wang and J.-Q.~Liang:
{Phys.~Rev.~B} \textbf{74} (2006) 144302.

\bibitem{Chakrabarti06}
A.~Chakrabarti:
{Phys.~Rev.~B} \textbf{74} (2006) 205315.

\bibitem{Zitko06}
R.~{\v{Z}}itko and J.~Bon{\v{c}}a:
{Phys.~Rev.~B} \textbf{73} (2006) 035332.

\bibitem{Li08}
T.~C.~Li and S.-P.~Lu:
{Phys.~Rev.~B} \textbf{77} (2008) 085408.

\bibitem{Porod92}
W.~Porod, Z.~Shao, and C.~S.~Lent:
{Appl.~Phys.~Lett.} \textbf{61} (1992) 1350.

\bibitem{Porod93}
W.~Porod, Z.~A.~Shao, and C.~S.~Lent:
{Phys.~Rev.~B} \textbf{48} (1993) 8495.

\bibitem{Shao94}
Z.~A.~Shao, W.~Porod, and C.~S.~Lent:
{Phys.~Rev.~B} \textbf{49} (1994) 7453.

\bibitem{Presilla96}
C.~Presilla and J.~Sj{\"{o}}strand:
{J.~Math.~Phys.} \textbf{37} (1996) 4816.

\bibitem{RBBM1997}
T.~N.~Rescigno, M.~Baertschy, D.~Byrum, and C.~W.~McCurdy:
{Phys.~Rev.~A} \textbf{55} (1997) 4253.

\bibitem{Landau77}
L.~D.~Landau and E.~M.~Lifshitz:
Quantum Mechanics (Non-relativistic Theory)
(Pergamon Press, Oxford, 1977) 3rd ed.

\bibitem{Shapiro06}
H.~Kunz and B.~Shapiro:
{J.~Phys.~A.~Math.~Gen.} \textbf{39} (2006) 10155.

\bibitem{Shapiro08}
H.~Kunz and B.~Shapiro:
{Phys.~Rev.~B} \textbf{77} (2008) 054203.

\bibitem{Nakanishi58}
N.~Nakanishi:
{Prog.~Theor.~Phys.} \textbf{19} (1958) 607.

\bibitem{PPT1991}
T.~Petrosky, I.~Prigogine, and S.~Tasaki:
{Physica A} \textbf{173} (1991) 175.

\bibitem{Ohanian74}
H.~O.~Ohanian and C.~G.~Ginsburg:
{Am.~J.~Phys.} \textbf{42} (1974) 310.

\bibitem{Friedrichs1948}
K.~O.~Friedrichs:
{Commun.~Pure Appl.~Math.} \textbf{1} (1948) 361.

\bibitem{Anderson1961}
P.~W.~Anderson:
{Phys.~Rev.} \textbf{124} (1961) 41.

\bibitem{Sudershan1962}
E.~C.~G.~Sudershan:
{Structure of Dynamical Theories}
(W.~A.~Benjamin, New York, 1962).

\bibitem{OPP2001}
G.~Ordonez, T.~Petrosky, and I.~Prigogine:
{Phys.~Rev.~A} \textbf{63} (2001) 052106.

\bibitem{Miyamoto2004}
M.~Miyamoto:
{Phys.~Rev.~A} \textbf{70} (2004) 032108.

\bibitem{Miyamoto2005}
M.~Miyamoto:
{Phys.~Rev.~A} \textbf{72} (2005) 063405.

\bibitem{FL1981}
D.~S.~Fisher and P.~A.~Lee:
{Phys.~Rev.~B} \textbf{23} (1981) 6851.

\bibitem{Khalfin}
L.~A.~Khalfin:
{Zh.~Eksp.~Teor.~Fiz.} \textbf{33} (1957) 1371
[Sov.~Phys.~JETP \textbf{6} (1958) 1053].

\bibitem{Misra}
B.~Misra and E.~C.~G.~Sudarshan:
{J.~Math.~Phys.} \textbf{18} (1977) 756.

\bibitem{Petrosky}
T.~Petrosky, G.~Ordonez, and I.~Prigogine:
{Phys.~Rev.~A} \textbf{64} (2001) 062101.

\bibitem{Peierls32} R. Peierls: {Z.~Phys.} \textbf{80} (1933) 763.

\bibitem{SH2008}
K.~Sasada and N.~Hatano:
{J.~Phys.~Soc.~Jpn.} \textbf{77} (2008) 025003.

\bibitem{Livshits57}
M.~S.~Livshits:
{Zh.~Eksp.~Teor.~Fiz.} \textbf{31} (1956) 121
[Sov.~Phys.~JETP, 4, 91-98 (1957)].

\bibitem{Feshbach58}
H.~Feshbach:
{Ann.~Phys.~(New York)} \textbf{5} (1958) 357.

\bibitem{Feshbach62}
H.~Feshbach:
{Ann.~Phys.~(New York)} \textbf{19} (1962) 287.

\bibitem{Albeverio96}
S.~Albeverio, F.~Haake, P.~Kurasov, M.~Ku{\'{s}}, and P.~{\v{S}}eba:
{J.~Math.~Phys.} \textbf{37} (1996) 4888.

\bibitem{Fyodorov97}
Y.~V.~Fyodorov and H.-J.~Sommers:
{J.~Math.~Phys.} \textbf{38} (1997) 1918.

\bibitem{Dittes00}
F.-M.~Dittes:
{Phys.~Rep.} \textbf{339} (2000) 215.

\bibitem{Pichugin01}
K.~Pichugin, H.~Schanz, and P.~{\v{S}}eba:
{Phys.~Rev.~E} \textbf{64} (2001) 056227.

\bibitem{Sadreev03}
A.~F.~Sadreev and I.~Rotter:
{J.~Phys.~A.~Math.~Gen.} \textbf{36} (2003) 11413.

\bibitem{Okolowicz03}
J.~Oko{\l}owicz, M.~P{\l}oszajczak, and I.~Rotter:
{Phys.~Rep.} \textbf{374} (2003) 271.


\end{thebibliography}
\end{document}